\documentclass[11pt]{article}

\usepackage{fullpage} 

\usepackage{latexsym,amsmath,amssymb,amsfonts,amsthm,graphicx,graphics,epsfig,epsf,lscape,subfigure}

\usepackage[normalem]{ulem}

\newcommand{\bei}{\begin{itemize}}
\newcommand{\eei}{\end{itemize}}
\newcommand{\bes}{\begin{small}}
\newcommand{\ees}{\end{small}}
\newcommand{\bec}{\begin{center}}
\newcommand{\eec}{\end{center}}

\newcommand{\beq}{\begin{equation}}
\newcommand{\eeq}{\end{equation}}
\newcommand{\beqn}{\begin{equation*}}
\newcommand{\eeqn}{\end{equation*}}
\newcommand{\beqr}{\begin{eqnarray}}
\newcommand{\eeqr}{\end{eqnarray}}
\newcommand{\beqrn}{\begin{eqnarray*}}
\newcommand{\eeqrn}{\end{eqnarray*}}
\newcommand{\brr}{\begin{array}}
\newcommand{\err}{\end{array}}
\newcommand{\bef}{\begin{figure}}
\newcommand{\eef}{\end{figure}}

\newcommand{\eqdef}{\stackrel{\rm{def}}{=}}

\newtheorem{prop}{Proposition}[section]

\newcommand{\subscr}[2]{#1_{\textup{#2}}}
\newcommand{\supscr}[2]{#1^{\textup{#2}}}

\newcommand{\map}[3]{#1: #2 \rightarrow #3}

\def \prob{\mathbb{P}}
\def \expt {\mathbb{E}}
\def \mc {\mathcal}

\def \real{\mathbb{R}}

\def \bs {\boldsymbol}

\DeclareMathOperator{\csch}{csch}

\DeclareMathOperator{\sinch}{sinch}

\graphicspath{{./fig/}}

\usepackage{color}

\definecolor{mygreen}{rgb}{0,0.9,0}

\title{Explicit moments of decision times for single- and
double-threshold drift-diffusion processes}

\author{V. Srivastava$^{1}$, P. Holmes$^{1,2}$ and P. Simen$^{3}$ \\
$^{1}$Department of Mechanical and Aerospace Engineering, \\
$^{2}$Program in Applied and Computational Mathematics and
Princeton Neuroscience Institute \\
Princeton University, NJ 08544. \\
$^{3}$Department of Neuroscience, Oberlin College, OH 44704. }

\begin{document}

\maketitle

\begin{abstract}

We derive expressions for the first three moments of the decision time
(DT) distribution produced via first threshold crossings by sample paths
of a drift-diffusion equation. The ``pure'' and ``extended'' diffusion
processes are widely used to model two-alternative forced choice
decisions, and, while simple formulae for accuracy, mean DT and
coefficient of variation are readily available, third and higher
moments and conditioned moments are not generally available. We
provide explicit formulae for these, describe their behaviors as drift rates and
starting points approach interesting limits, and, with the support of
numerical simulations, discuss how trial-to-trial variability of drift
rates, starting points, and non-decision times affect these behaviors
in the extended diffusion model. Both unconditioned moments and those
conditioned on correct and erroneous responses are treated. We argue
that the results will assist in exploring mechanisms of evidence
accumulation and in fitting parameters to experimental data.

\vspace{0.3cm}  

\textbf{Classification:} Decision theory

\textbf{Running title:} Explicit moments for diffusion processes 

\end{abstract}

\section{Introduction}
\label{s.intro}

In this paper we derive explicit expressions for the mean, variance,
coefficient of variation and skewness of decision times (DTs)
predicted by the stochastic differential equation (SDE)
\beq
dx = a \, dt + \sigma \, dW, \ x(0) = x_0 ,
\label{e.dd1}
\eeq
which models accumulation of the difference $x(t)$ between the streams of
evidence in two-alternative forced-choice tasks. 
 An example of such a perceptual decision-making task is one in which a participant determines if the image on the screen has more white or black pixels (e.g., \cite{RatcliffRouderExtendedDDM})
Here
drift rate $a$ and standard deviation $\sigma$ are constants, $dW$ denotes independent random (Wiener) increments, and $dx$ is the change in evidence during the time interval $(t, t+ dt)$. Decision times (DTs) are determined by first passages
through upper and lower thresholds $x = +z$ and $-z$ that respectively correspond to correct responses and errors, between which the starting point $x_0$ is assumed to lie. Thus, without loss of generality we may set $a > 0$, although we will also consider limits $a \to 0$. Predictions of response times (RTs) for comparison to
behavioral data are obtained by adding to DTs a non-decision latency,
$\subscr{\rm T}{nd}$, to account for sensory and motor processes.

SDEs like Eqn.~\eqref{e.dd1} are variously called diffusion or
drift-diffusion models (DDMs); in \cite{boga03} Eqn.~\eqref{e.dd1} was
named the pure DDM to distinguish it from Ratcliff's extended
diffusion model \cite{Rat78}, which allows trial to trial variability
in drift rates and starting points $x_0$. See
\cite{Rat78,RatSmith04,boga03} for background on diffusion models, and
note that several different variable-naming conventions are used in
parameterizing DDMs, e.g. in \cite{Rat78,RatSmith04,Wagenm-JMPsy05}
$v$ and $s$ replace $a$ and $\sigma$, and thresholds are set at $x =
0$ and $x = a$ with $x_0 \in [0, a]$; in \cite{boga03} $a$ and
$\sigma$ are named $A$ and $c$.

Many of the following results have appeared in the stochastic process
literature, or are implicit in it, and some have appeared in the
psychological literature (e.g.
\cite{Rat78,Wagenm-JMPsy05,grasman2009mean}). However, their
dependence on key parameters such as threshold and starting point and
behaviors in the limits of low and high drift rates have not been
fully explored (see \cite{Wagenm-JMPsy05} for some cases of $a \to
0$). Nor are we aware of explicit derivations of third order
moments. Here we provide these, and also prove a Proposition that
describes the structure of the coefficient of variation (CV) for  DTs 
predicted by Eqn.~\eqref{e.dd1}, relating it to
the CV for a single-threshold DDM. 
We summarize the expressions for moments of decision times in Table~\ref{tab:summary}.
The MatLab and R code for these expressions is available at: \verb|https://github.com/PrincetonUniversity/higher_moments_ddm|.
We end by considering the extended
DDM, introduced in \cite{Rat78}, showing how trial-to-trial
variability of drift rates and starting points affects the results for
the pure DDM and examining the effects of non-decision latency
on response times.

\subsection*{Notation and units}

We start by reviewing definitions and dimensional units and establishing notation.
For a random variable $\xi$, we define the $n$-th non-central moment by $\expt[\xi^n]$ and the $n$-th central moment by $\expt[(\xi - \expt[\xi])^n]$. The first central moment is zero and the second central moment is the variance. The coefficient of variation (CV) of $\xi$ is defined as the ratio of standard deviation to mean of $\xi$, i.e., $\rm CV = \sqrt{\expt[(\xi - \expt[\xi])^2]}/\expt[\xi]$. Similarly, the skewness of $\xi$ is defined as the ratio of the third central moment to the cube of the standard deviation of $\xi$: 
\[
\rm skew = \frac{\expt[(\xi -\expt[\xi])^3]}{\expt[(\xi - \expt[\xi])^2]^{3/2}}.
\]
The variable $x(t)$ and thresholds $\pm z$ in Eqn.~\eqref{e.dd1} are dimensionless, while the parameters $a$ and $\sigma$ have dimensions [time]$^{-1}$ and [time]$^{-\frac{1}{2}}$ respectively. 
When providing numerical examples we will work in secs.
For $a>0$ we define the
normalized threshold $k_z$ and  starting point $k_x$:
\beq
k_z = \frac{a z}{\sigma^2} \ge 0 \ \mbox{ and } \ 
k_x = \frac{a x_0}{\sigma^2} \in (-k_z, k_z) ; 
\label{e.ks}
\eeq
these nondimensional parameters will allow us to give relatively
compact expressions.

\section{The single-threshold DDM}
\label{ss.sddm}

Eqn.~\eqref{e.dd1} with a single upper threshold $z > 0$ necessarily produces only correct responses in decision tasks, but it is of interest because it provides a simple approximation of the two-threshold DDM when accuracy is at ceiling and errors due to passages through the lower threshold are rare. Specifically, for $a>0$, DTs of this model with starting point $x_0$ are described by the Wald (inverse-Gaussian) distribution~\cite[Eq. (2.0.2)]{ANB-PS:02},\cite{Wald-bk47,Luce-bk86}. 

\beq 
p(t) = \frac{z- x_0}{\sigma}\sqrt{\frac{1}{2 \pi t^3}} \exp \left( \frac{-(z -x_0 - a t )^2}{2 \sigma^2 t} \right) 
\label{e.wald1} 
\eeq 

The mean DT, its variance, and CV are:
\beq
\mathbb{E}[{\rm DT}]  = \frac{\sigma^2}{a^2 }(k_z -k_x), \ \  \mathrm{Var}[{\rm DT}] = \frac{\sigma^4}{a^4} (k_z -k_x),
\ \mbox{and} \ \ {\rm CV} = \frac{\sqrt{\mathrm{Var}[{\rm DT}]}} {\mathbb{E}[{\rm DT}]}
= \frac{1}{\sqrt{k_z- k_x}}, 
\label{e.wald2}
\eeq
and the skewness is
\beq
\frac{3}{\sqrt{k_z- k_x}} \ \ (= 3 \, {\rm CV} ).
\label{e.wald3}
\eeq
In the limit $a \to 0^+$, the distribution~\eqref{e.wald1} converges to the L{\' e}vy distribution, and in this limit none of the moments exist. However, as shown below, moments of the double threshold DDM exist in this limit.

The single threshold process has been proposed as a model for interval timing \cite{Simenetal-JNSci11,LuzardoLudvigRivest,BalciSimenActa2014,SimenVP}.   Interval timing, loosely defined, is the capacity either to make a response or judgment at a specific time relative to some event in the environment, or simply to estimate inter-event durations. Classic timing tasks include ``production'' tasks, such as the Fixed Interval (FI) task, in which a participant receives a reward for any response produced after a delay of a given duration since the last reward was received \cite{FersterSkinner}, and discrimination tasks, in which two different stimulus durations are compared to see which is longer (see \cite{Creelman1962} and \cite{TreismanTiming1963} for historical reviews of early human timing research). Production tasks can be modeled similarly to decision tasks by a diffusion model: instead of accumulating evidence about a perceptual choice, a timing diffusion model accumulates a steady ``clock signal'' toward a threshold for responding \cite{Creelman1962,GibbonChurchMeck1984,KilleenFettermanBehavioralTiming,TreismanTiming1963}. The resulting production times, relative to stimulus onset, are then comparable to perceptual decision-making response times, typically yielding a slightly positively skewed Gaussian density \cite{GibbonChurchCognition1990}. 
 Simen~\emph{et al.}~\cite{SimenRivestLudvigBalciKilleen2013} show that the single-threshold DDM can fit RT data from a variety of interval timing experiments when the starting point is set to 0, drift is set equal to threshold over duration  ($a = z/T$, with $T =$ target duration), and  normalized thresholds $k_z$ are set to high values, typically of order $20$ (see \cite{Simenetal-JNSci11}).
In contrast, $k_z$ is usually much lower in fits of typical two-choice decision data, typically of order 1. Noise $\sigma$ is typically fixed at 0.1 in the  literature \cite{VandekTuerl2007} and fitted thresholds typically range from 0.05 to 0.15; see e.g. \cite{Balci+APP11,Bogacz+QJExpP10,DutilWag-Psychon09,RatSuperDiscrim2014}. Despite this difference,  DDM can be fitted to both two-choice decision RTs and timed production RTs in humans  with suitably larger thresholds for timing \cite{SimenVP}, suggesting that both tasks may be accomplished by similar accumulation  processes.


\section{The double-threshold DDM: Unconditional moments of decision time}
\label{ss.dddm}
 
 We now turn to the two-threshold DDM and derive unconditional moments of decision time. 
The DT distribution for the double-threshold DDM may be expressed as a
convergent series \cite[Appendix]{Rat78}, and successive moments of
the unconditional DT (i.e. averaged over correct responses and errors)
may be obtained by solving boundary value problems for a sequence of
linear ordinary differential equations (ODEs) derived from the
backwards Fokker-Planck or Kolmogorov equation \cite[Chap.~5]{gard09}.

\subsection{Error rate and expected decision time}

The expressions for error rate and mean decision time are well known,
although the following forms are more compact than those given in
\cite{boga03}, for example:
\beqr
\mathrm{ER} &=& \frac{e^{-2k_x} - e^{-2k_z}}{e^{2k_z} - e^{-2k_z}} ,
\label{e.er} \\
\mathbb{E}[{\rm DT}] &=& \frac{\sigma^2}{a^2} \left[ k_z \coth(2k_z)
 - k_z e^{-2k_x} \csch(2k_z) - k_x \right] .
\label{e.mdt}
\eeqr
In Appendix~\ref{app-mean} we show that these expressions agree with the analogous ones of \cite{boga03}.

\noindent
For an unbiased starting point $k_x=0$ the mean decision time 
becomes
\begin{equation}
\mathbb{E}[{\rm DT}] = \frac{\sigma^2 k_z}{a^2} \tanh(k_z) ,
\label{e.mdtx0}
\end{equation} 
and in the limit $a \to 0$ ($k_z \to 0, k_x \to 0$) we have
\beq
\mathrm{ER} = \frac{k_z -k_x}{2k_z} = \frac{z-x_0}{2z}
\ \mbox{ and } \
\mathbb{E}[{\rm DT}] = \frac{\sigma^2 (k_z^2 - k_x^2)}{a^2}
 = \frac{z^2 - x_0^2}{\sigma^2} . 
\label{e.mdt0}
\eeq
Expressions for the error rate and unconditional moments of decision time are illustrated in Figs.~\ref{fig:ddm-numerics-1}~and~\ref{fig:ddm-numerics-2} below.

\subsection{Variance and coefficient of variation of  decision time}

We derive the following expression for the unconditional variance of decision time in Appendix~\ref{app-variance}:
\begin{align}
\mathrm{Var} &= \frac{\sigma^4}{a^4} \left[3 k_z^2 \csch^2(2k_z)
 - 2 k_z^2 e^{-2 k_x} \csch(2k_z) \coth(2k_z) - 4 k_z k_x e^{-2k_x}
 \csch(2k_z) \right. \nonumber \\ 
 & \left. - k_z^2 e^{-4 k_x} \csch^2(2k_z) + k_z \coth(2k_z) 
 - k_z e^{-2k_x} \csch(2k_z) - k_x \right] .
\label{e.var}
\end{align}

For an unbiased starting point $k_x = 0$ Eqn.~\eqref{e.var} reduces to 
\begin{align}
\mathrm{Var} &=  \frac{\sigma^4}{a^4} \left[ 2k_z^2 ( \csch^2(2k_z)
 - \csch(2k_z) \coth(2k_z) ) + k_z ( \coth(2k_z) - \csch(2k_z) ) \right]
\nonumber \\
&= \frac{\sigma^4}{a^4} \left[ k_z \tanh(k_z) - k_z^2 \mathrm{sech}^2(k_z) \right] 
 = \frac{\sigma^4}{a^4} \left[ \frac{k_z(1 - 4k_z e^{-2k_z} - e^{-4k_z})}
{(1 + e^{-2k_z})^2} \right]
\label{e.varkx0}
\end{align}
(cf.~\cite[Eqns.~(10-12)]{Wagenm-JMPsy05}), and in the limit $a = 0$
we have
\beq
\mathrm{Var} = \frac{2 \sigma^4 (k_z^4 - k_x^4)}{3 a^4} 
 = \frac{2(z^4 - x_0^4)}{3 \sigma^4} \, . 
\label{e.var0}
\eeq

The coefficient of variation can be determined from 
Eqns.~\eqref{e.var} and \eqref{e.mdt}:
\beq \mathrm{CV} = \frac{[\mathrm{Var}]^\frac{1}{2}}{\mathbb{E}[{\rm DT}]}
 = \frac{ \left[ 3 k_z^2 \csch^2(2k_z) - 2 k_z^2 e^{-2 k_x}
 \csch(2k_z) \coth(2k_z) -  \ldots - k_x \right]^\frac{1}{2}}
 {k_z \coth(2k_z) - k_z e^{-2k_x} \csch(2k_z) - k_x} ;
\label{e.cv}
\eeq
the complete numerator appears in brackets in Eqn.~\eqref{e.var}.
For $k_x = 0$ Eqn.~\eqref{e.cv} reduces to
\beq
\mathrm{CV} = \sqrt{\frac{1 - 2 k_z \csch(2k_z)}{k_z 
[ \coth(2k_z) - \csch(2k_z) ]}} = \sqrt{ \frac{1 - 4k_z e^{-2k_z}
 - e^{-4k_z}}{k_z(1 - e^{-2k_z})^2}} \, ,  
\label{e.cvx0}
\eeq
and in the case $a = 0$, from Eqs.~\eqref{e.var0} and \eqref{e.mdt0}
we have
\beq
\mathrm{CV} = \sqrt{\frac{2 (z^2 + x_0^2)}{3 (z^2 - x_0^2)}} 
\ \to \sqrt{\frac{2}{3}} \ \mbox{ as } \ z \to \infty \ 
\mbox{ or } \ x_0 \to 0 . 
\label{e.cv0}
\eeq
Note that the multiplicative factors $\sigma^2/a^2$ cancel and that
CV depends only upon the nondimensional threshold and starting point
$k_z, k_x$ (or $x_0/z$ in case $a = 0$).

If $a > 0$, as the threshold $z$ increases, $\mathbb{E}[{\rm DT}]$ and
Var both increase, but CV decreases, with the following behaviors in
the limit $z \to \infty$ ($k_z \to \infty$) for $k_x$ fixed:
\beq
\frac{\mathbb{E}[{\rm DT}]}{k_z }\to \frac{\sigma^2}{a^2}  , \quad 
\frac{\mathrm{Var}}{k_z } \to \frac{\sigma^4}{a^4}  \quad 
\mbox{and} \ \mathrm{CV} \to k_z ^{-\frac{1}{2}} ;
\label{e.zinf}
\eeq
these behaviors follow from the facts that $k_z^m \csch^n (2k_z) \sim
k_z^m e^{-2nk_z}$ and $\coth(2k_z) \sim 1$. For $a = 0$,
$\mathbb{E}[{\rm DT}]$ and Var also increase with $z$, as one sees from
Eqns.~\eqref{e.mdt0} and \eqref{e.var0}, but CV approaches the limit
$\sqrt{2/3}$ (Eqn.~\eqref{e.cv0}). In \S\ref{s.cvs} we describe the
behavior of the CV with unbiased starting point $k_x = 0$ throughout
the range $k_z \in (0,\infty)$, and show that the CV of the
single threshold DDM provides an upper bound for  Eqn.~\eqref{e.cvx0}.

\subsection{Third moment and skewness of decision time}

We end this section by computing the expression for skewness.  The third moment of decision time can be computed by solving a boundary value problem analogous to that in Appendix~\ref{app-variance}. 
However, this computation is very tedious. Instead we obtain
skewness from the non-central third moments of DTs conditioned on
correct responses and errors derived in \S\ref{s.cond} below (this
also illustrates the relationships between unconditioned and
conditioned moments). Introducing the notation $\tau$ for DT, the
non-central third moments can be written as 
\begin{align}
\mathbb{E}[\tau^3| x(\tau)=z] &= \mathrm{Skew}_{+} \mathrm{Var}_{+}^{3/2}
 + 3 \mathrm{Var}_{+} \, \mathbb{E}[{\rm DT}]_{+} + \mathbb{E}[{\rm DT}]_{+}^3,
 \quad \text{and} \label{e.nc1} \\
\mathbb{E}[\tau^3| x(\tau)=-z] &= \mathrm{Skew}_{-} \mathrm{Var}_{-}^{3/2}
 + 3 \mathrm{Var}_{-} \, \mathbb{E}[{\rm DT}]_{-} + \mathbb{E}[{\rm DT}]_{-}^3,
\label{e.nc2}
\end{align}
where $\mathbb{E}[{\rm DT}]_\pm$, $\mathrm{Var}_\pm$,
$\mathrm{Skew}_\pm$ denote expected value, variance, and skewness of
DT conditioned on correct responses and errors, respectively. Summing
appropriate fractions of these conditional moments gives the 
unconditioned third moment
\beq
\mathbb{E}[\tau^3] = (1 - \text{ER}) \times \mathbb{E}[\tau^3| x(\tau)=z]
 + \text{ER} \times \mathbb{E}[\tau^3| x(\tau)=- z] ,
\label{e.nc3}
\eeq
from which skewness can be derived as follows:
\beq
\mathrm{Skew} = \mathbb{E} \left[ \left( \frac{\tau - \mathbb{E}[{\rm DT}]}
 {\mathrm{Var}^{\frac{1}{2}}} \right)^3 \right] = \frac{\mathbb{E}[\tau^3]
 - 3 \mathrm{Var} \, \mathbb{E}[{\rm DT}]
 - \mathbb{E}[{\rm DT}]^3}{\mathrm{Var}^\frac{3}{2}} .
\label{e.skew}
\eeq

Substituting the expressions~\eqref{e.er} for ER and \eqref{eq:edtc1},
\eqref{eq:evarc1} and \eqref{eq:skwc1} for conditional moments
from~\S\ref{s.cond} into Eqns.~(\ref{e.nc1}-\ref{e.nc3}), and using
the expressions \eqref{e.mdt}~and~\eqref{e.var} for the mean and
variance of DT, we obtain
\begin{align}
&\mathbb{E}[\tau^3] - 3 \mathrm{Var} \, \mathbb{E}[{\rm DT}] - \mathbb{E}[{\rm DT}]^3 \nonumber \\
&= \frac{\sigma^6}{a^6} \bigg[\Big((24 k_x k_z^2 + 6 k_z^2 - 12 k_z^3) e^{-2 k_z - 4k_x} + (24 k_x^2 k_z + 24 k_x k_z - 16 k_z^3 + 6 k_z)e^{- 2k_x} \nonumber \\  
&- (12 k_x^2 k_z + 12 k_x k_z^2 + 12 k_x k_z + 4 k_z^3 + 6 k_z^2 + 3 k_z) e^{4 k_z - 2 k_x}  
  - (24 k_x k_z^2 + 6 k_z^2 + 12 k_z^3) e^{2 k_z - 4 k_x} \nonumber  \\ 
  &- 8 k_z^3 e^{- 6k_x} - 3 k_z \cosh(2 k_z)  + 3 k_z \cosh(6 k_z) 
   + 9 k_x  \sinh(2 k_z) - 3 k_x  \sinh(6 k_z) + 56 k_z^3  \cosh(2 k_z)  \nonumber  \\
   & + 36 k_z^2 \sinh(2 k_z) 
   -( 3 k_z - 6 k_z^2   + 4 k_z^3  + 12 k_x k_z  - 12 k_x k_z^2  + 12 k_x^2 k_z)  e^{-4k_z-2 k_x} 
\Big) \frac{\csch^3(2k_z)}{4 }  \bigg].
\label{e.num.skew}
\end{align}
Finally, skewness may be obtained by substituting Eqns.~\eqref{e.var}
and (\ref{e.num.skew}) into Eqn.~\eqref{e.skew}. After substitution,
the $\sigma^6 / a^6$ factors cancel out so that, like CV,
skewness depends only on $k_z$ and $k_x$.

For an unbiased starting point $x_0=k_x=0$, Eqn.~\eqref{e.num.skew} can be simplified to
\begin{align}
\mathbb{E}[\tau^3] - 3 \mathrm{Var} \, \mathbb{E}[{\rm DT}] - \mathbb{E}[{\rm DT}]^3 
&= \frac{\sigma^6}{a^6} \Big[ 3 k_z \tanh(k_z) -3 k_z^2 \mathrm{sech}^2(k_z) - 2 k_z^3 \tanh(k_z) \mathrm{sech}^2(k_z) \Big].  \label{e.num.skew0}
\end{align}
We also note that the limits of the double-threshold moments approach
those of the single-threshold moments as $k_z \to \infty$
with $k_x$ fixed. Specifically:
\beq
\frac{\mathbb{E}[{\rm DT}]}{k_z} \to \frac{\sigma^2}{a^2} , \
\frac{\mathrm{Var}}{k_z^2}\to \frac{\sigma^4}{a^4}  , \
\mathrm{CV} \to  k_z^{-\frac{1}{2}} \quad 
\mbox{and} \quad \mathrm{Skew} \to
3  k_z^{-\frac{1}{2}}  = 3 \, \mathrm{CV} .
\label{e.zinf2}
\eeq
In the limit $a =0$, we obtain
\begin{align}\label{eq:driftless-skew1}
\mathbb{E}[\tau^3] - 3 \mathrm{Var} \, \mathbb{E}[{\rm DT}]
 - \mathbb{E}[{\rm DT}]^3 
& =  \frac{16(z^6 -x_0^6)}{\sigma^6}, \quad \text{and} \quad {\rm Skew}
 = \sqrt{\frac{96}{25} } \frac{(z^6 -x_0^6)}{(z^4 - x_0^4)^{3/2}},
\end{align}
and the skewness to CV ratio is $12/5$ as $z \to \infty$ or $x_0 \to 0$.

Two further limits are of interest, those in which the starting point
approaches either threshold: $k_x \to \pm k_z$ with $k_z$
fixed and finite. In this case ER $\to 0$ or 1, $\mathbb{E}[{\rm DT}]$
$\to 0$,  CV $\to \infty$, and Skew $\to \infty$. Letting $k_x = \pm 
k_z(1 - \epsilon)$ and expanding for small $\epsilon \ge 0$, we have
\begin{align}
\mathbb{E}[{\rm DT}] &= \frac{\sigma^2}{a^2}  \left[ k_z \coth(2k_z)
 - k_z e^{ \mp 2k_z (1 - \epsilon)} \csch(2k_z) \mp k_z (1 - \epsilon) \right]
  \nonumber \\ 
 &= \frac{\sigma^2}{a^2} \left[\pm 1 - \frac{4 k_z}{e^{\pm 4 k_z} - 1} \right]
 (k_z \mp k_x) + {\mathcal{O}}(| k_z \mp k_x |^2) \to 0^+ .
\label{e.kx-kz}
\end{align}
Similarly, for the variance and third central moment, we have
\begin{align}
\mathrm{Var} &= \frac{\sigma^2}{a^2}  \left[\frac{\mp 8 k_z^2(1+3 e^{\pm 4 k_z})}{(e^{\pm 4 k_z}-1)^2} + \frac{4 k_z}{e^{\pm 4 k_z}-1} \pm 1\right]
 (k_z \mp k_x) + {\mathcal{O}}(| k_z \mp k_x |^2) \to 0^+ ,
\label{e.var.kx-kz} \\ 
\expt[(\tau -\expt[\tau])^3]& = 
\mp \frac{\sigma^3}{a^3} \left[ 18 \sinh(2 k_z) - 6 \sinh(6k_z) +
 e^{\pm 2k_z}(112k_z^3 -12 k_z) + 24 k_z e^{\mp k_z} \right. \nonumber \\ 
& \left.  -12 k_z e^{-6 k_z} + 256 k_z^3 e^{\mp 2 k_z} +16 k_z^3 e^{\mp 6 k_z} \right] (k_z \mp k_x) + {\mathcal{O}}(| k_z \mp k_x |^2) \to 0^+, \label{e.skew.kx-kz} 
\end{align}
so that both CV and skewness diverge like $| k_z \mp k_x |^{-1/2}$. However, the ratio of skewness to CV remains finite as $k_x \to \pm k_z$.

Examples of the functions $\mathbb{E}[{\rm DT}]$, Var, CV,
skewness and the third central moment of DT are plotted vs. threshold
$z$ in the left hand columns of Figs.~\ref{fig:ddm-numerics-1} and
\ref{fig:ddm-numerics-2} below.

\section{The double-threshold DDM: Conditional moments of decision time} \label{s.cond}

We now turn to moments of DTs conditioned on correct and
incorrect responses, deriving them from cumulant and moment generating
functions using a method detailed in Appendix~\ref{app-cumgenfctn}
that requires only successive differentiation (see \cite[Chap 4,
\S6]{Gut:07} and \cite[\S2.6]{gard09}). It suffices to consider only
correct decisions, because the moments conditioned on errors can be
obtained by replacing $x_0$ by $-x_0$, or equivalently, $k_x$ by
$-k_x$ in the following expressions, as demonstrated by the moment
generating functions \eqref{eq:mgfc} and \eqref{eq:mgfe} in
Appendix~\ref{app-cumgenfctn}.  The following expressions for the conditional moments of decision time are illustrated in Figs.~\ref{fig:ddm-numerics-1}~and~\ref{fig:ddm-numerics-2}.

\subsection{Conditional cumulant generating function and expected decision time}

As derived there from
Eqn.~\eqref{eq:mgfc}, the cumulant-generating function of DTs conditioned on correct decisions is 
\begin{equation}
\label{eq:cgfc}
K_+(\alpha)       
= C(a,\sigma,z,x_0) + \log \sinh\Big(\frac{(z+x_0)\sqrt{a^2 -2 \alpha \sigma^2}}{\sigma^2}\Big) - \log \sinh\Big(\frac{2 z\sqrt{a^2 -2 \alpha \sigma^2}}{\sigma^2}\Big),
\end{equation}
where $C(a,\sigma,z,x_0)$ is a function independent of $\alpha$ that
will disappear when the cumulants are computed by successive
differentiation of $K_+(\alpha)$ with respect to $\alpha$.

The expected DT conditioned on correct decisions is the first derivative
of $K_+(\alpha)$ evaluated at $\alpha=0$: 
\begin{align}
\mathbb{E}[{\rm DT}]_+ = \expt[\tau | x(\tau) =z] &= \frac{d}{d \alpha} K_+(\alpha) \Big|_{\alpha=0} = \frac{2z}{a} \coth\Big( \frac{2 a z}{\sigma^2} \Big) - \frac{z+x_0}{a} \coth\Big( \frac{ a (z +x_0)}{\sigma^2} \Big)
\nonumber \\
& = \frac{\sigma^2}{a^2} \Big( 2k_z \coth(2 k_z) - (k_x + k_z) \coth(k_x+k_z)\Big) ,
\label{eq:edtc1}
\end{align}
and it can be verified that in the limit $a \to 0^+$
\begin{equation}
\label{eq:edtc2}
\mathbb{E}[{\rm DT}]_+ = \frac{4z^2 - (z+x_0)^2}{3 \sigma^2}.
\end{equation}

\subsection{Conditional variance and coefficient of variation of decision time}
The variance of DT conditioned on correct decisions is the second
derivative of $K_+(\alpha)$ at $\alpha=0$:
\begin{align}
\mathrm{Var}_+ &= \text{Var}[\tau | x(\tau) =z] = \frac{d^2}{d \alpha^2}
K_+ (\alpha) \Big|_{\alpha=0} \nonumber \\
&= 
\frac{4z^2}{a^2} \csch^2 \Big( \frac{2 z a}{\sigma^2} \Big) + \frac{2 \sigma^2 z}{a^3} \coth\Big( \frac{2 z a}{\sigma^2 }\Big) 
- \frac{(z+ x_0)^2}{a^2} \csch^2 \Big( \frac{a (z+x_0)}{\sigma^2}\Big) \nonumber \\
& \qquad \qquad \qquad \qquad \qquad \qquad  \qquad \qquad \qquad \qquad \qquad
 - \frac{\sigma^2 (z+x_0)}{a^3} \coth\Big( \frac{a(z+x_0)}{\sigma^2 } \Big) \nonumber \\
& = \frac{\sigma^4}{a^4} \left[ 4 k_z^2 \csch^2(2 k_z) + 2 k_z \coth(2k_z) -(k_x + k_z)^2 \csch^2(k_x + k_z) - (k_x + k_z) \coth(k_x + k_z) \right] ;
\label{eq:evarc1}
\end{align}
in the limit $a \to 0^+$: 
\begin{equation}
\label{eq:evarc2}
\mathrm{Var}_+ = \frac{32z^4 - 2(z+x_0)^4}{45 \sigma^4}.
\end{equation}

The CV of DT conditioned on correct decisions is therefore
\begin{align}
\mathrm{CV}_+ &= \frac{\mathrm{Var_+}^{\frac{1}{2}}}{\mathbb{E}[{\rm DT}]_+} = \frac{\left[4 k_z^2 \csch^2(2 k_z) + 2 k_z \coth(2k_z) -(k_x + k_z)^2 \csch^2(k_x + k_z) - (k_x + k_z) \coth(k_x + k_z) \right]^{1/2}}{2k_z \coth(2 k_z) - (k_x + k_z) \coth(k_x+k_z)};
\label{eq:cvc}
\end{align}
again, the factors $\sigma^2 / a^2$ cancel and the conditional CV
depends only on $k_z$ and $k_x$.

As in \S\ref{ss.dddm} Eqns.~(\ref{e.kx-kz}-\ref{e.var.kx-kz}), it can
be shown that $\mathrm{CV}_+$ diverges as $k_x \to k_z$ (and hence, by
the $k_x \leftrightarrow -k_x$ symmetry, $\mathrm{CV}_-$ diverges as
$k_x \to -k_z$). However, the behavior as $k_x \to -k_z$ is more
interesting and quite subtle, especially as $k_z$ also becomes
small. 
To study this double limit we first set $k_x = \beta k_z$, where $\beta \in (-1,1)$, and expand the hyperbolic functions in Taylor
series for $k_z \ll 1$ (e.g. \cite[Eqns.(4.5.65-66]{abra84}) to obtain 
\begin{align}
\mathrm{CV}_+ 
& = \frac{\left[ \frac{2}{45} (\beta^2 + 2 \beta +5) (3 -2\beta -\beta^2) k_z^4 + O(k_z^6) \right]^{1/2}}{\frac{1}{3} (3 -2\beta - \beta^2) k_z^2 + O(k_z^4)} \nonumber \\
&= \left[  \frac{2(\beta^2 + 2 \beta +5)  + O(k_z^2) }{5(3 -2\beta - \beta^2) + O(k_z^2)}\right]^{1/2}. 
\label{eq:cvclim0}
\end{align}
It follows that 
\begin{equation}
\mathrm{CV}_+ \to \sqrt{\frac{2(\beta^2 + 2 \beta +5)}{5(3- \beta^2 -2\beta)}}
 \ \mbox{ as } \ k_z \to 0^+ . 
\label{eq:cvclim2}
\end{equation}
In these distinguished limits, $\mathrm{CV}_+$ can approach any value
in the range $(\sqrt{2/5}, \infty)$. For $\beta = 0$ ($k_x = 0$) the
starting point is unbiased (or $a = 0$), and we obtain the limit
$\mathrm{CV}_+ = \sqrt{2/3}$, as for the unconditioned CV; cf.
Eqn.~\eqref{e.cv0} and see Proposition~\ref{CVprop} below. For $\beta
\to  1^-$ the starting point lies on the correct threshold and
$\mathrm{CV}_+$ diverges as noted above. Aspects of this limiting
behavior are illustrated in Fig.~\ref{fig:cov} below.

\subsection{Conditional third moment and skewness of decision time}
The third central moment of DT conditioned on correct decisions is the third
derivative of $K_+(\alpha)$, evaluated at $\alpha=0$. 
The skewness of DT is obtained  by dividing the third central moment with
the cube of standard deviation. 
Thus, the third central moment of DT is
\begin{align}
\mathrm{Skew}_+& \mathrm{Var}_+^{\frac{3}{2}} = \mathrm{Var}_+^{\frac{3}{2}}\times \text{skewness}[\tau | x(\tau) =z] = \frac{d^3}{d \alpha^3} K_+(\alpha) \Big|_{\alpha=0} \nonumber \\
& = \frac{12\sigma^2 z^2}{a^4} \csch^2\Big( \frac{2az}{\sigma^2} \Big) + 
\frac{16z^3}{a^3} \coth\Big( \frac{2az}{\sigma^2} \Big)\csch^2 \Big( \frac{2az}{\sigma^2}\Big) + \frac{6\sigma^4 z}{a^5} \coth \Big( \frac{2az}{\sigma^2} \Big) \nonumber \\
&  - \frac{3\sigma^2 (z+x_0)^2}{a^4} \csch^2\Big( \frac{a(z+x_0)}{\sigma^2} \Big) 
 - \frac{2(z+x_0)^3}{a^3} \coth\Big( \frac{a(z+x_0)}{\sigma^2} \Big)\csch^2 \Big( \frac{a(z+x_0)}{\sigma^2}\Big) \nonumber \\
& - \frac{3\sigma^4 (z+x_0)}{a^5} \coth \Big( \frac{a(z+x_0)}{\sigma^2} \Big) \nonumber \\
& = \frac{\sigma^6}{a^6} \left[ 12 k_z^2 \csch^2 ( 2 k_z) + 16 k_z^3 \coth( 2k_z)\csch^2(2 k_z) + 6k_z\coth ( 2 k_z) - 3(k_z+ k_x)^2 \csch^2( k_z + k_x) \right. \nonumber \\
& \qquad \qquad  \left.  - 2(k_z + k_x)^3 \coth(k_z+k_x) \csch^2 (k_z+k_x) - 3(k_z+k_x) \coth (k_z+k_x) \right] .
\label{eq:skwc1}
\end{align}
An expression for $\mathrm{Skew}_+$ is obtained by dividing
Eqn.~\eqref{eq:skwc1} by the $3/2$ power of Eqn.~\eqref{eq:evarc1}.
In the limit $a \to 0^+$ Eqn.~\eqref{eq:skwc1} becomes
\begin{equation}
\label{eq:skwc2}
\mathrm{Skew}_+ \mathrm{Var}_+^{\frac{3}{2}}  = 
\frac{1024 z^6-16(z+x_0)^6}{945\sigma^6} \ \mbox{ and } \ 
\mathrm{Skew}_+ = \sqrt{\frac{45}{2}} \left[
\frac{8 (64z^6 - (z+x_0)^6)}{21 (16 z^4 - (z+x_0)^4)^{\frac{3}{2}}} \right].
\end{equation}

Similar to $\rm{CV}_+$, $\rm{Skew}_+$ diverges as $k_x \to k_z$. For 
$k_x =  \beta k_z$ and $\beta \in (-1,1)$,
\begin{equation}\label{eq:skew-wrong-side-alpha}
{\rm{Skew}_+} \to \frac{4\sqrt{10}}{7} \frac{(\beta^2 +3)(\beta^2 + 4 \beta +7)}{(\beta^2+ 2 \beta +5)^{3/2}(3 -2\beta -\beta^2)^{1/2}}, \quad \text{as} \quad k_z \to 0^+.
\end{equation}
In these distinguished limits $\rm{Skew}_+$ can approach any value
in the range $(4 \sqrt{10}/7, \infty)$.

\begin{figure}[ht!]
\subfigure[Expected DT]{\includegraphics[width=0.33\textwidth]{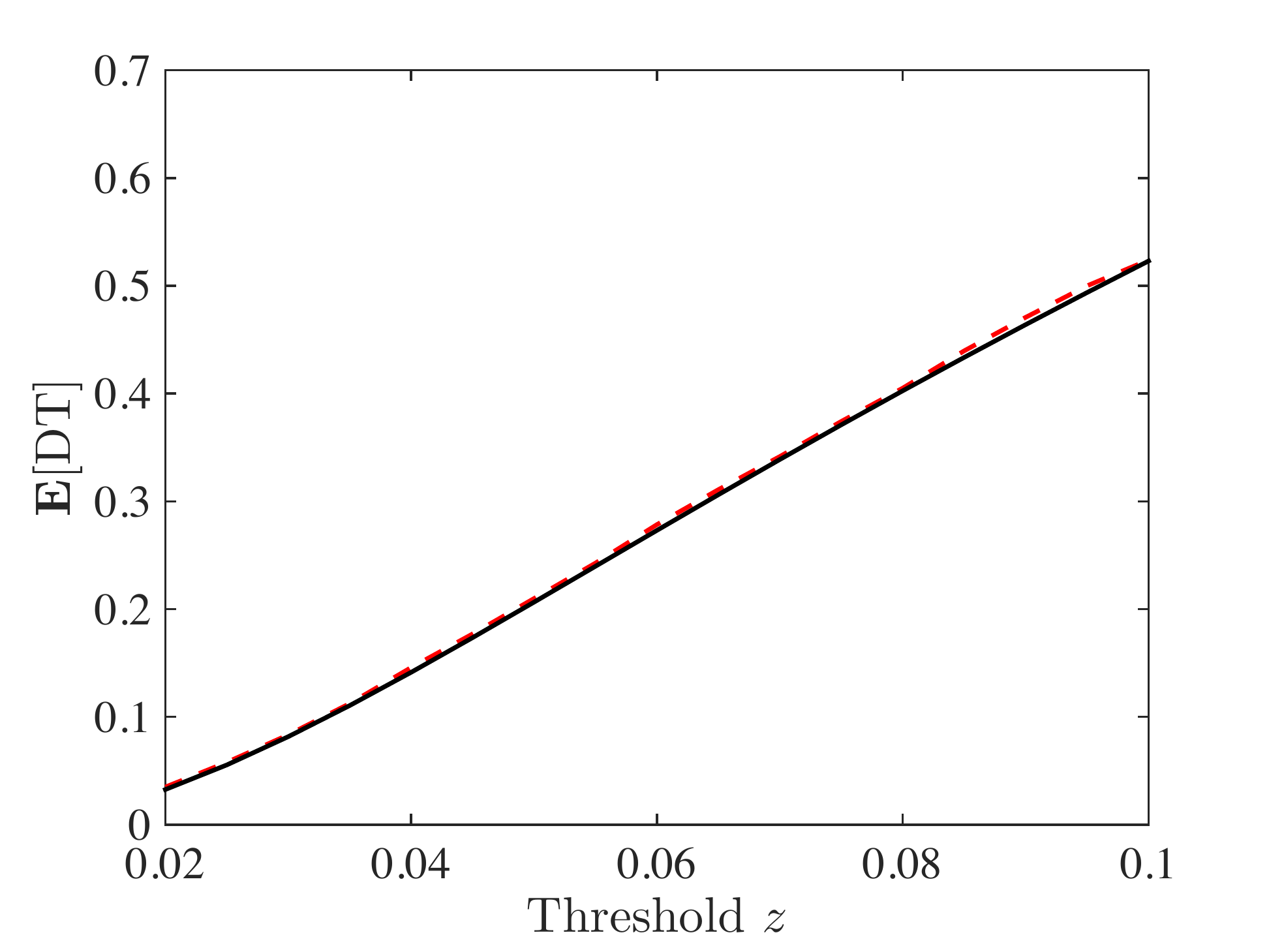}}
\subfigure[Expected DT conditioned on correct decision]{\includegraphics[width=0.33\textwidth]{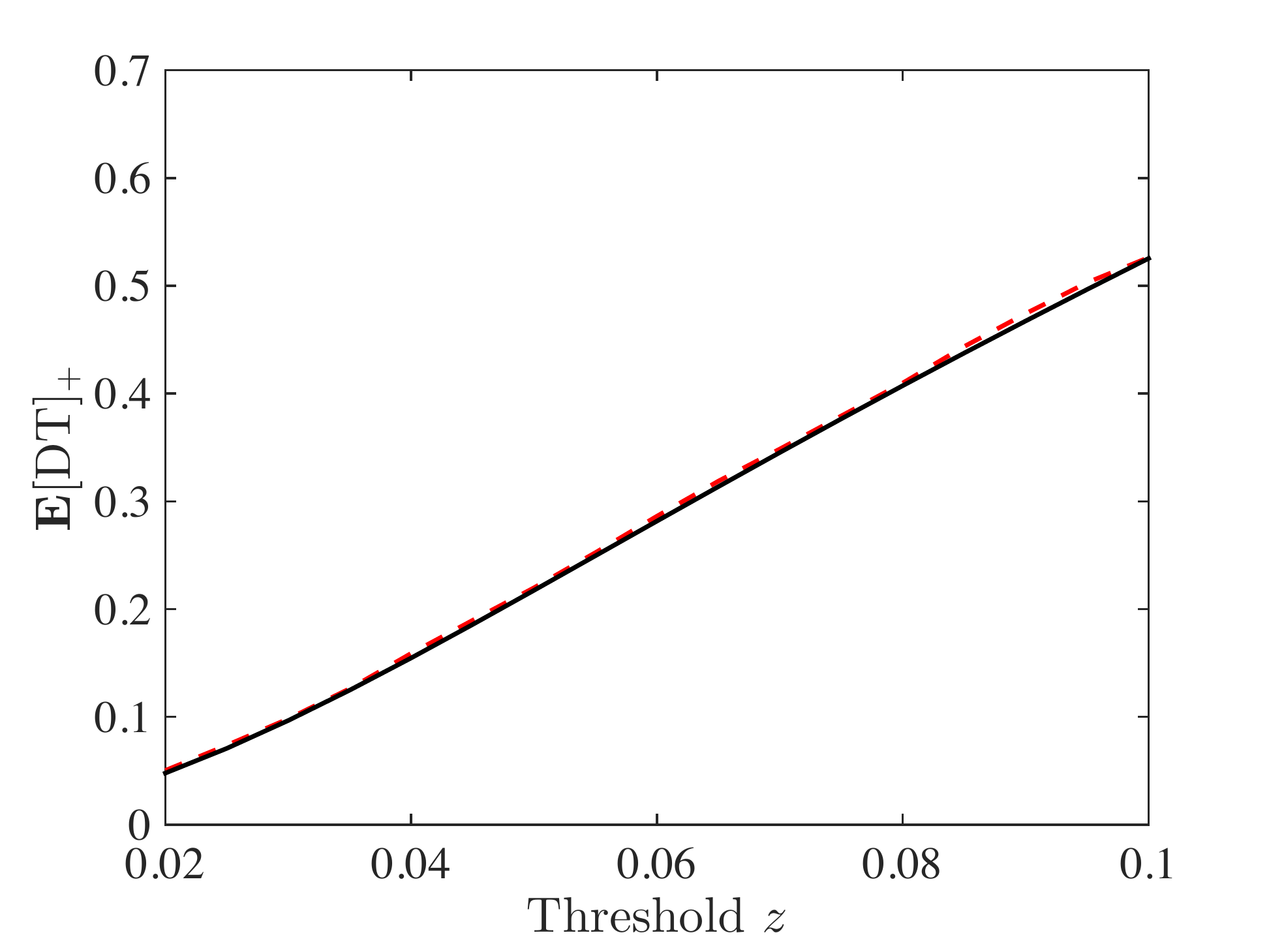}}
\subfigure[Expected DT conditioned on error]{\includegraphics[width=0.33\textwidth]{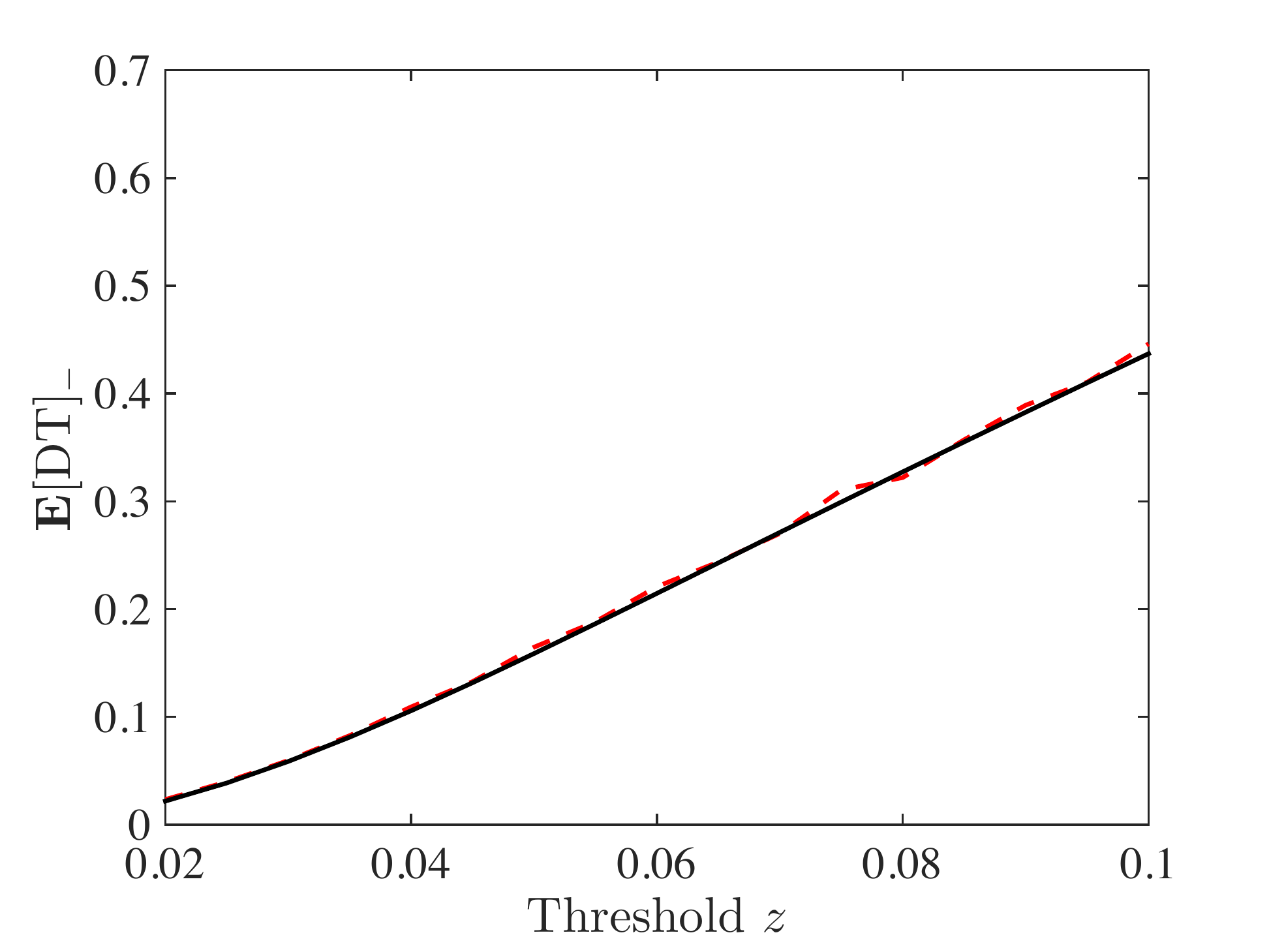}}

\subfigure[Variance of DT]{\includegraphics[width=0.33\textwidth]{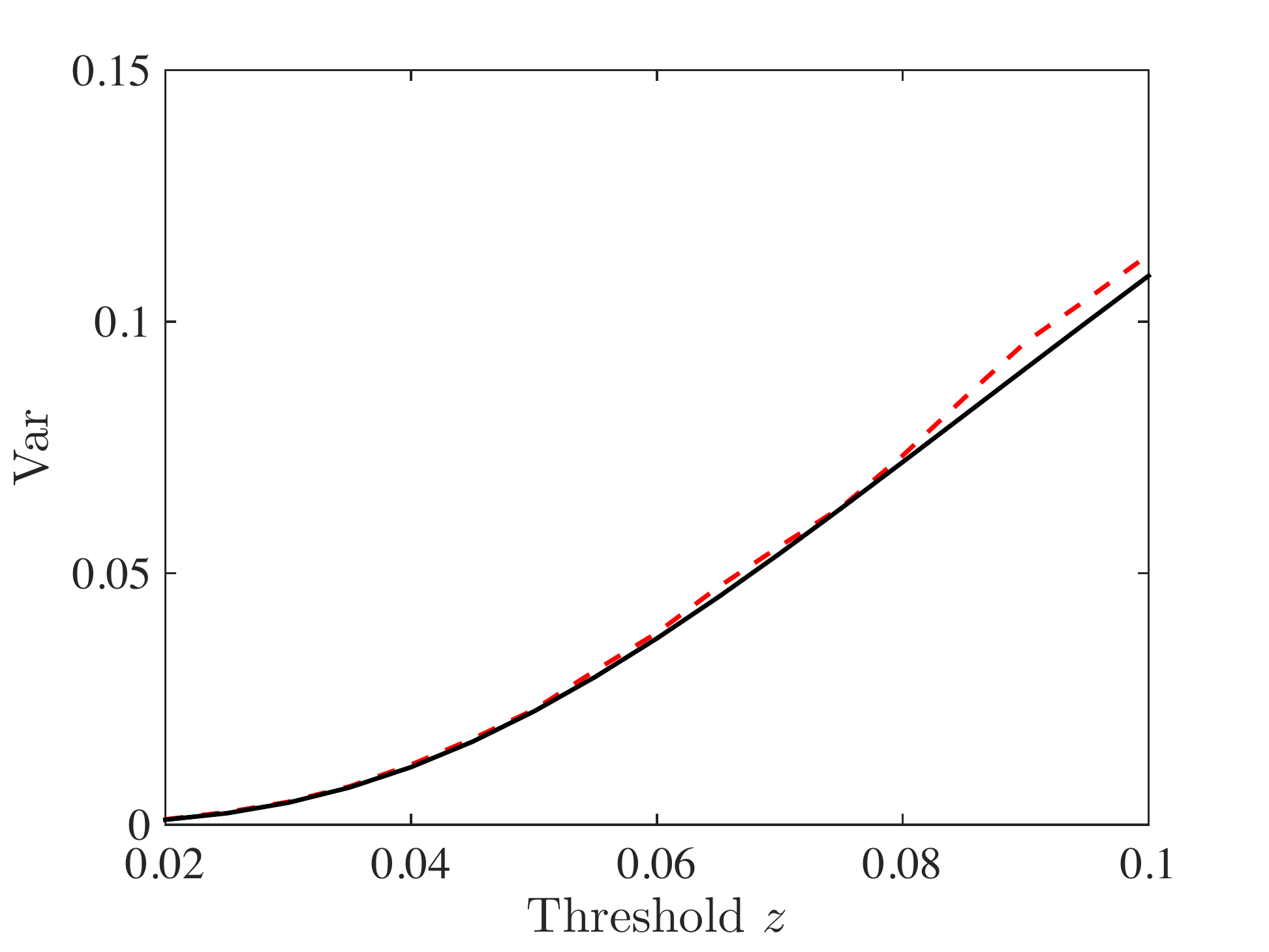}}
\subfigure[Variance of DT conditioned on correct decision]{\includegraphics[width=0.33\textwidth]{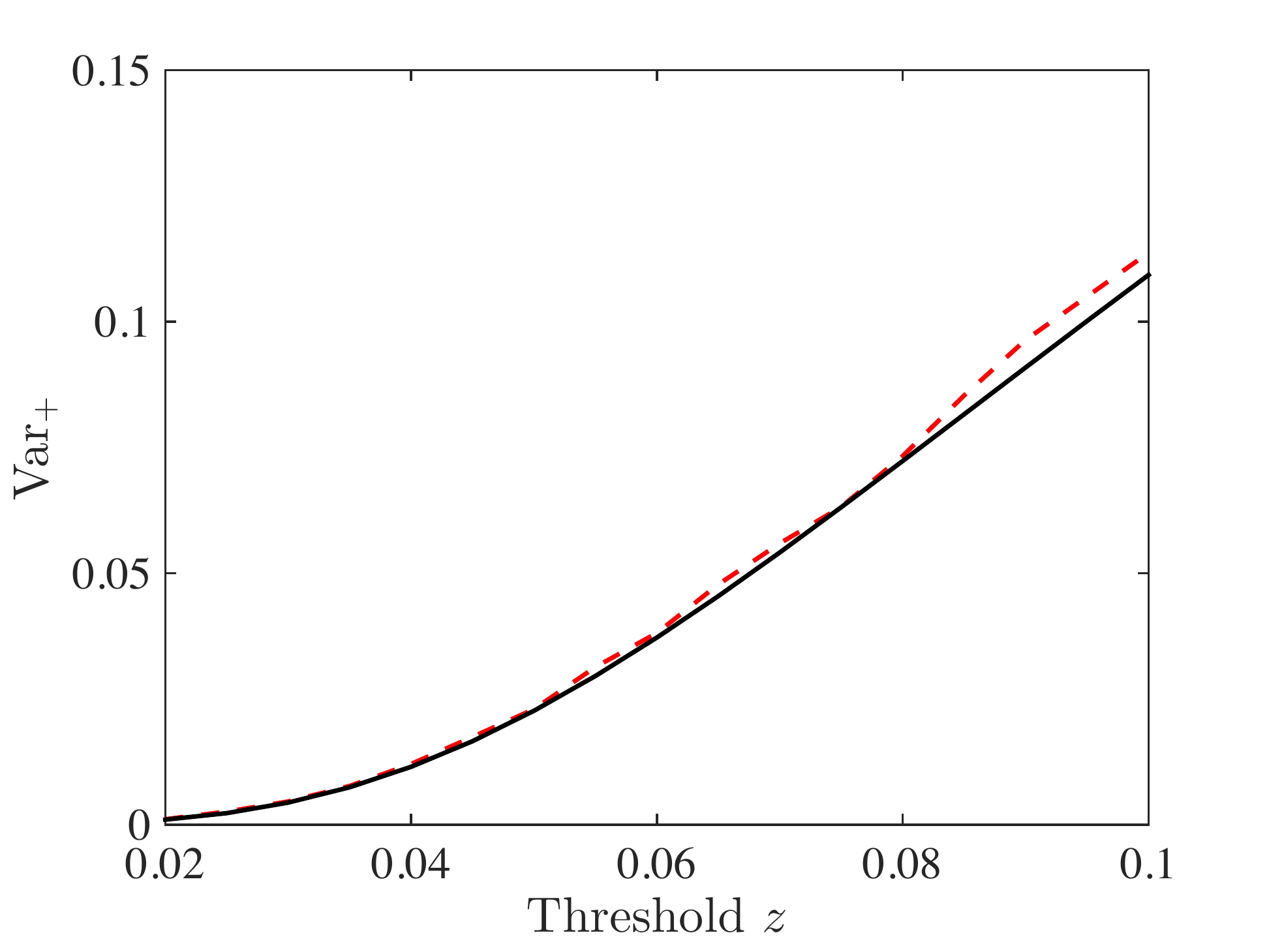}}
\subfigure[Variance of DT conditioned on error]{\includegraphics[width=0.33\textwidth]{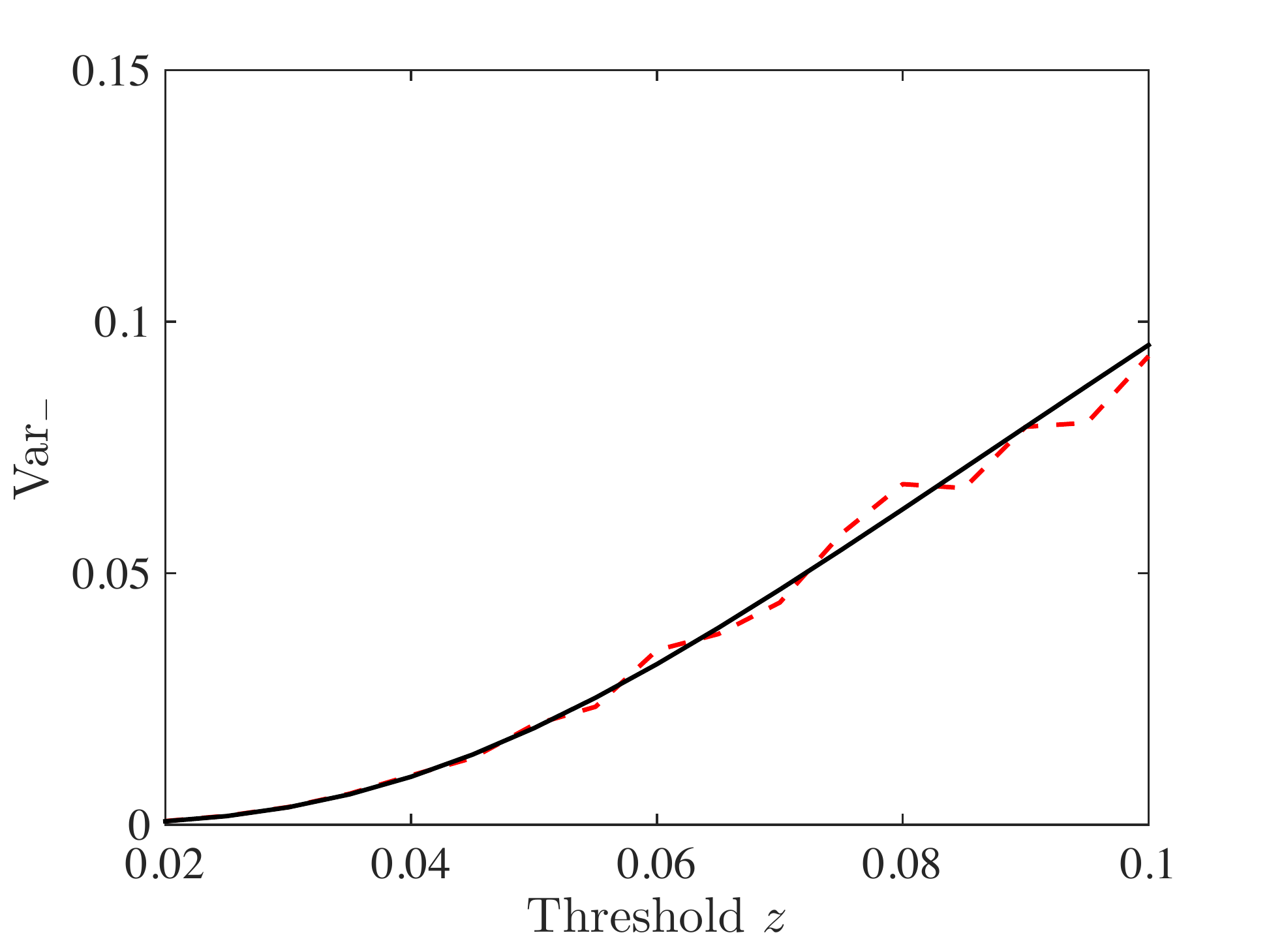}}

\subfigure[CV of DT]{\includegraphics[width=0.33\textwidth]{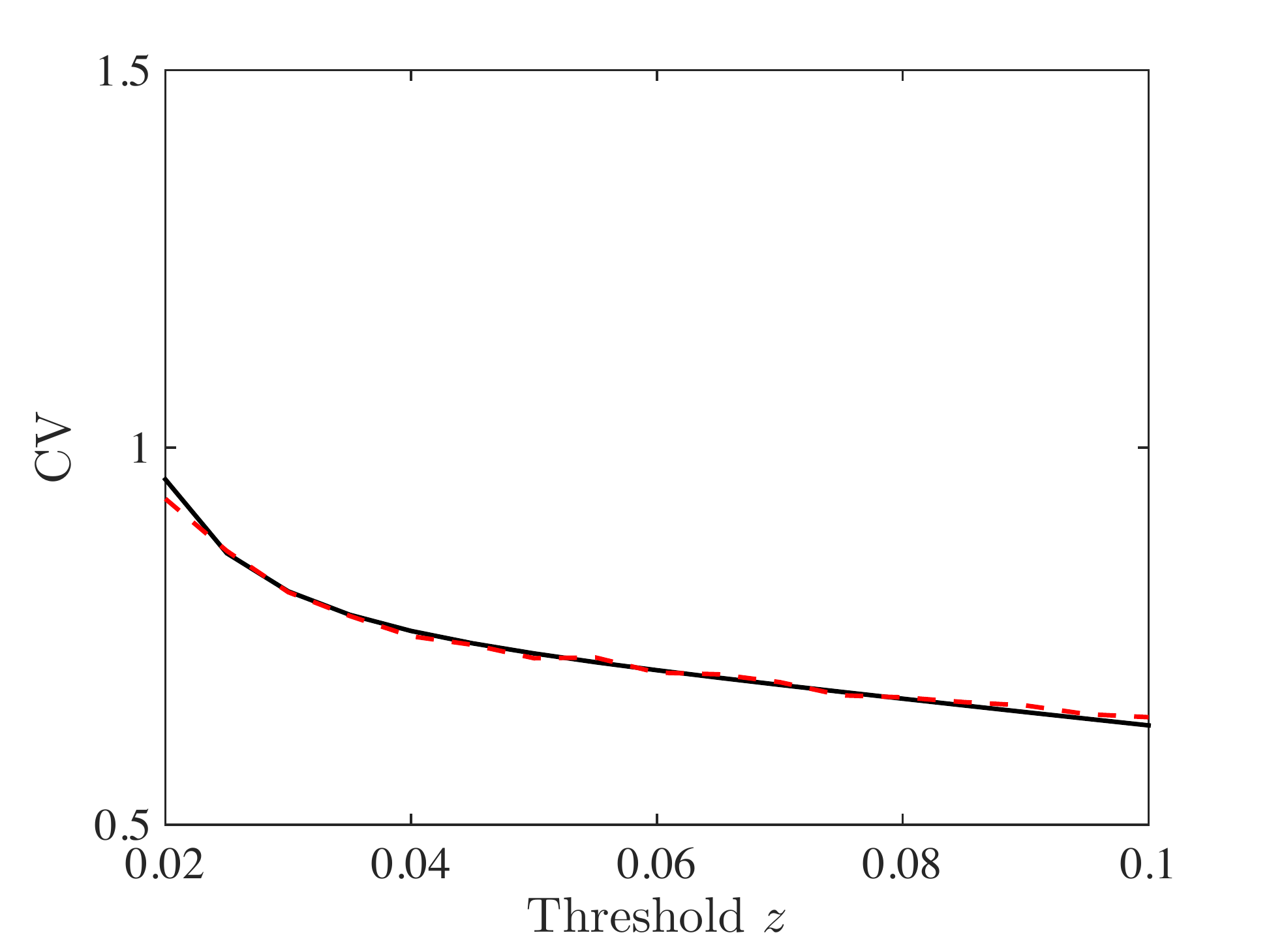}}
\subfigure[CV of DT conditioned on correct decision]{\includegraphics[width=0.33\textwidth]{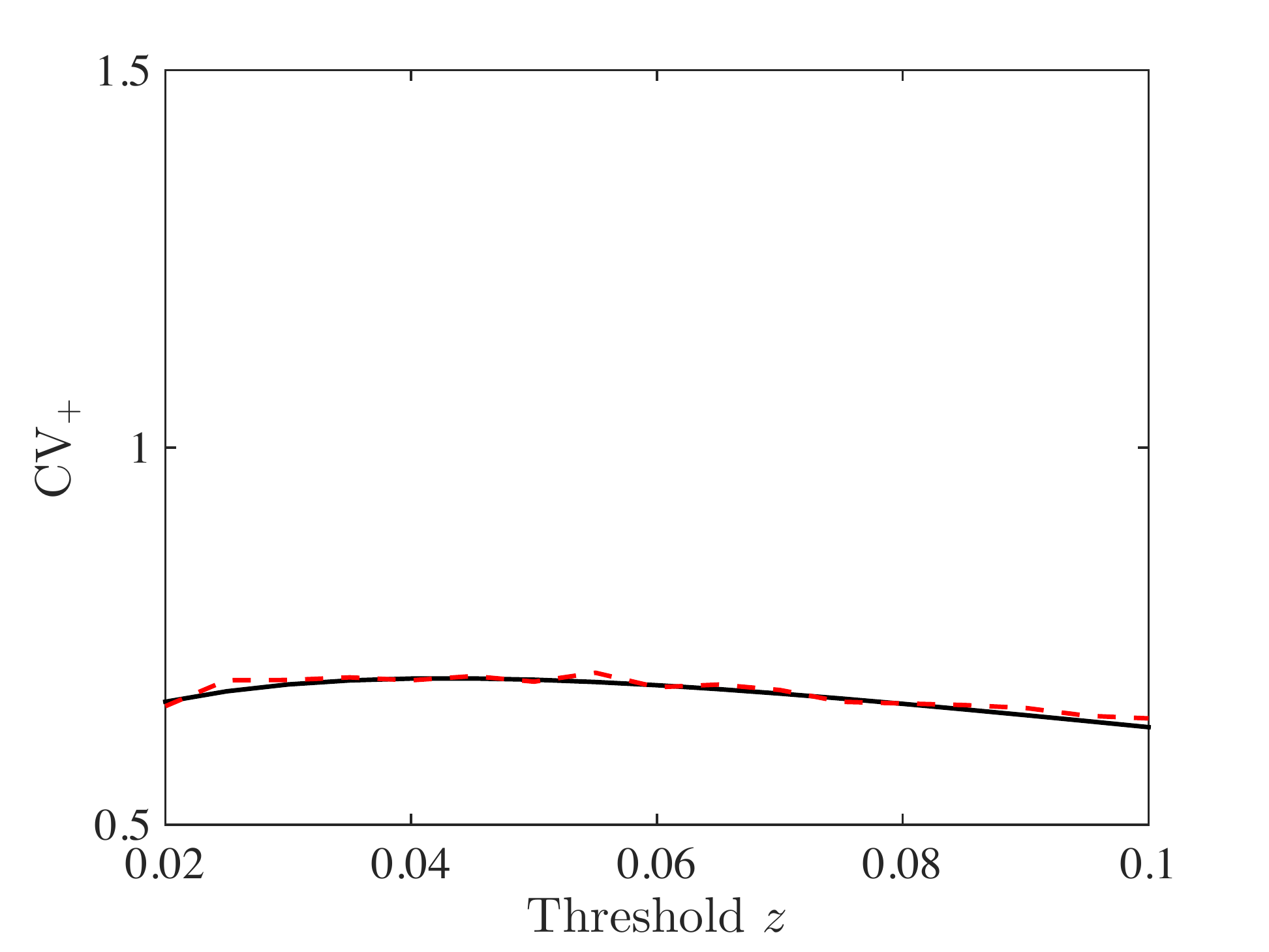}\label{fig:non-monotone-cv}}
\subfigure[CV of DT conditioned on error]{\includegraphics[width=0.33\textwidth]{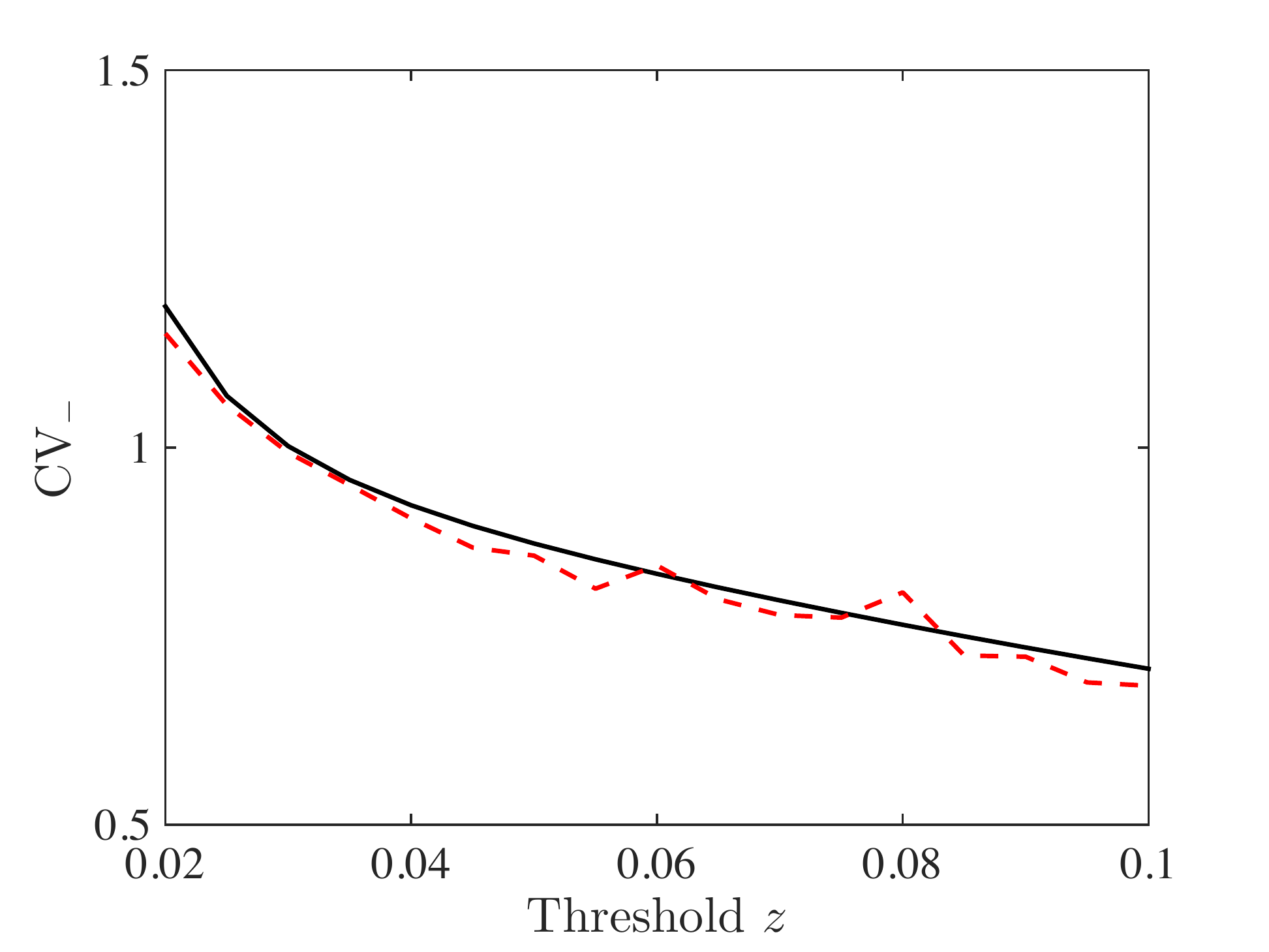}}
\caption{Expected decision times, variances and CVs of decision
times for a DDM with $a=0.2$, $\sigma=0.1$, and $x_0=-0.01$, showing
dependence on threshold $z$. Solid curves represent functions derived
in \S\ref{ss.dddm} and \S\ref{s.cond}; dashed line segments connect
point values obtained by 10,000 Monte-Carlo simulations of 
Eqn.~\eqref{e.dd1}. Note the non-monotonicity evident in panel h.}
\label{fig:ddm-numerics-1}  
\end{figure}

\begin{figure}[ht!]
\subfigure[Third central moment of DT]{\includegraphics[width=0.33\textwidth]{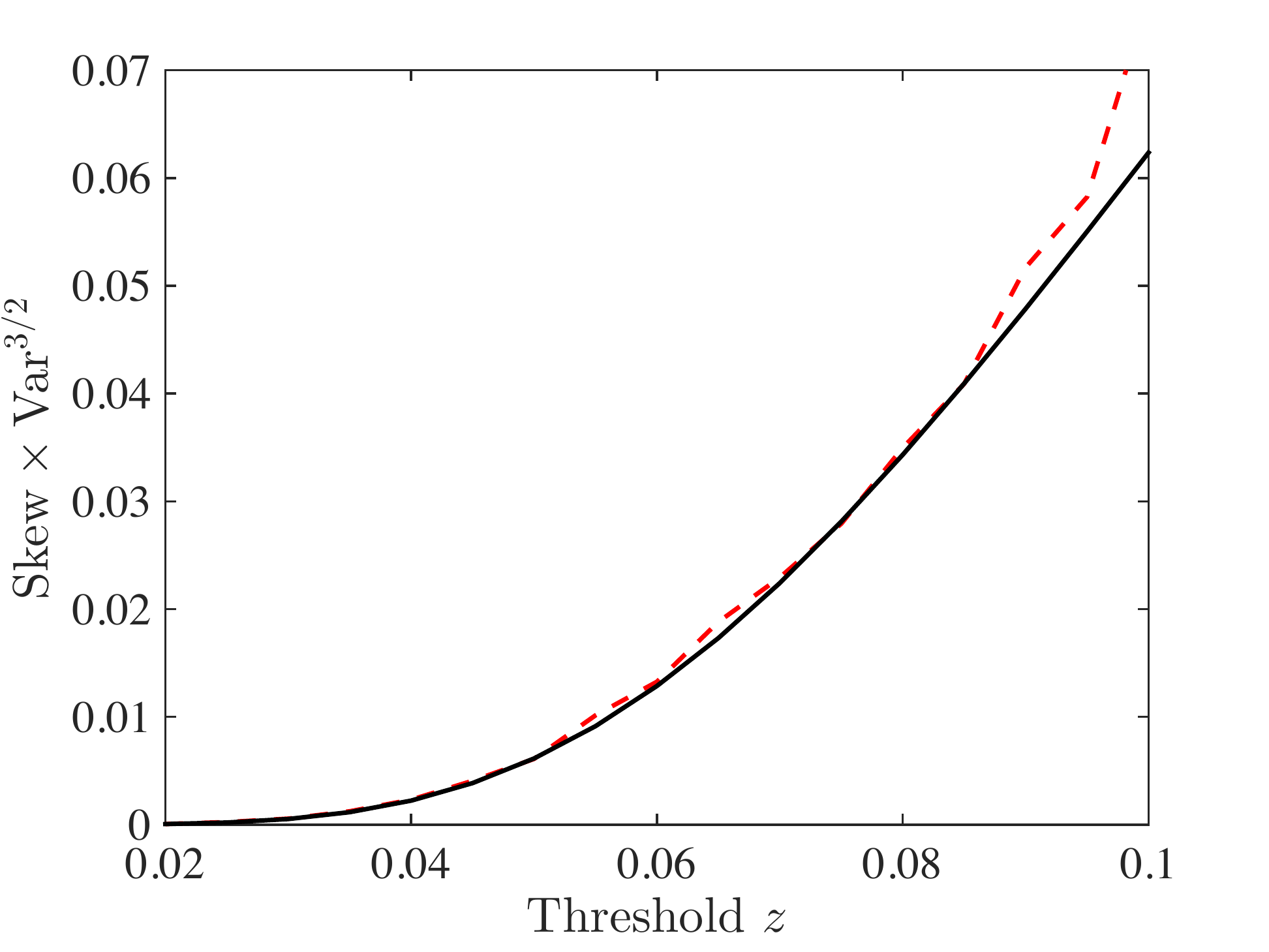}}
\subfigure[Third central moment of DT conditioned on correct decision]{\includegraphics[width=0.33\textwidth]{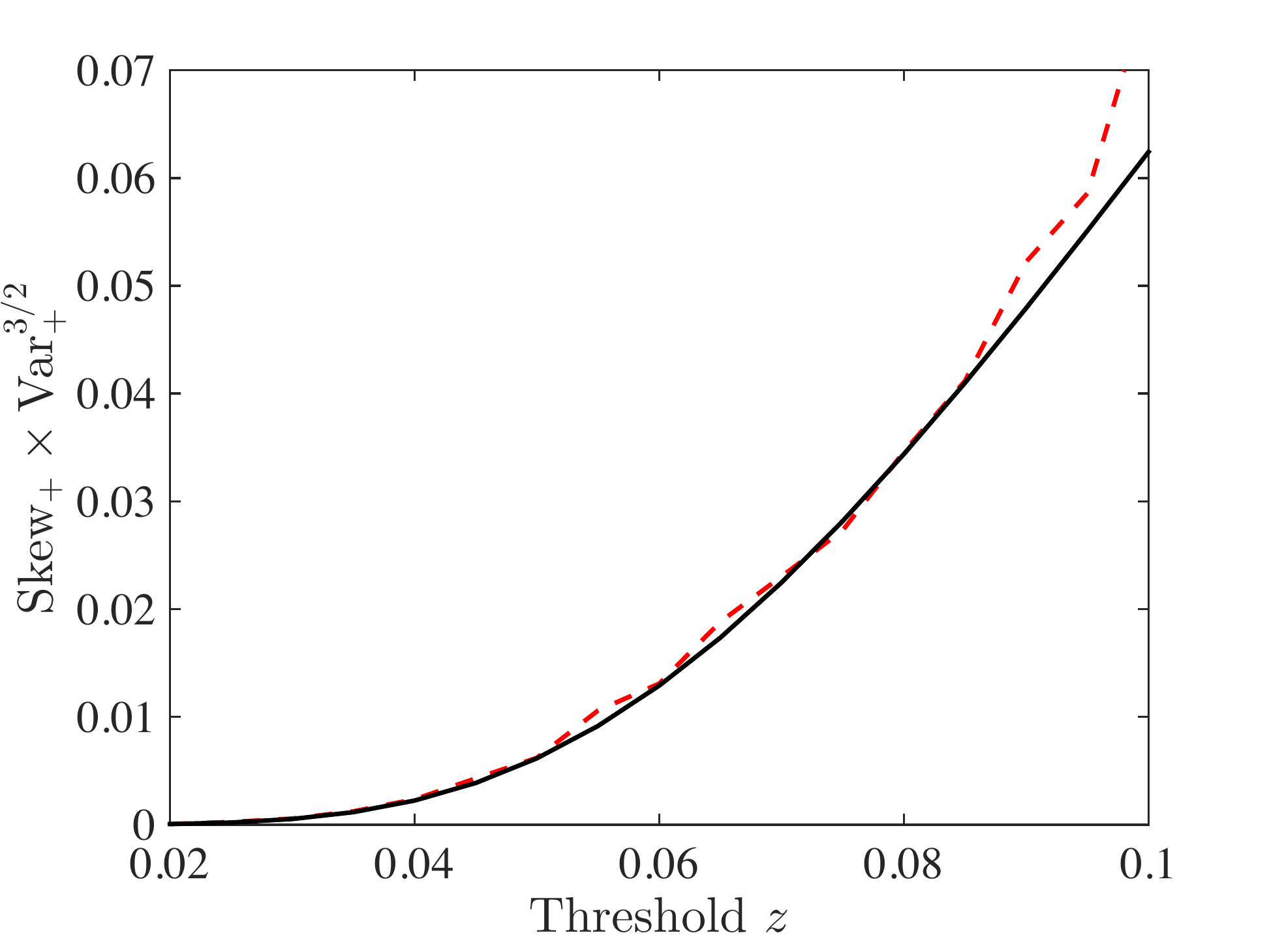}}
\subfigure[Third central moment of DT conditioned on incorrect decision]{\includegraphics[width=0.33\textwidth]{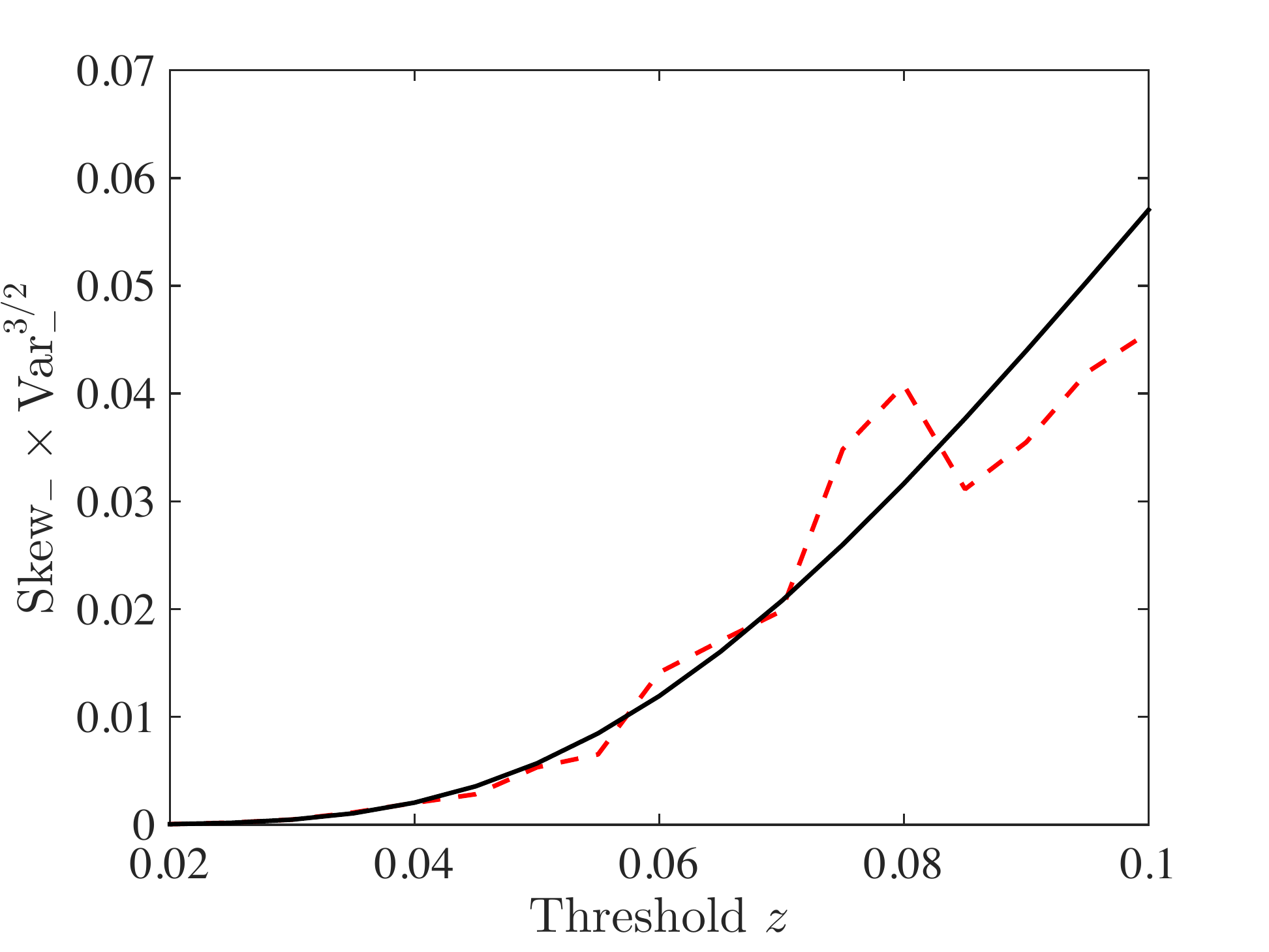}}

\subfigure[Skewness of DT]{\includegraphics[width=0.33\textwidth]{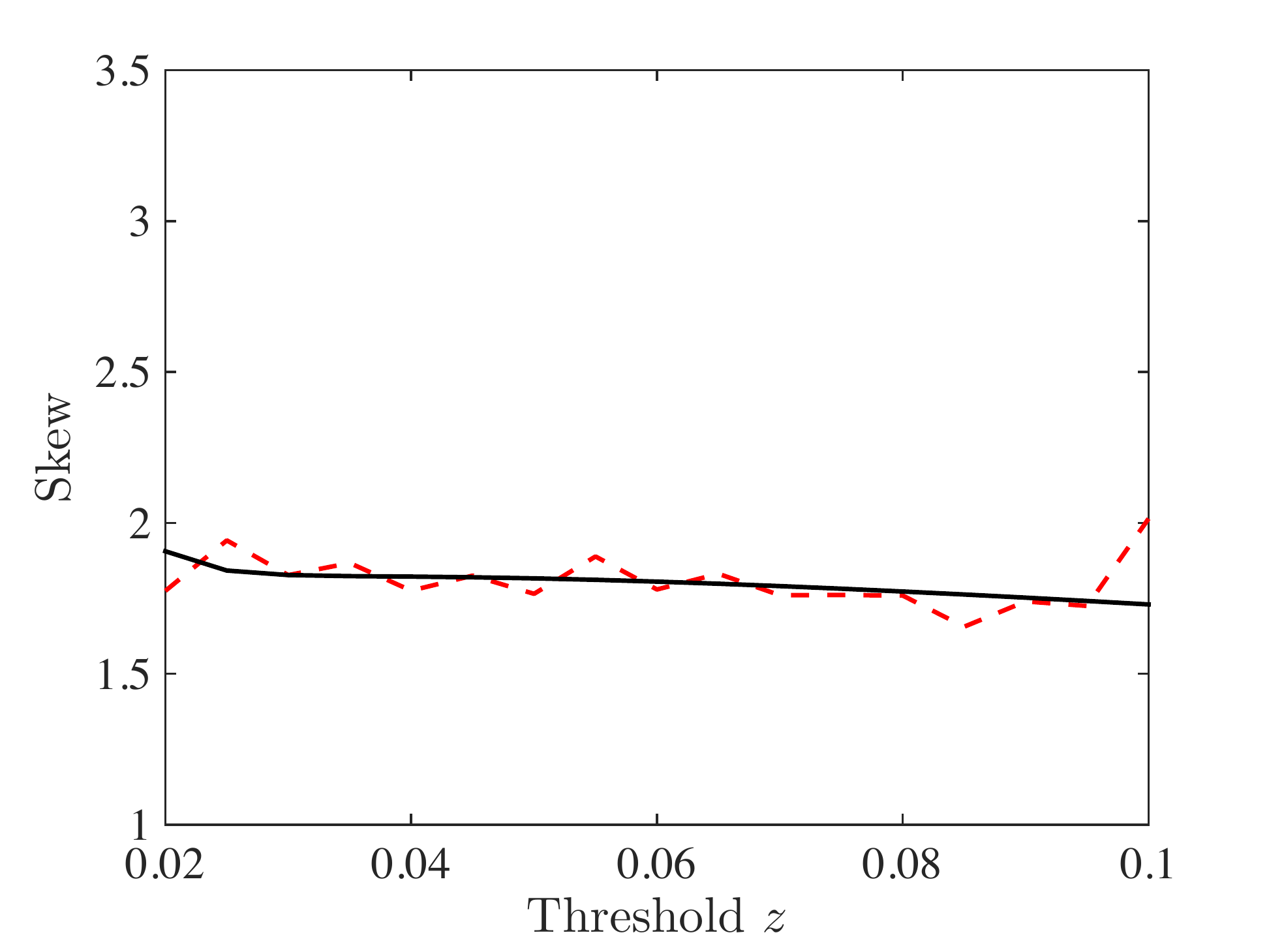}}
\subfigure[Skewness of DT conditioned on correct decision]{\includegraphics[width=0.33\textwidth]{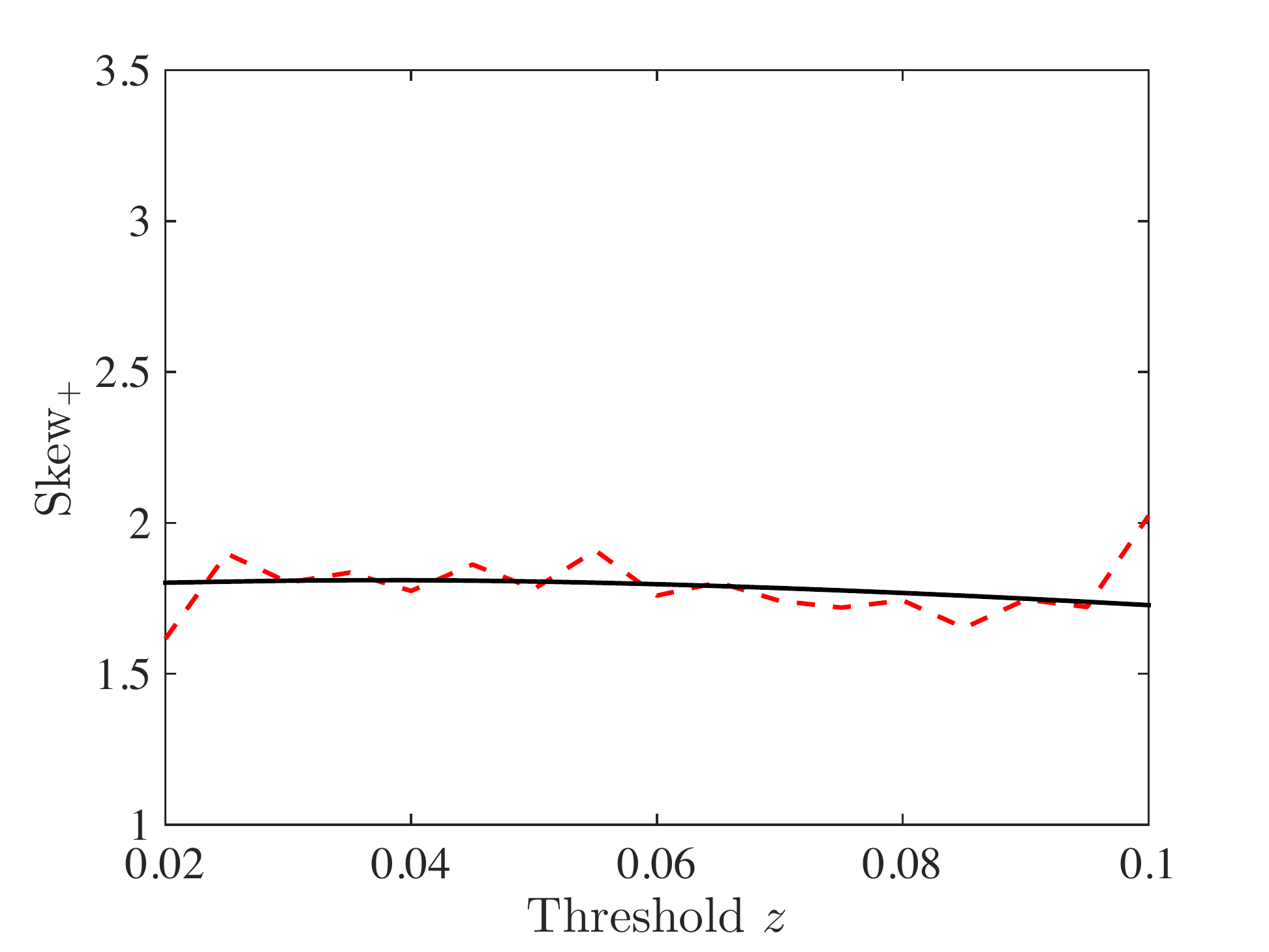}}
\subfigure[Skewness of DT conditioned on incorrect decision]{\includegraphics[width=0.33\textwidth]{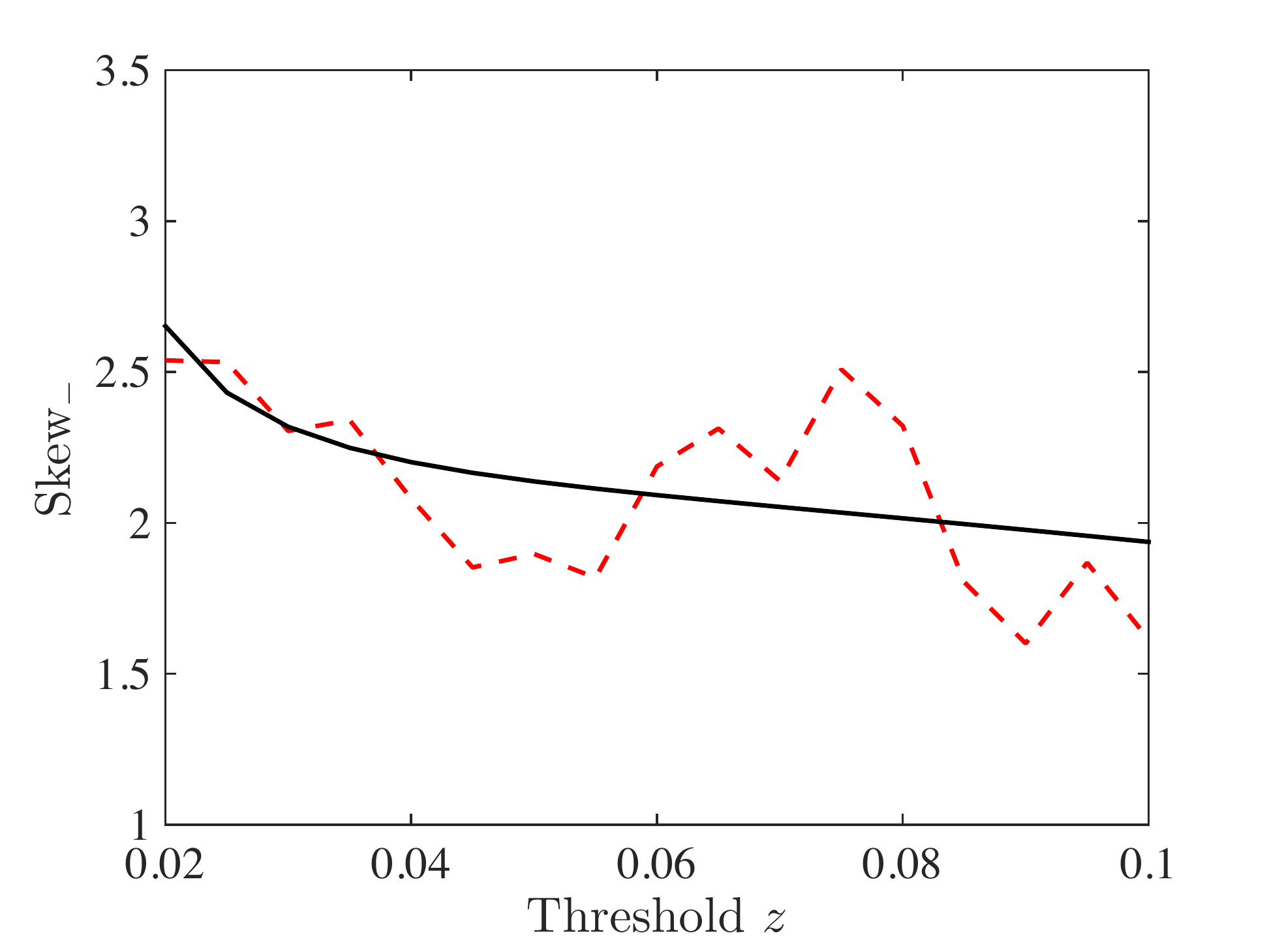}}
\caption{Third central moments and skewnesses of decision times for a
DDM with $a=0.2$, $\sigma=0.1$, and $x_0=-0.01$, showing dependence on
threshold $z$. Solid curves represent functions derived in
\S\ref{ss.dddm} and \S\ref{s.cond}; dashed line segments connect point
values obtained by 10,000 Monte-Carlo simulations of Eqn.~\eqref{e.dd1}.}
\label{fig:ddm-numerics-2} 
\end{figure}

In Figs.~\ref{fig:ddm-numerics-1} and \ref{fig:ddm-numerics-2} key
expressions derived above are plotted vs. threshold $z$ for the
DDM~\eqref{e.dd1} with $a = 0.2$, $\sigma = 0.1$, and $x_0 =
-0.01$. These parameter values were chosen as representative of fits
to human data (e.g. \cite{Simen+JEPHPP09}), and to illustrate the
general forms of the functions.  Drift values in this case might be
expected to range from -0.4  to 0.4
(e.g. \cite{RatSuperDiscrim2014}). See also, among many others,
\cite{Balci+APP11,BalciSimenActa2014,Bogacz+QJExpP10,DutilWag-Psychon09},
for similar ranges of fitted parameter values.  The results of
Monte-Carlo simulations of Eqn.~\eqref{e.dd1} using the Euler-Maruyama
method \cite{high01} with step size $10^{-4}$ are also shown for
comparison.  Note that, even with 10,000 sample paths, numerical
estimates of the third moment and skewness have not converged very
well.

\section{Behavior of CVs}
\label{s.cvs}

We first consider the unconditional CV with unbiased starting point
$x_0 =  k_x = 0$, for which we can prove the following result. 

\begin{prop} \label{CVprop}
{\em Behavior of CVs of decision times for the DDM.} The CV for the
double-threshold DDM with $k_x = 0$, Eqn.~\eqref{e.cvx0}, is
bounded above by the CV for  the  single-threshold DDM,
Eqn.~\eqref{e.wald2}:
\beq
\frac{ \sqrt{ \frac{1}{k_z} (1 - e^{-4 k_z} - 4 k_z e^{-2 k_z})}}
{1-e^{-2 k_z}} \eqdef F(k_z) < \sqrt{ \frac{1}{k_z} } .
\label{e.cv00} 
\eeq
Moreover, $F(0) = \sqrt{2/3}$ and $F(k_z)$ decays monotonically as $k_z$
increases. 
\end{prop}

\bef[!htb]
\centering
{\includegraphics[width=0.5\textwidth]{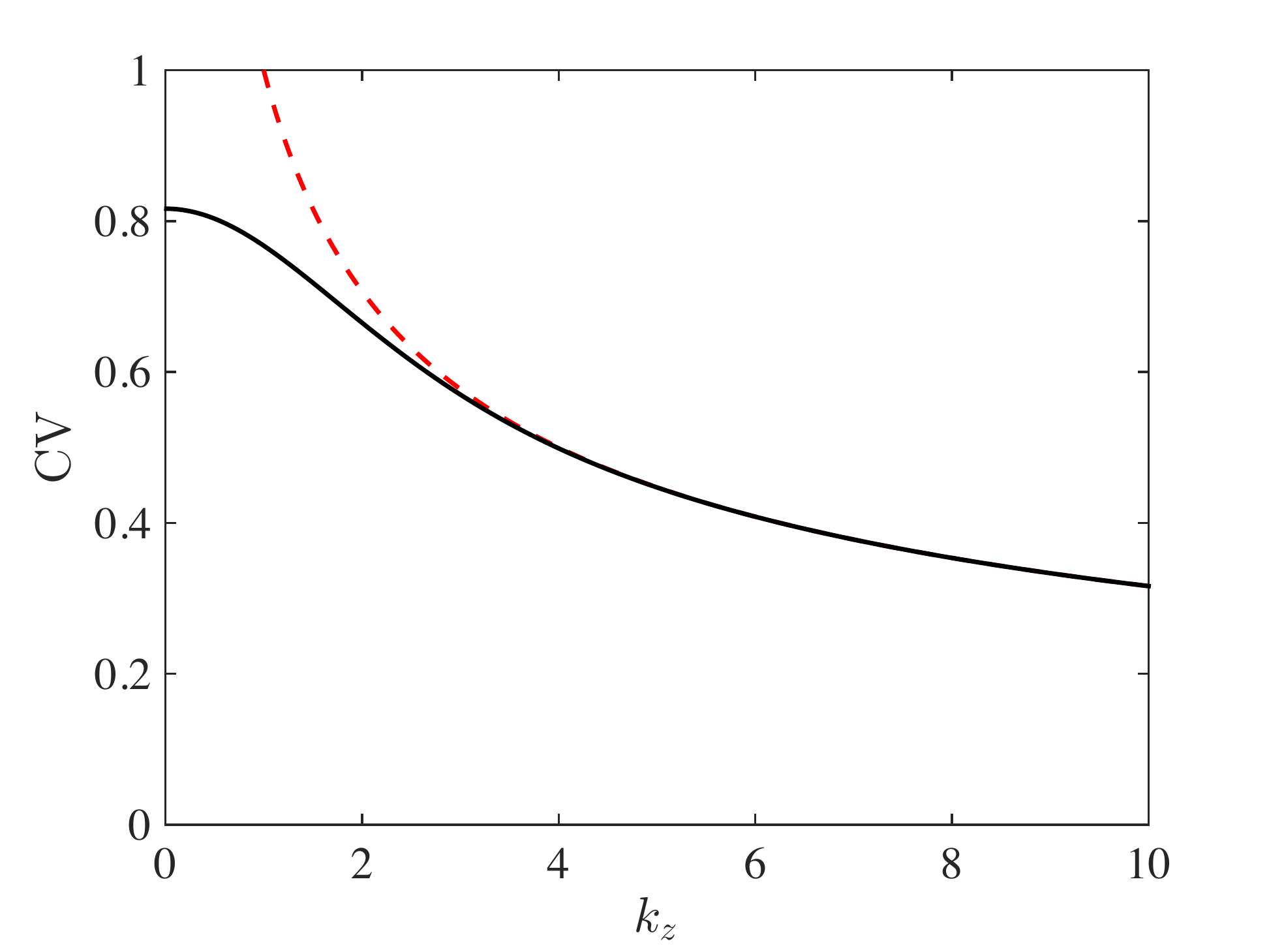}}
\caption{Coefficients of variation as functions of $k_z$ for the
single threshold DDM (dashed) and the DDM with double thresholds and
$k_x = 0$ (solid).}
\label{fig:cvs}
\eef

For the proof of the above proposition see Appendix~\ref{app-proof-CVprop}.
Fig.~\ref{fig:cvs} illustrates the proposition by plotting both CV 
functions over the range $0 \le k_z \le 10$.

\begin{figure}[t!]
\centering 
\subfigure[Unconditioned CV vs. $k_x$]{\includegraphics[width=0.475\textwidth]{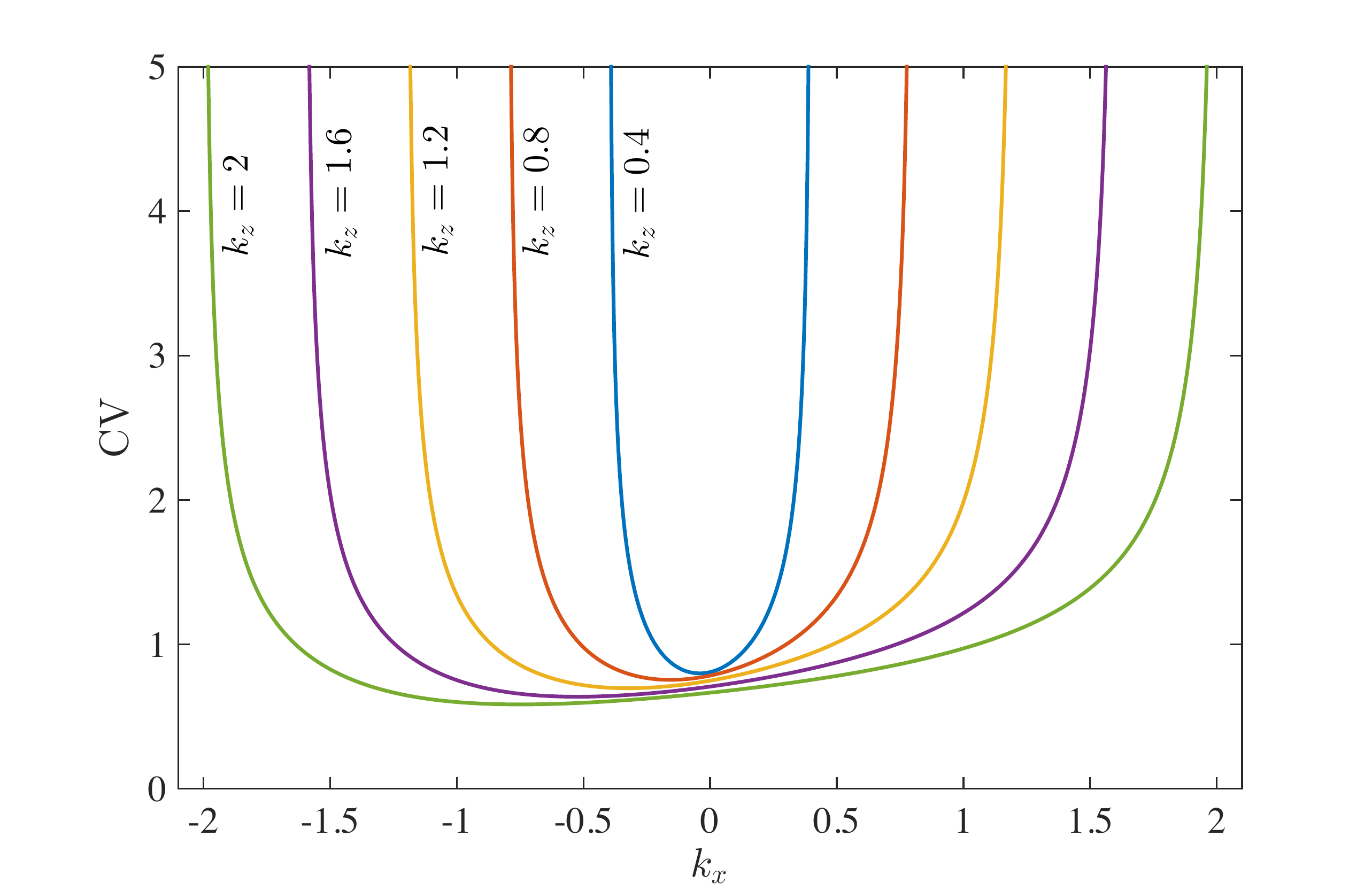}}
\subfigure[Unconditioned CV vs. $k_z$]{\includegraphics[width=0.475\textwidth]{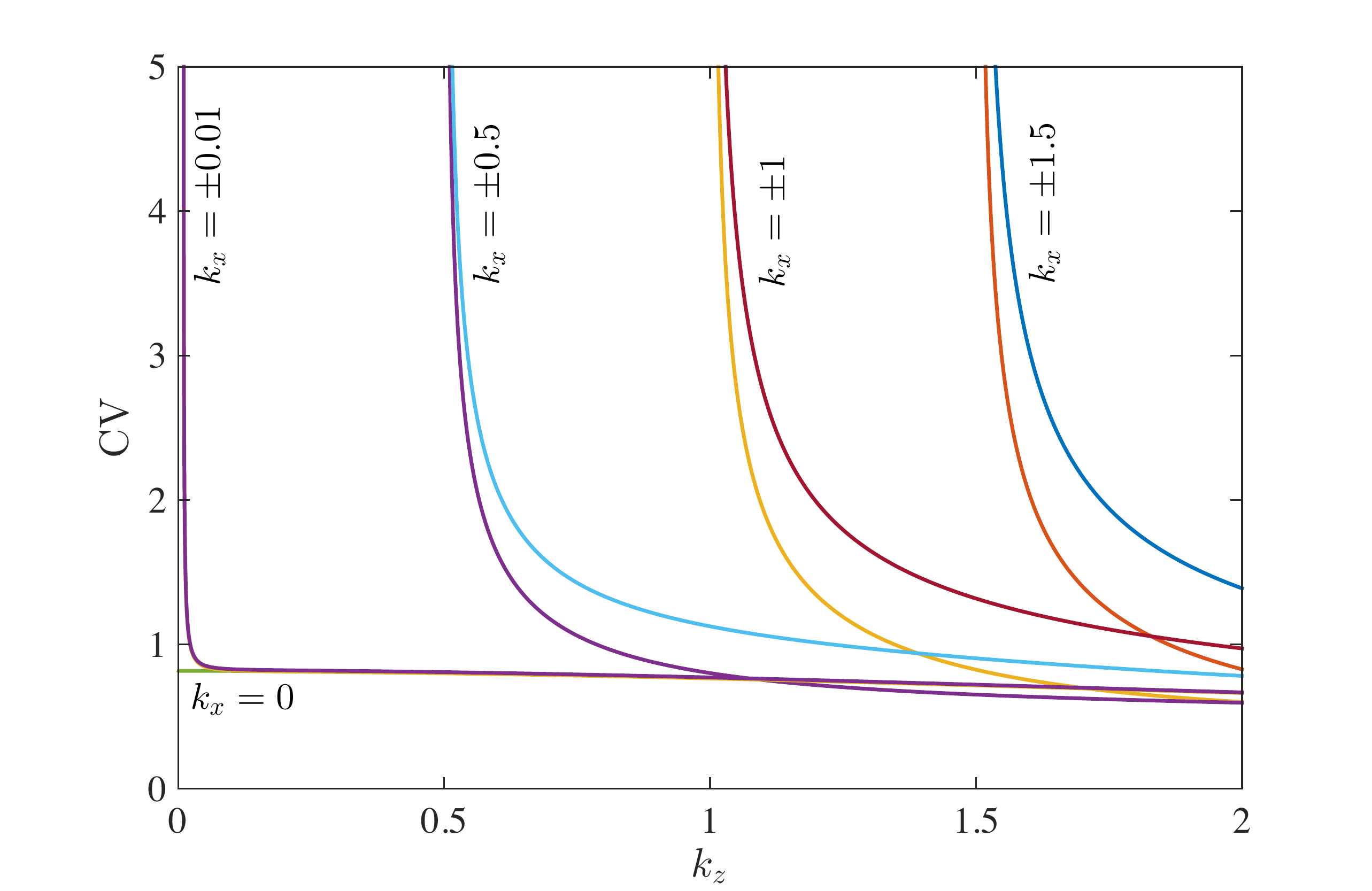}}
\subfigure[CV conditioned on correct decision vs. $k_x$]{\includegraphics[width=0.475\textwidth]{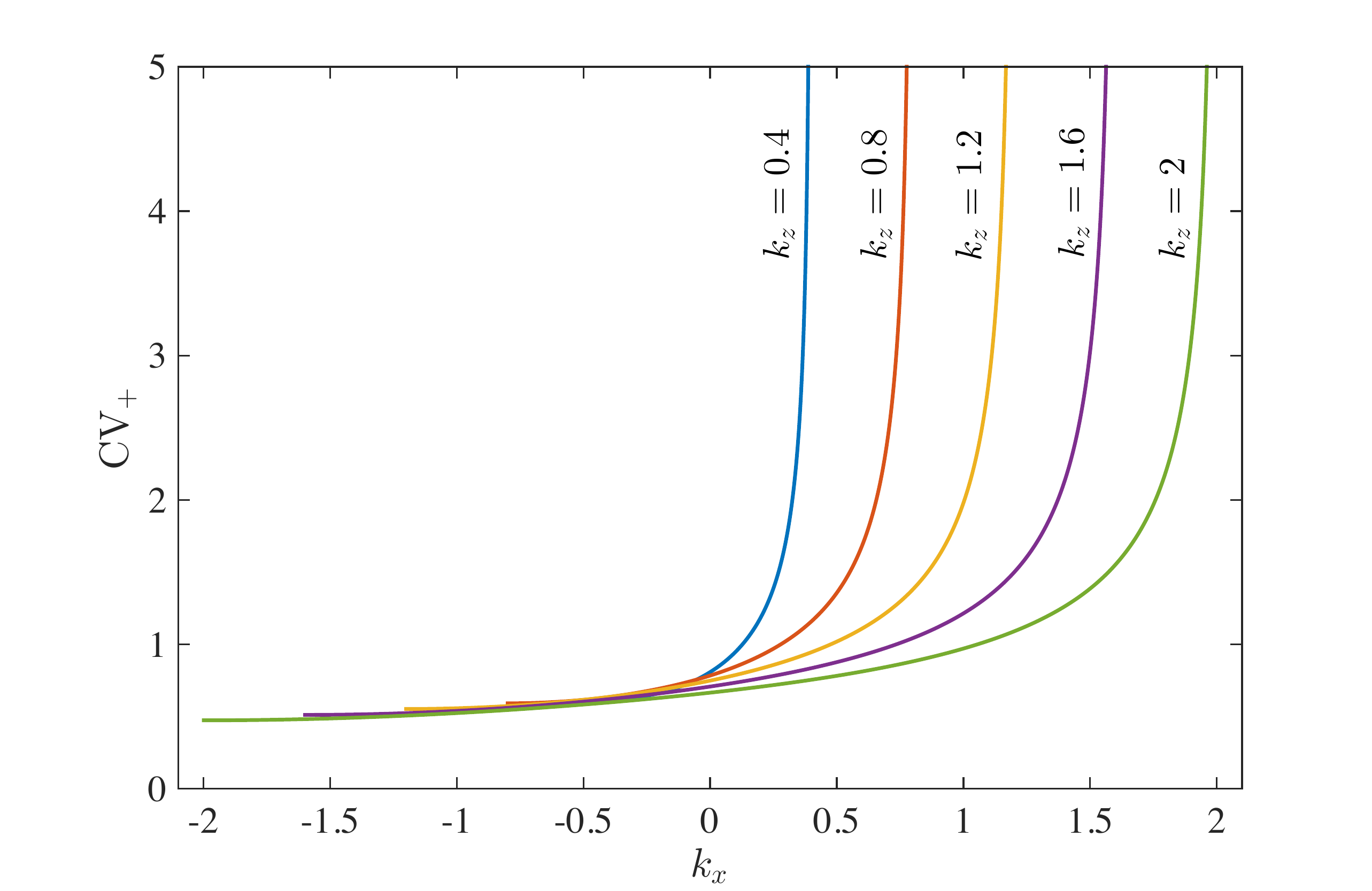}}
\subfigure[CV conditioned on correct decision vs. $k_z$]{\includegraphics[width=0.475\textwidth]{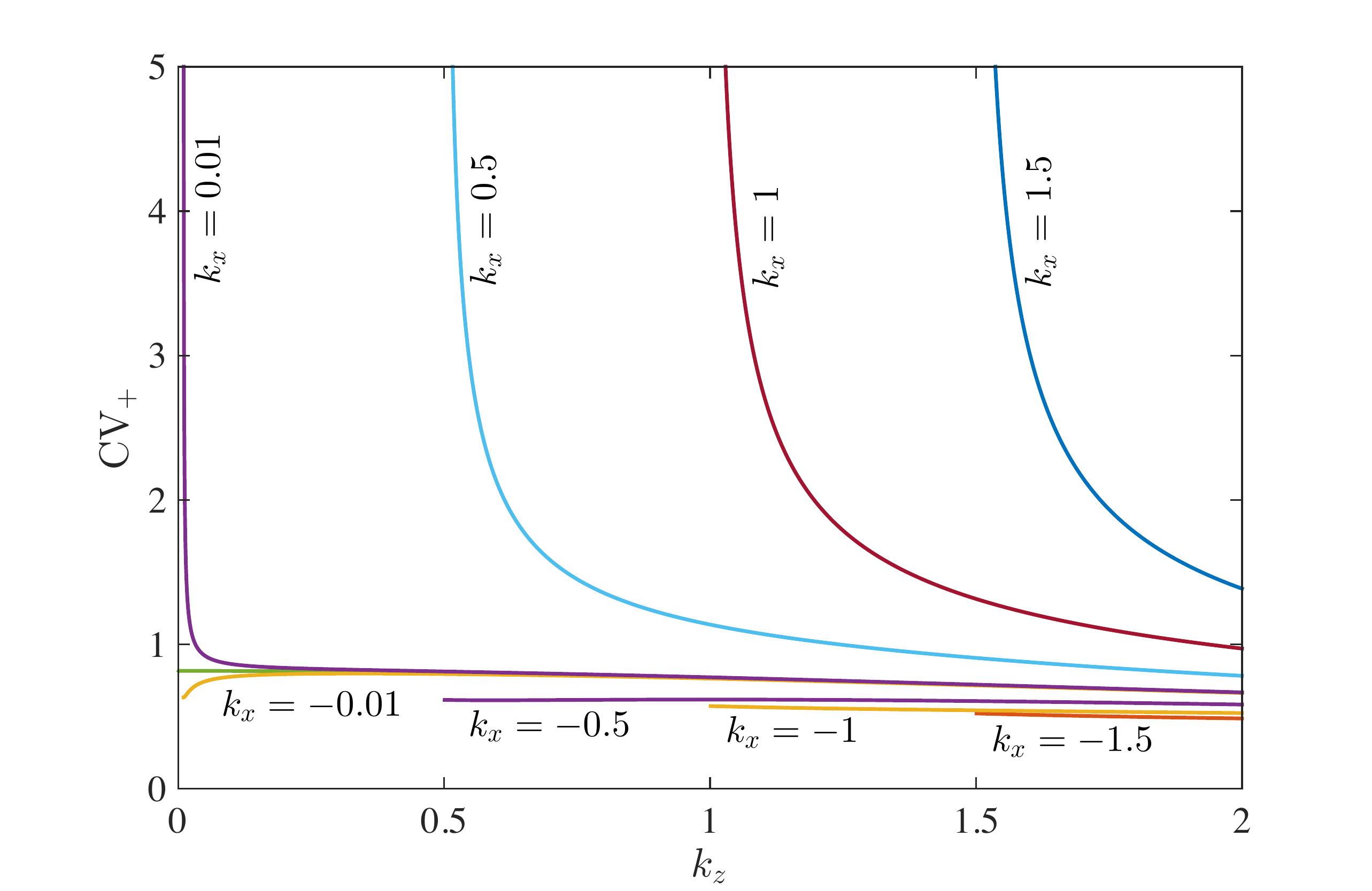}\label{fig:non-monotone-1}}
\subfigure[CV conditioned on error vs. $k_x$]{\includegraphics[width=0.475\textwidth]{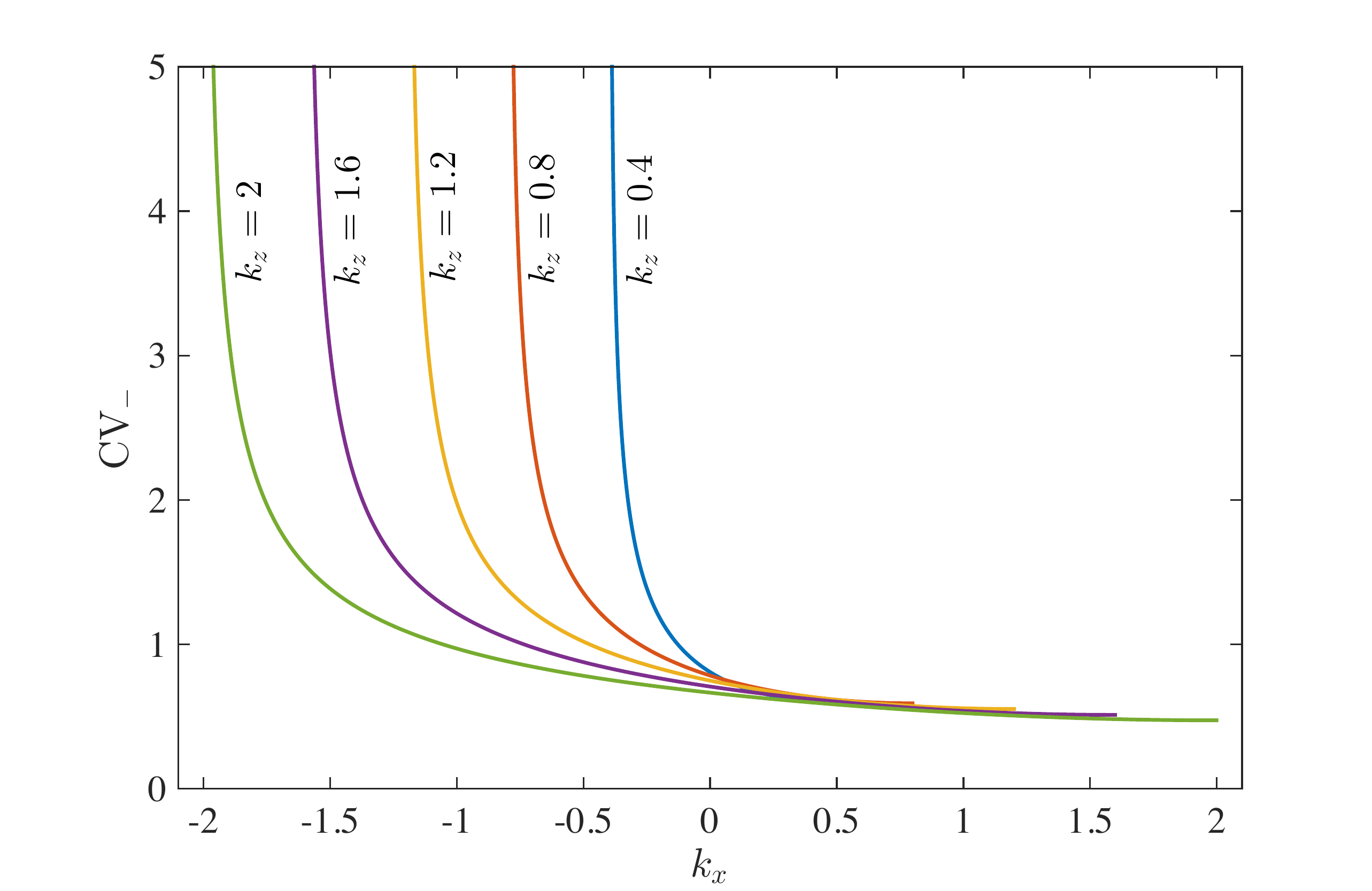}}
\subfigure[CV conditioned on error vs. $k_z$]{\includegraphics[width=0.475\textwidth]{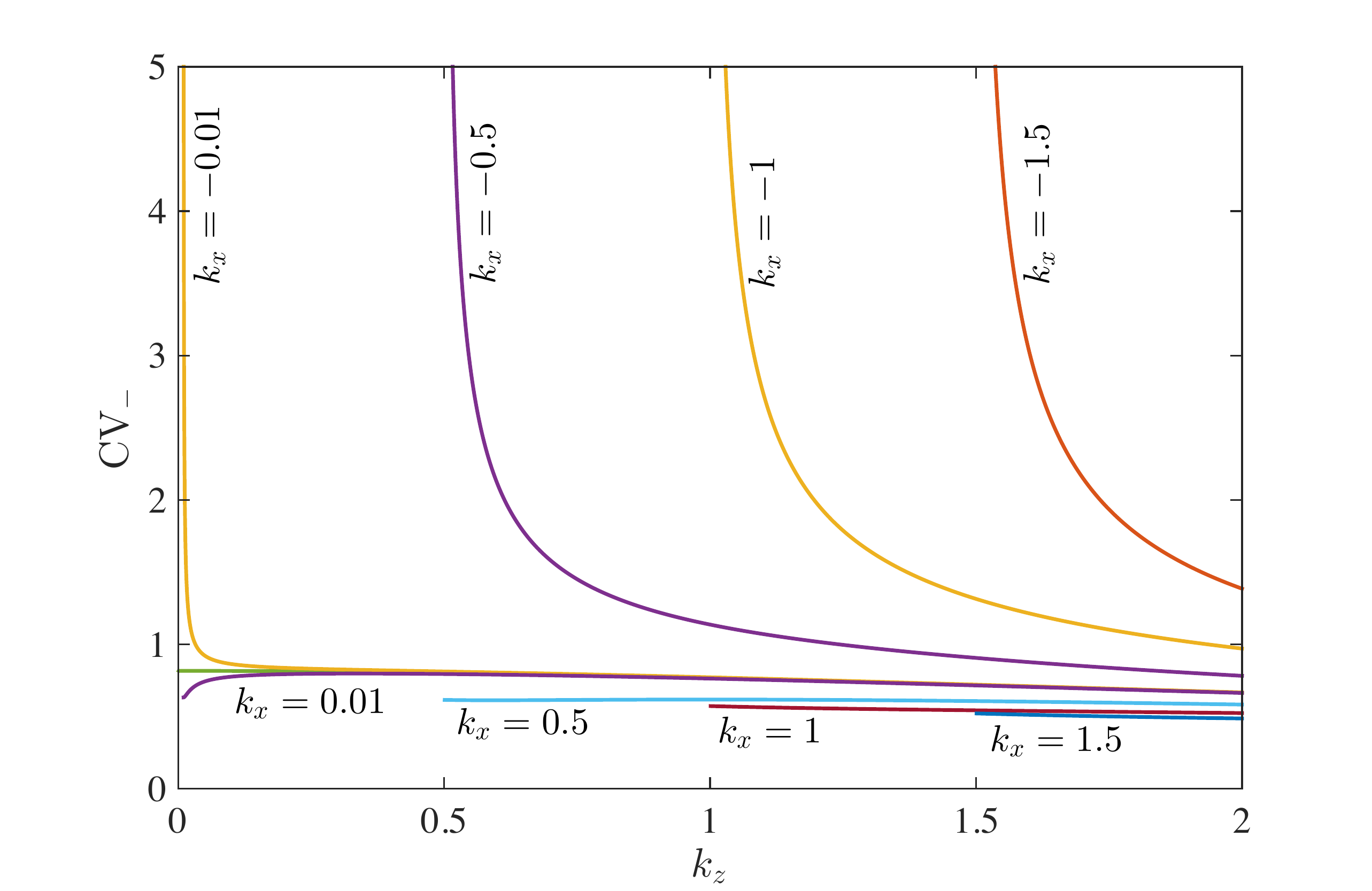}\label{fig:non-monotone-2}}
\caption{Coefficients of variation of decision time as functions of
$k_x=ax_0/\sigma^2$ for several $k_z$'s (left column) and
$k_z=az/\sigma^2$ for several $k_x$'s (right column).  Unconditioned
CVs are shown in top row, conditioned CVs in middle and bottom
rows. Observe the symmetry $k_x \mapsto -k_x$ relating the latter, as
noted at the beginning of \S\ref{s.cond}.} 
\label{fig:cov}
\end{figure}

\vspace{0.3cm}

It seems difficult to prove a result analogous to Proposition
\ref{CVprop} for the general CV expressions of Eqns.~\eqref{e.var} and
\eqref{eq:evarc1} due to their complexity. However, plots of the
unconditional and conditional CVs as functions of the normalized
threshold and starting point $k_z=az/\sigma^2$ and $k_x=ax_0/\sigma^2$
shown in Fig.~\ref{fig:cov} illustrate their behavior over the
($k_z,k_x $)-plane. 

Here, as shown in Proposition~\ref{CVprop} and
Eqns.~(\ref{eq:cvclim0}-\ref{eq:cvclim2}), for $k_x=0$ both
conditioned and unconditioned $\textrm{CV}$s converge to $\sqrt{2/3}$
from below as $k_z \to 0^+$ (see right column).  However, for $k_x \ne
0$, the behavior is significantly different. In particular, as shown
in \S\ref{ss.dddm}, Eqns.~(\ref{e.kx-kz}-\ref{e.var.kx-kz}), the
unconditioned \textrm{CV}s diverge as $k_x \to \pm k_z$ (see left column).  \textrm{CV}s
for symmetric starting points $\pm k_x$ diverge along different
curves as $|k_x| \to k_z$; however, these curves converge to each other
as $k_z \to 0^+$ (see left column).  Similarly, \textrm{CV}s conditioned on correct
responses and errors diverge as $k_x \to k_z$ and $k_x \to -k_z$
respectively. Interestingly, \textrm{CV}s conditioned on correct
responses and errors converge to finite limits \emph{smaller than}
$\sqrt{2/3}$ as  $k_x \to - k_z < 0$ and $k_x \to k_z > 0$
respectively. In
Fig.~\ref{fig:non-monotone-1}, as shown in \S\ref{s.cond},
$\mathrm{CV}_+ $ converges to $\sqrt{2/5}$ as $k_x \to -k_z$ and $k_z
\to 0^+$.   It is interesting to note that this convergence is not
monotone. 

The bottom four panels of Fig.~\ref{fig:cov} illustrate the symmetry
of moments conditioned on correct responses and errors with respect to
$k_x \mapsto -k_x$, noted at the beginning of \S\ref{s.cond}. 
 Unlike the case $k_x = 0$ for which CV is monotonic in 
$k_z$, as shown in Proposition~\ref{CVprop}, conditioned {\rm CV}s
are not monotone functions
of $z$ or $k_z$ in general. Some instances of non-monotonicity
appear in Figs.~\ref{fig:non-monotone-cv},~\ref{fig:non-monotone-1}
and \ref{fig:non-monotone-2} above.

\section{Behavior of moments for the extended DDM}
\label{s.extddm}

We end by describing some results for the extended DDM introduced by
Ratcliff \cite{Rat78}, specifically, the effects of drawing drift
rates and starting points for Eqn.~\eqref{e.dd1} from Gaussian and
uniform distributions $\mathcal{N}(a, \sigma_a)$ and $\mathcal{U}(x_0
- \frac{s_x}{2},  x_0 + \frac{s_x}{2})$ respectively, where $x_0 \pm  \frac{s_x}{2}  \in
[-z, z]$, and standard deviation $\sigma_a$ and range $s_x$
characterize trial-to-trial variability of drift rates and starting
points.  Complete analytical results on moments for this extended
model are not known, and we therefore perform numerical studies. In
particular we investigate departures from the analytical results
derived above as the variance/range of the distributions $\mathcal{N}$ and
$\mathcal{U}$ increase from zero. We also consider the effects
of non-decision time.

\subsection{Analytical and semi-analytical expressions}
\label{ss.analytix}

\begin{figure}[ht!]
\includegraphics[width=0.33\textwidth]{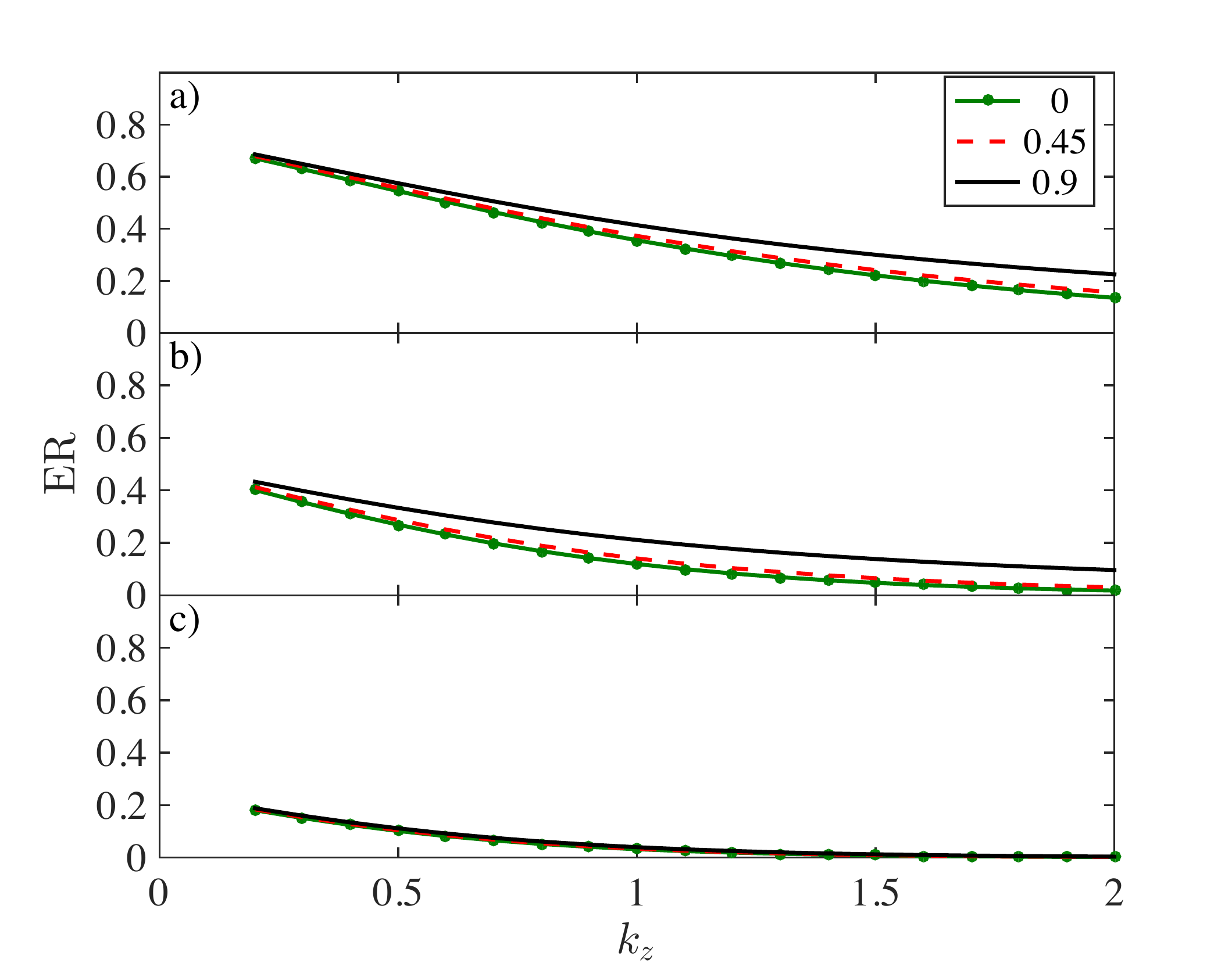}
\includegraphics[width=0.33\textwidth]{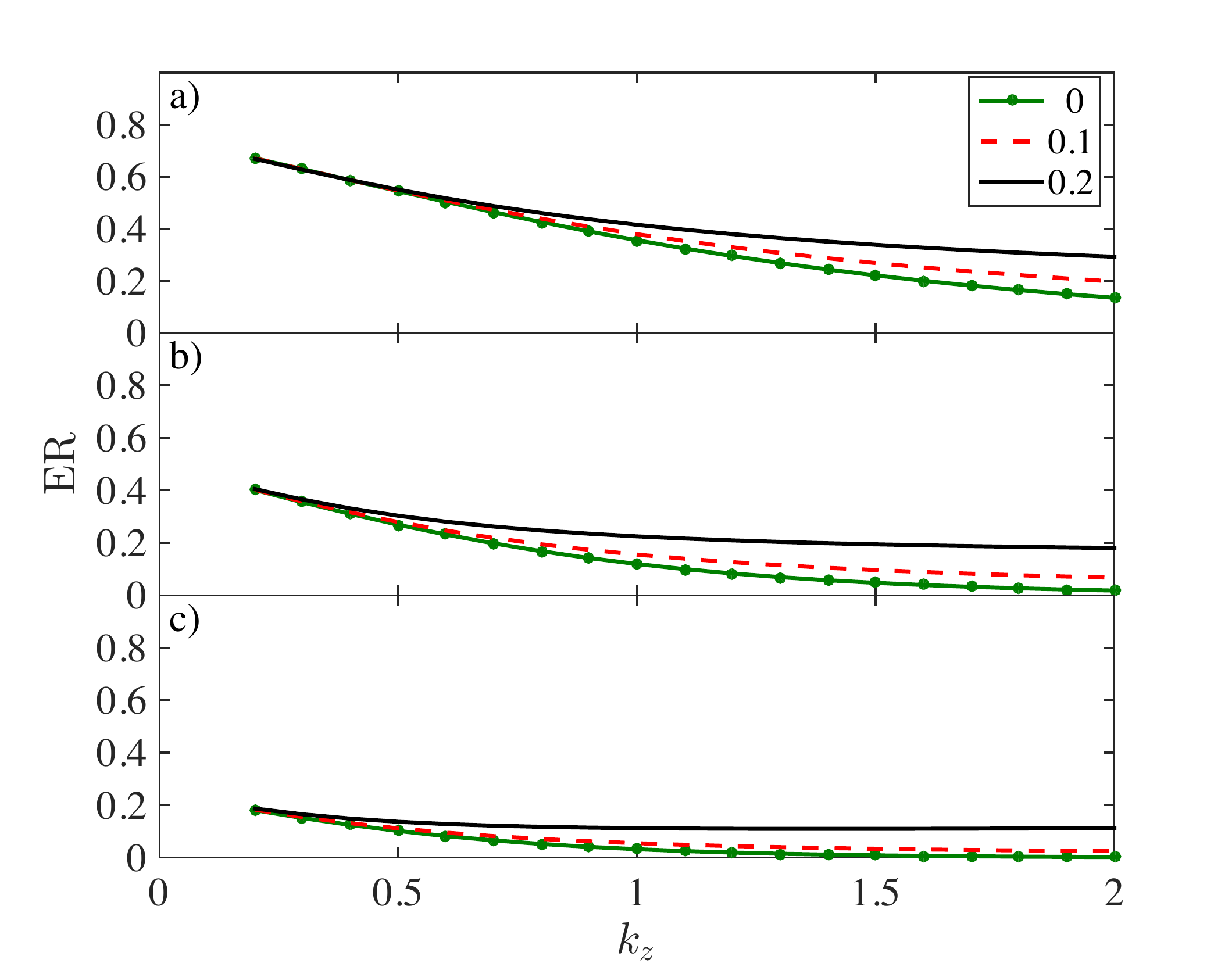}
\includegraphics[width=0.33\textwidth]{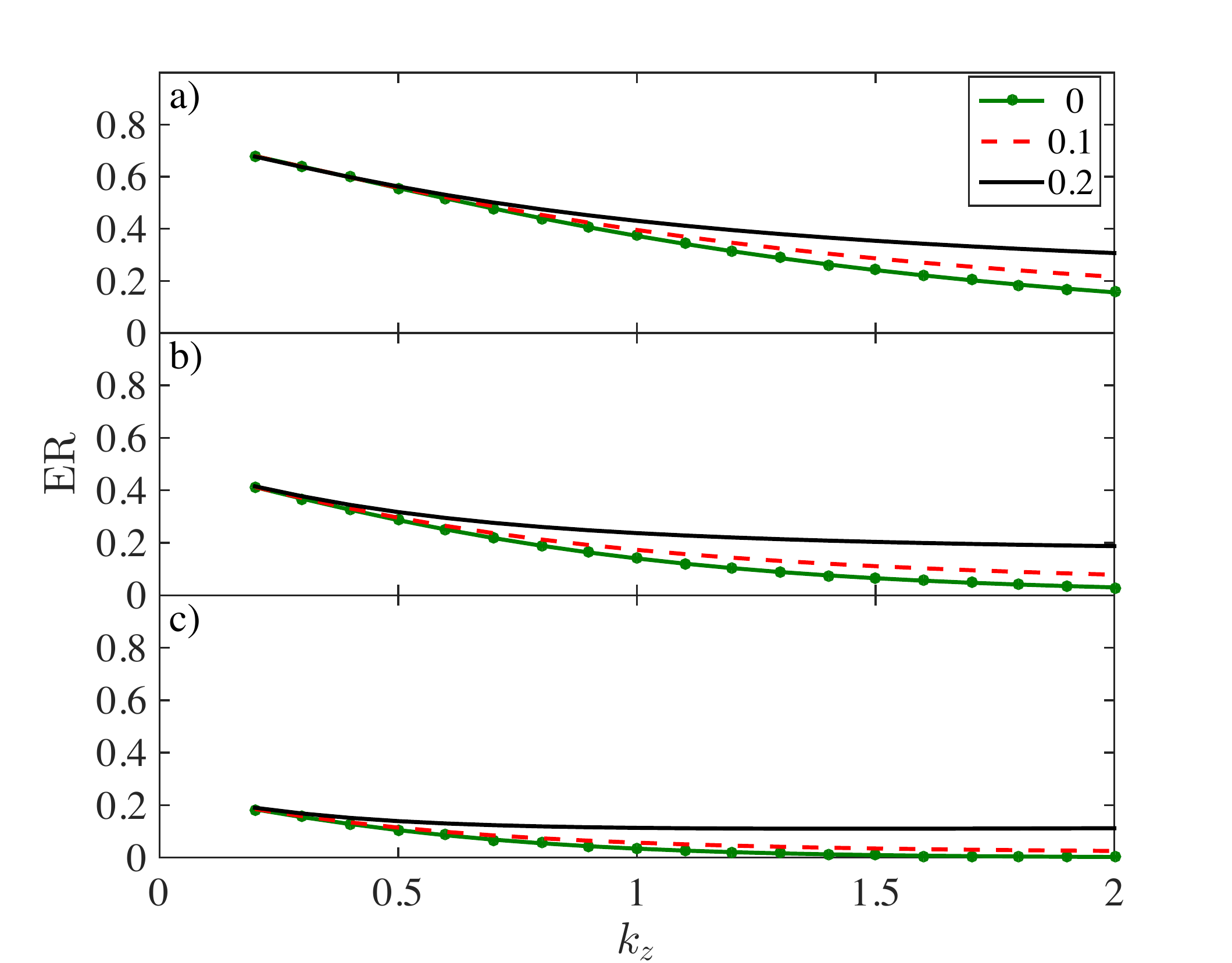}\\
\includegraphics[width=0.33\textwidth]{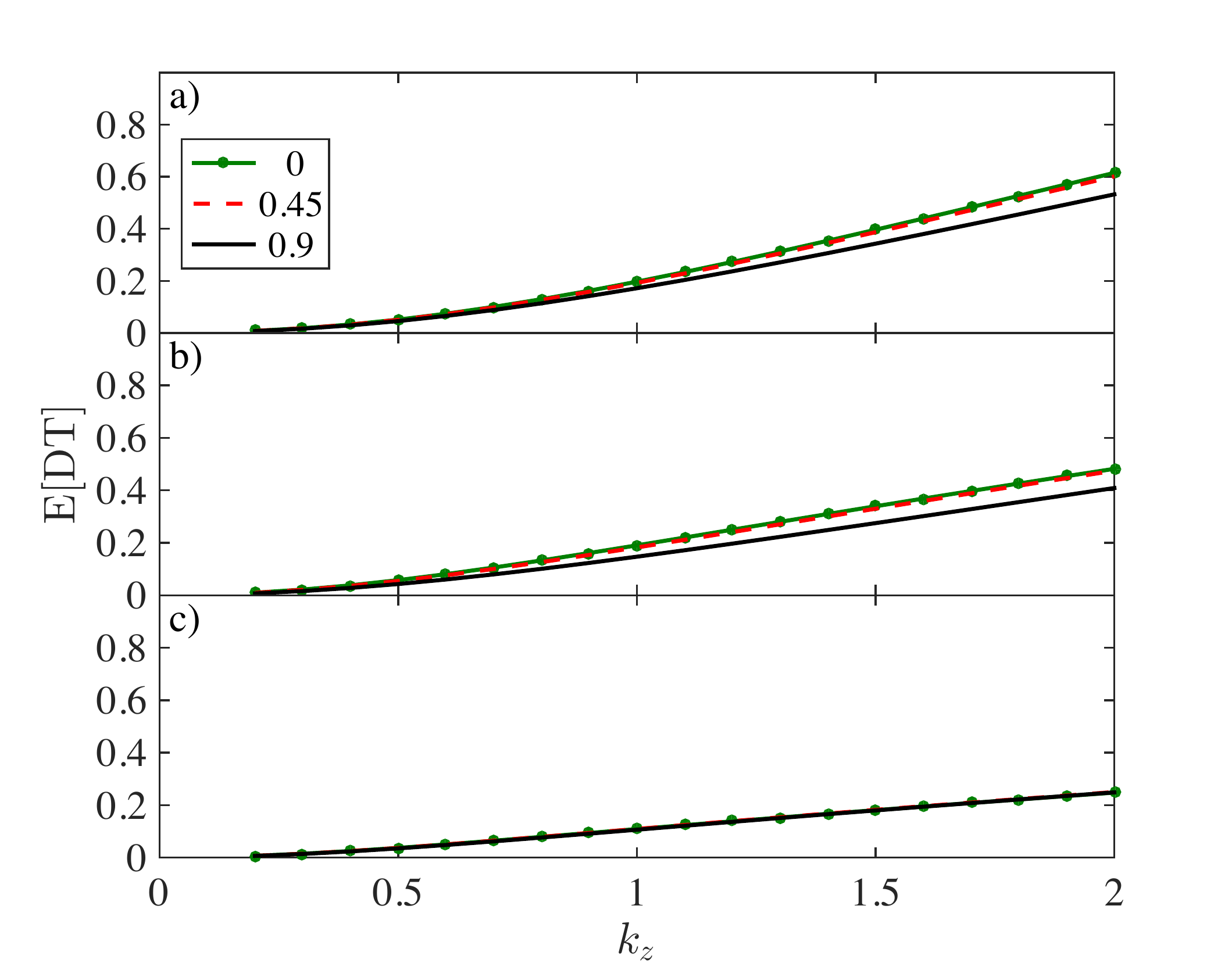}
\includegraphics[width=0.33\textwidth]{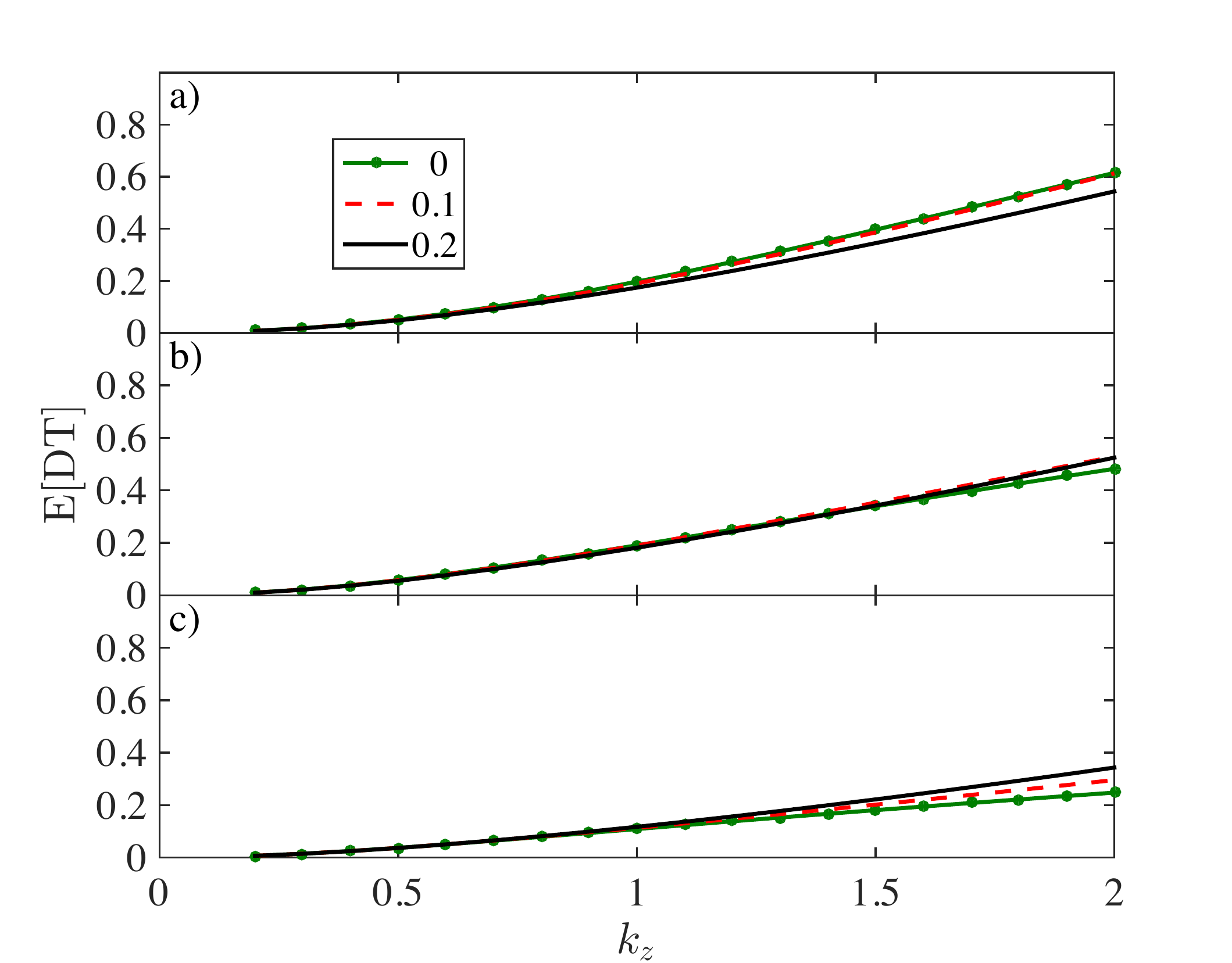}
\includegraphics[width=0.33\textwidth]{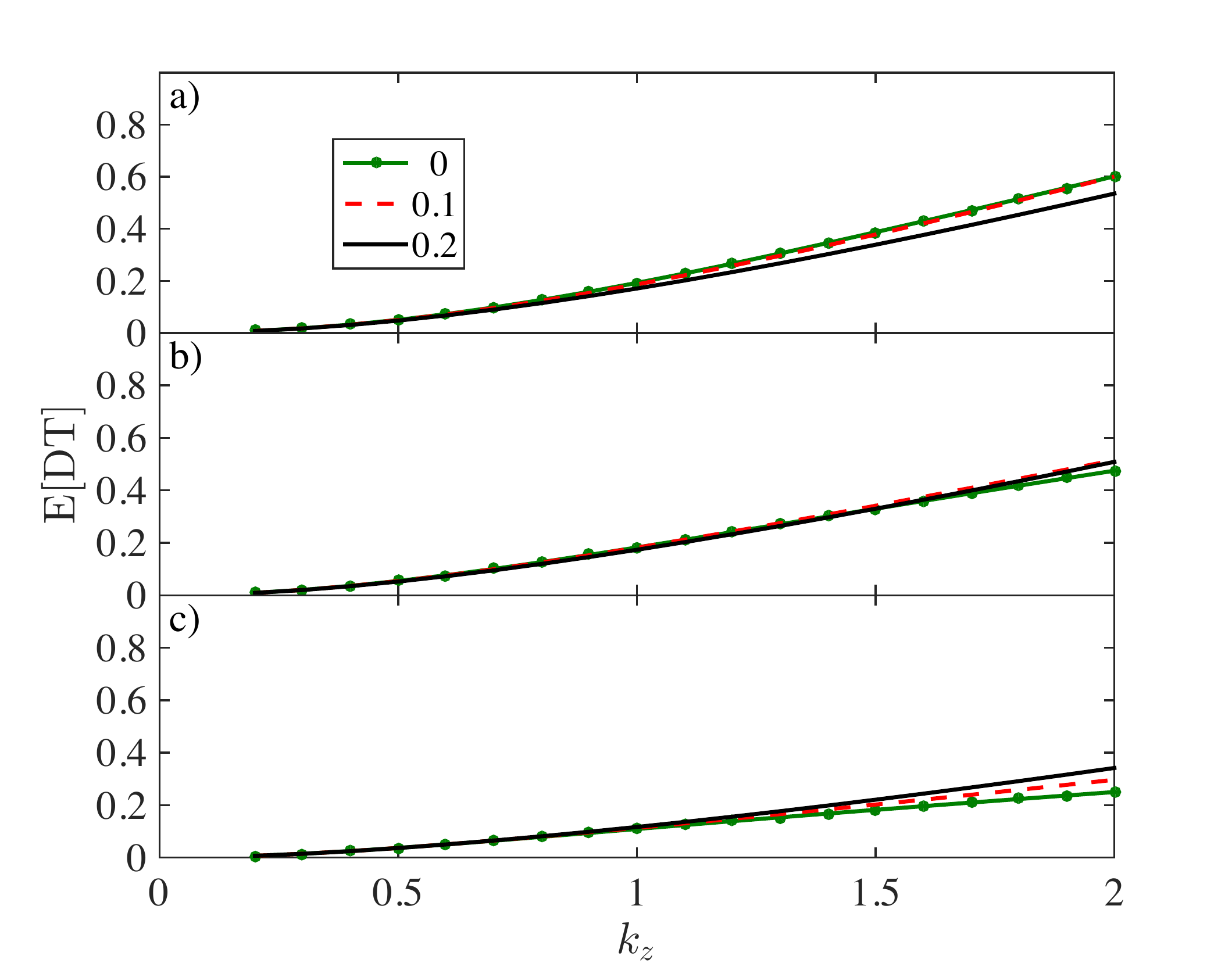} \\
\includegraphics[width=0.33\textwidth]{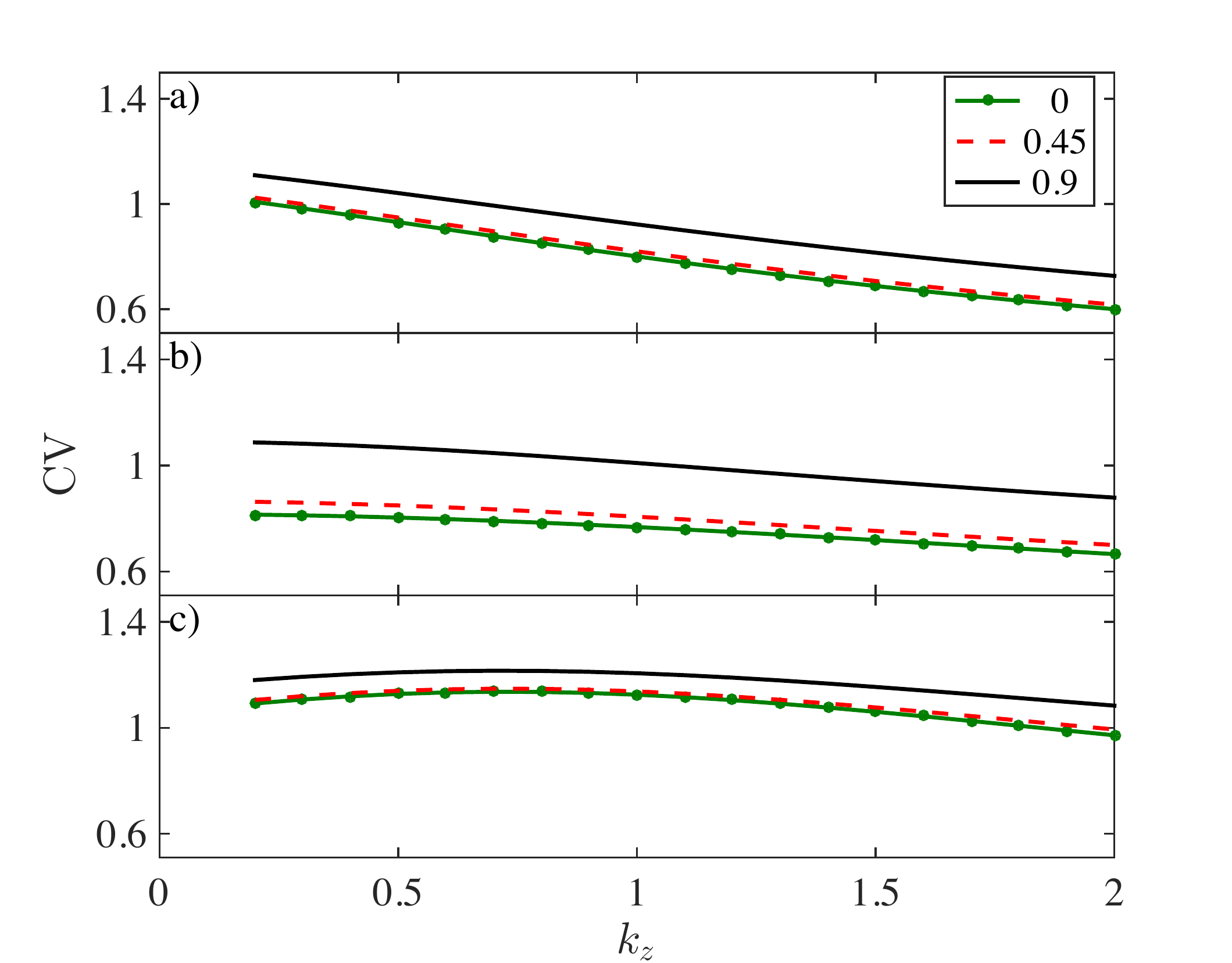}
\includegraphics[width=0.33\textwidth]{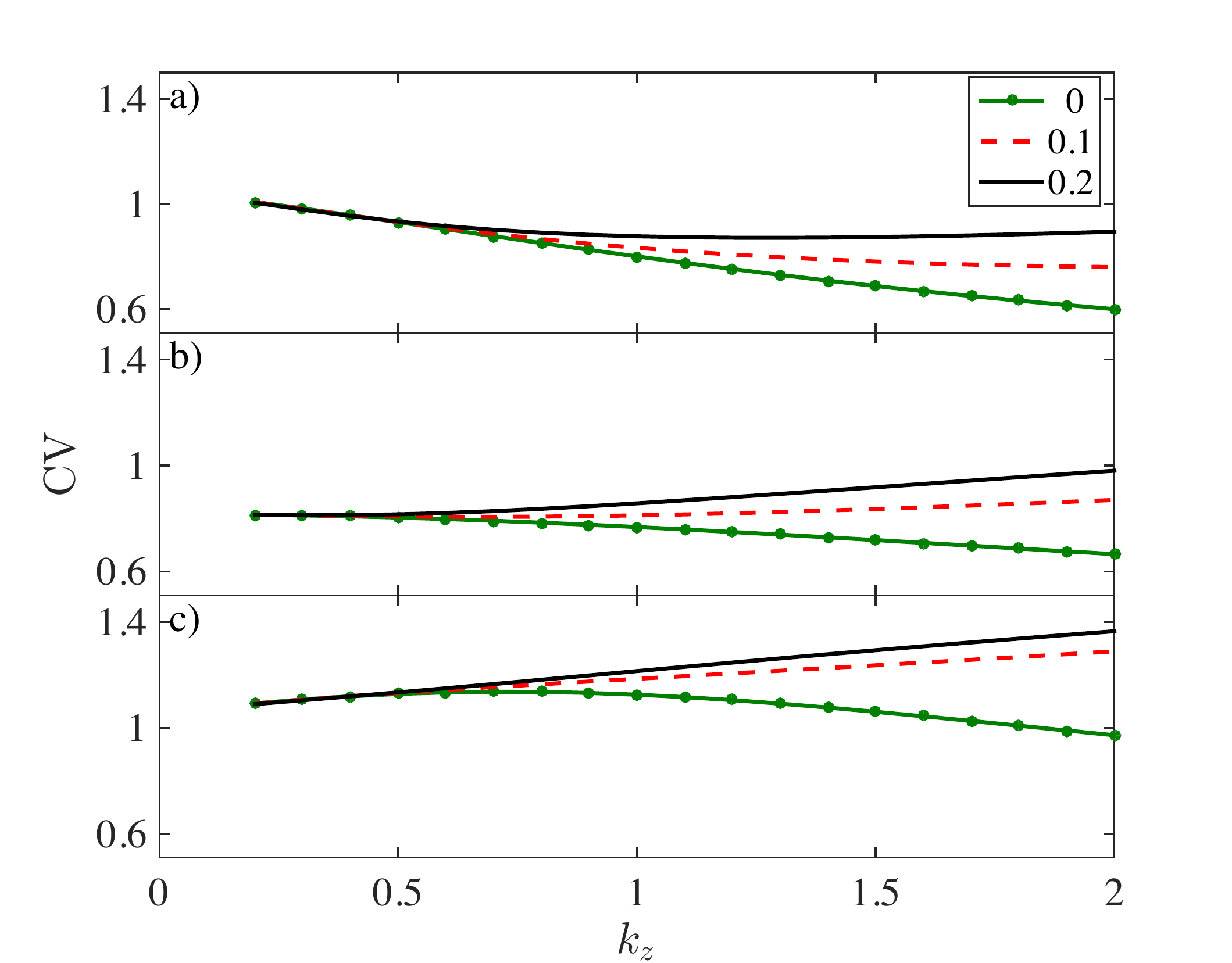}
\includegraphics[width=0.33\textwidth]{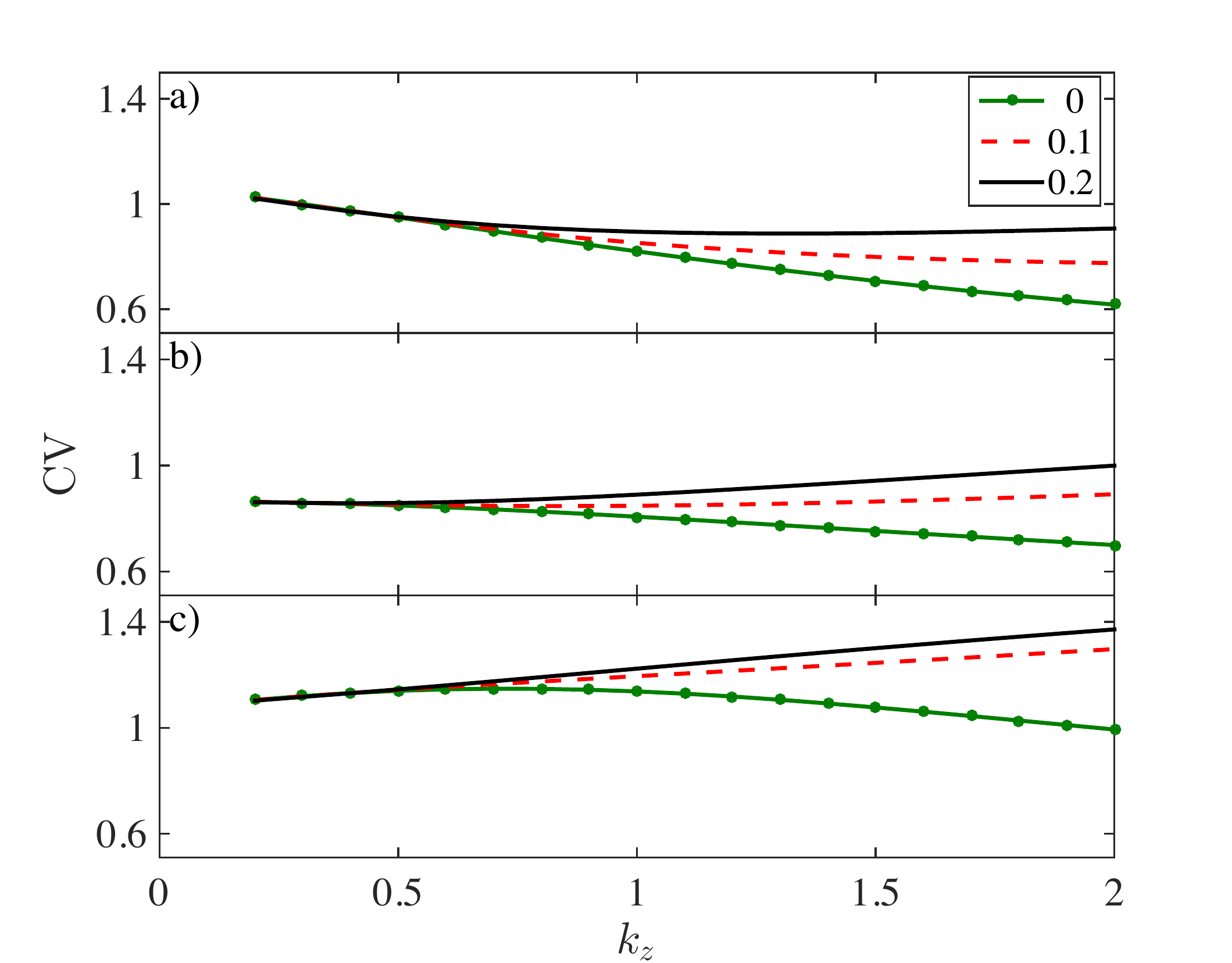} \\
\includegraphics[width=0.33\textwidth]{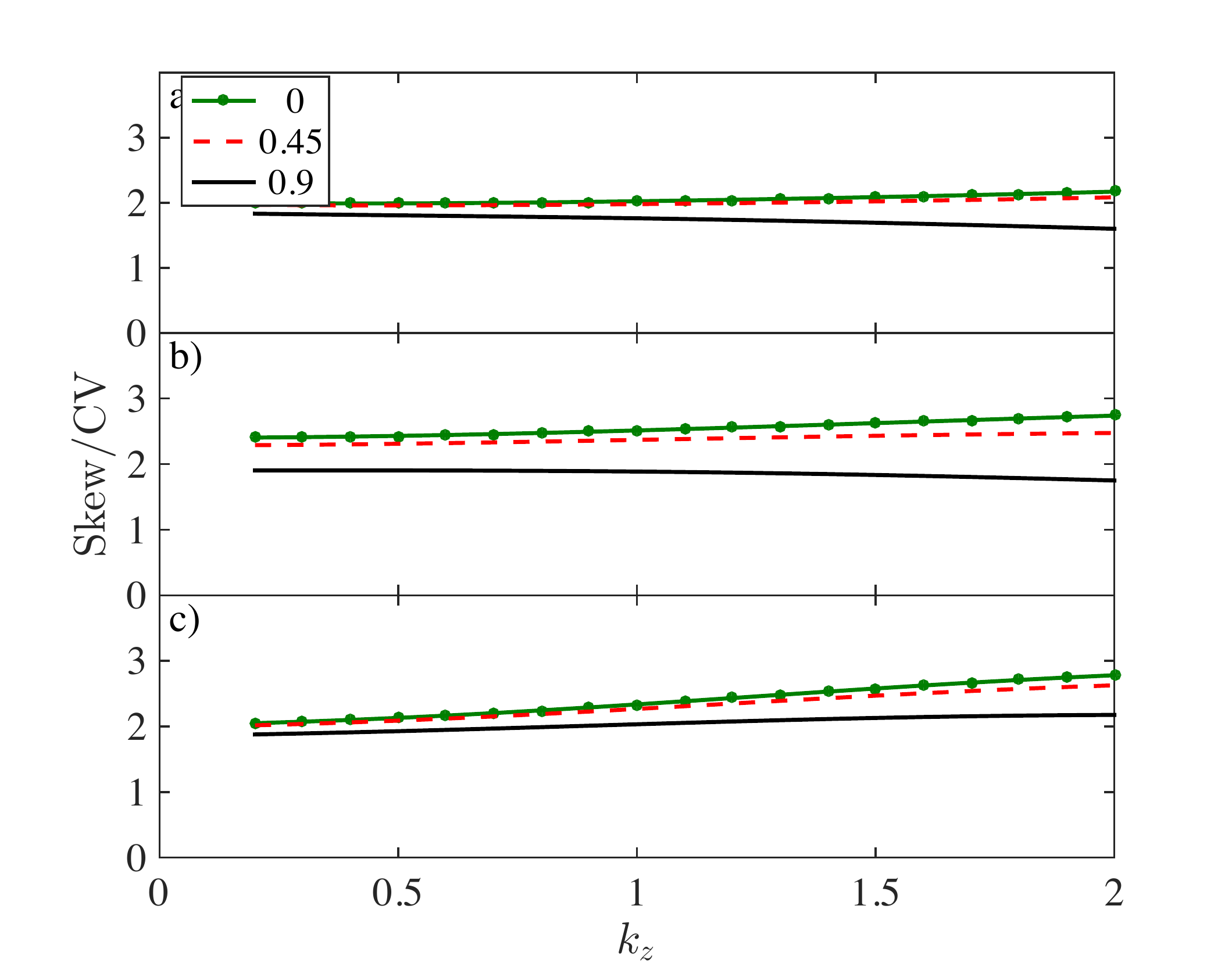}
\includegraphics[width=0.33\textwidth]{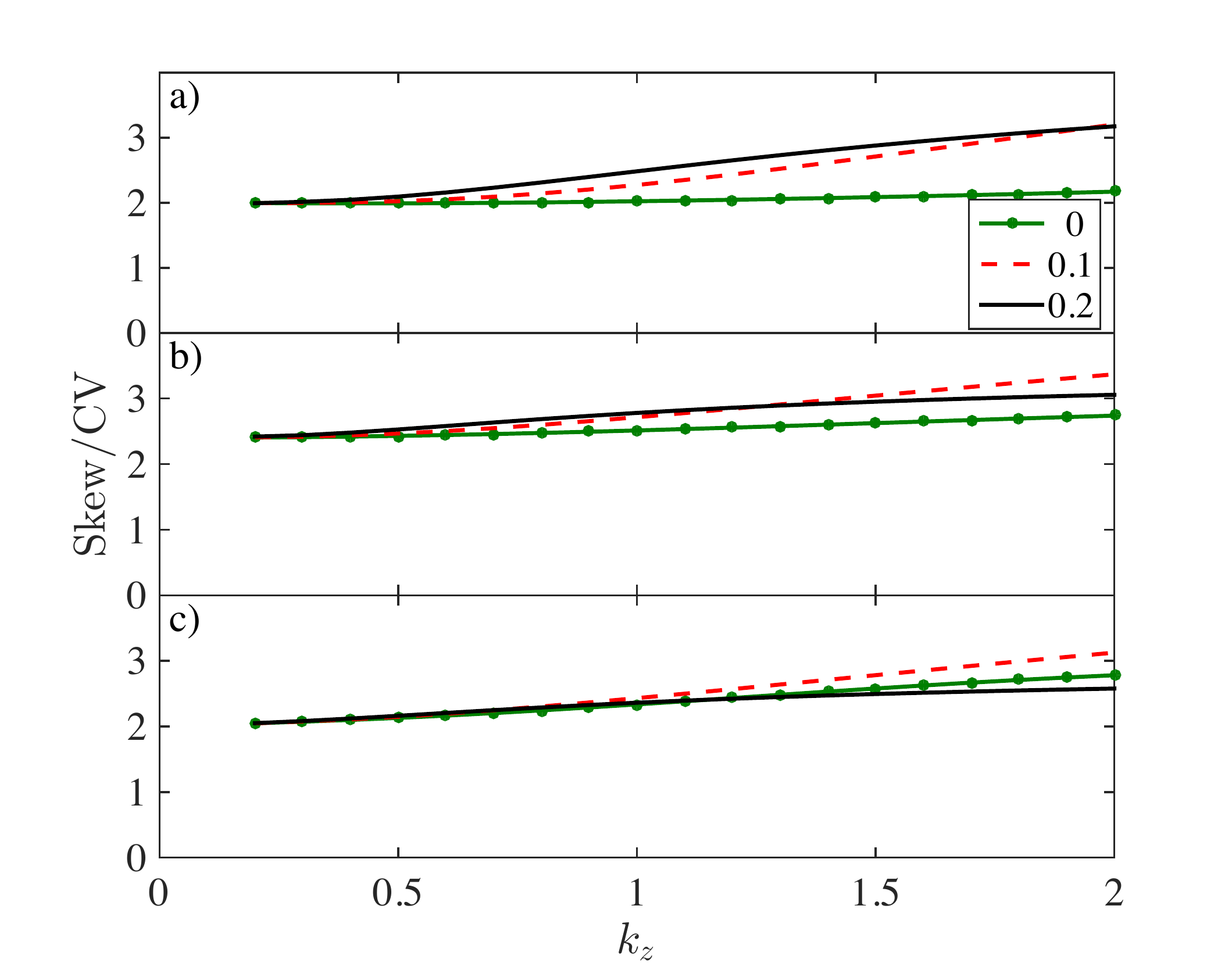}
\includegraphics[width=0.33\textwidth]{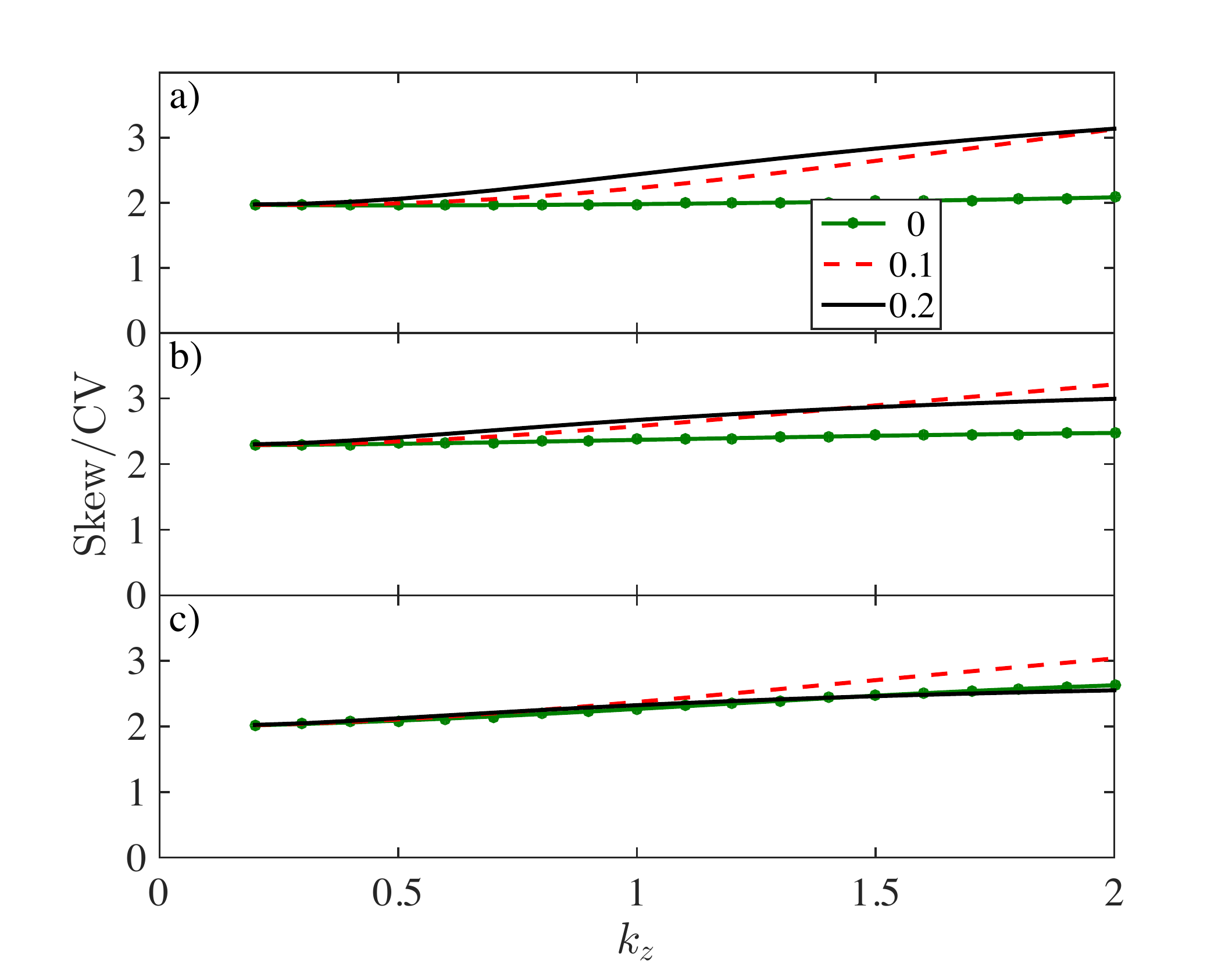}
\caption{Behavior of moments for the extended DDM. 
In all panels $a=0.2$ and $\sigma =0.1$. Three sub-panels in
each panel correspond to $x_0 = - z/2 ,0$ and $z/2$, respectively,
from top to bottom. Left panels correspond to $\sigma_a=0$
and green solid with dots, red dashed, and black solid curves to $\frac{s_x}{2}=0, 0.45 \min \{z -x_0,
x_0+z\}$ and $0.9\min \{z -x_0, x_0+z\}$, respectively. Middle
panels correspond to $s_x=0$ and green solid with dots, red dashed, and black solid curves to
$\sigma_a = 0, 0.1$ and $0.2$, respectively. Right panels are
analogous to middle panels and correspond to $\frac{s_x}{2}= 0.45 \min
\{z -x_0, x_0+z\}$.}
\label{fig:extended-ddm-plots}
\end{figure}

We first discuss how expressions for the moments of decision times
and error rate for the pure DDM can be leveraged to efficiently
compute analogous explicit expressions for the extended DDM. For
clarity, we denote the decision time of the pure DDM for a given drift
rate $a$ and starting point $x_0$ by $\tau(a, x_0)$, and the error
rate by ${\rm ER}(a, x_0)$. The following expressions for the extended DDM are illustrated in Fig.~\ref{fig:extended-ddm-plots}.

The error rate of the extended DDM is the expected value of the error
rate of the pure DDM averaged over the distributions of drift rates and
starting points:
\begin{equation}\label{eq:er-ext-ddm}
\overline {\rm ER} = \expt_{A} \big[ \expt_{X_0} \big[ {\rm ER}
(A, X_0) \big]\big],
\end{equation}
where $\expt_{Y}[\cdot]$ denotes the expected value computed over the
distribution of random variable $Y$. The expectation over the random 
starting point $X_0$ in~\eqref{eq:er-ext-ddm} can be computed 
explicitly as
\begin{equation}\label{eq:Eer-ext-ddm}
\expt_{X_0}[{\rm ER}(a, X_0)] = \frac{e^{-2 k_x} \sinch(2 k_{\delta})
 - e^{-2 k_z} }{ e^{2k_z} - e^{-2k_z}},
\end{equation}
where $k_{\delta} = a \delta / \sigma^2$ and $\sinch(\cdot) : = 
\sinh(\cdot) / (\cdot)$.  Note that this expression 
reduces to Eqn.~\eqref{e.er} for $\delta = 0$, using $\sinch(0) = 1$.

The non-central moments of the decision times can be computed
similarly. In particular, if $T_n (a, x_0)$ is the non-central $n$-th
moment of the decision time for the pure DDM, then the non-central
$n$-th moment for the extended DDM is
\begin{equation}\label{eq:non-cen-n-mom-extnd}
\bar T_n = \expt_{A} \big[ \expt_{X_0} \big[ T_n (A, X_0) \big]\big]. 
\end{equation}
The non-central moments obtained using
Eqn.~\eqref{eq:non-cen-n-mom-extnd} can be used with 
Eqns.~\eqref{e.var}~and~\eqref{e.num.skew} to compute variance and
skewness of decision time for the extended DDM.
Eqn.~\eqref{eq:non-cen-n-mom-extnd} is valid for both unconditional
and conditional moments. The above expressions for the error rate and
expected decision time for the extended DDM can be found
in~\cite[Appendix, pp 761--763]{boga03}.

For unconditional moments, the expectation over $X_0$
in~\eqref{eq:non-cen-n-mom-extnd} can be computed in closed form for
first two moments, which may be written as 
\begin{align} \label{eq:ET-ext-ddm1}
\expt_{X_0}[T_1(a, X_0)] &= \frac{\sigma^2}{a^2} \left( k_z \coth(2 k_z)
 - k_z e^{-2 k_x} \sinch(2 k_{\delta}) \csch(2 k_z) - k_x \right) ; \\
\expt_{X_0}[T_2(a, X_0)] &= \frac{\sigma^4}{a^4}\Big(k_z^2 + 4 k_z^2
 \csch^2(2k_z) + k_z\coth(2k_z) - 4 k_z^2 e^{-2kx_0} \sinch(2k_{\delta})
 \csch(2k_z) \coth(2k_z) \nonumber \\
 & - k_z e^{-2kx_0} \sinch(2k_{\delta})\csch(2k_z)    - k_x  
+ k_x^2 +\frac{k_{\delta}^2}{3} -  2 k_z  k_x \coth(2k_z) \nonumber \\
&  -  2 k_z k_x e^{-2 k_x} \Big(\sinch(2k_{\delta})  + \frac{\sinch(2k_{\delta})
 - \cosh(2k_d)}{2k_x}\Big) \csch(2k_z) \Big).
\label{eq:ET-ext-ddm2}
\end{align}
Expected values in Eqn.~\eqref{eq:non-cen-n-mom-extnd},
involving integrals over the Gaussian distribution that are not
tractable in closed form, can easily be computed numerically, for
example, using Simpson's rule.

Fig.~\ref{fig:extended-ddm-plots} illustrates the behavior of the
unconditional moments of the extended DDM, computed as described
above. The introduction of variability in starting points results in
increase in error rate, decrease in expected decision time, increase
in CV, and decrease in skewness to CV ratio. Introduction of
variability in drift rate also causes increase in error rate,
decrease in expected decision time and increase in CV, but the
skewness to CV ratio increases (compare bottom panels).
Interestingly, for high values of drift rate variability CV is a
monotonically increasing function of $k_z$, in contrast to the
behavior of CV for pure DDM discussed in \S\ref{s.cvs}. The effect of
drift rate variability seems to dominate when both initial condition
and drift rate variability are present.

\subsection{Effect of non-decision time}
\label{ss.nondestime}

Before returning to the extended DDM, we investigate the role
of the non-decision part of the reaction time, the sensory-motor
latency, on its $\rm CV$ and skewness. Recall that $\rm RT = \rm DT +
\subscr{\rm T}{nd}$, where $\subscr{\rm T}{nd}$ is the non-decision
time. We define the following coefficients to characterize the
dependence of $\rm DT$ and $\subscr{\rm T}{nd}$:
\begin{align}
\rho_{11} & = \frac{\expt[(\rm DT - \expt[\rm DT]) (\subscr{\rm T}{nd}
 - \expt[\subscr{\rm T}{nd}])]}{\sqrt{\rm Var[\rm DT] \;
 Var[\subscr{\rm T}{nd}] }} \label{eq:rho11} , \\
\rho_{12} & = \frac{\expt[(\rm DT - \expt[\rm DT]) (\subscr{\rm T}{nd}
 - \expt[\subscr{\rm T}{nd}])^2]}{\sqrt{\rm Var[\rm DT] \;
 \expt[(\subscr{\rm T}{nd} - \expt[\subscr{\rm T}{nd}])^4]}}
 \label{eq:rho12} , \\
\rho_{21} & = \frac{\expt[(\rm DT - \expt[\rm DT])^2 (\subscr{\rm T}{nd}
 - \expt[\subscr{\rm T}{nd}])]}{\sqrt{\expt[(\rm DT - \expt[\rm DT])^4
 \; Var[\subscr{\rm T}{nd}] }}.
\label{eq:rho21}
\end{align}
Note that $\rho_{11}$ is the standard correlation coefficient between
$\rm DT$ and $\subscr{\rm T}{nd}$, and $\rho_{12}$, $\rho_{21}$ can be
interpreted as higher order correlation coefficients. If $\rm DT$ and
$\subscr{\rm T}{nd}$ are independent, then all these correlation
coefficients are zero. In this case, it follows from
the definition of RT that
\begin{align*}
\expt[\rm RT] &= \expt[\rm DT] + \expt[\subscr{\rm T}{nd}] \\
{\rm Var} [\rm RT] & = {\rm Var}[\rm DT] + {\rm Var}[\subscr{\rm T}{nd}]
 + 2 \rho_{11} \sqrt{ {\rm Var}[\rm DT] \; {\rm Var}[\subscr{\rm T}{nd}]} \\
\expt[({\rm RT} -\expt[{\rm RT}])^3 ] & = 
{\rm Skew}[\rm DT] {\rm Var}[\rm DT]^{3/2} + {\rm Skew}[\subscr{\rm T}{nd}]
 {\rm Var}[\subscr{\rm T}{nd}]^{3/2} \\
&+ 3 \rho_{12} \sqrt{{\rm Kur} [\subscr{T}{nd}] {\rm Var}[\rm DT]}
 {\rm Var}[\subscr{\rm T}{nd}] + 
3 \rho_{21} \sqrt{{\rm Kur} [\rm DT] {\rm Var}[\subscr{\rm T}{nd}]}
 {\rm Var}[\rm DT],
\end{align*}
where $\rm Kur[\cdot]$ is the kurtosis\footnote{We consider kurtosis
as the ratio of the fourth central moment and the square of the
variance. This is in contrast to the convention of subtracting $3$
from the above ratio so that the kurtosis of the standard normal
random variable is zero.}. The conditional mean decision time and
variance can be defined similarly  by introducing conditional equivalents of correlation coefficients~(\ref{eq:rho11}-\ref{eq:rho21}). However,  for simplicity of exposition, in the following we assume that non-decision time and decision time are independent; accordingly, the above correlation coefficients are zero.
 Formulae for $\rm CV$ and  $\rm
Skew$ for $\rm RT$'s follow immediately from above
expressions. For use below, we assume $\subscr{\rm T}{nd}$ is uniformly distributed with mean $\expt[\subscr{\rm T}{nd}]$ and range $s_t$.

\subsection{Effects of trial-to-trial variability}
\label{ss.ttotvar}

\begin{figure}[ht!]
\vspace{-0.75cm}
\includegraphics[width=\textwidth]{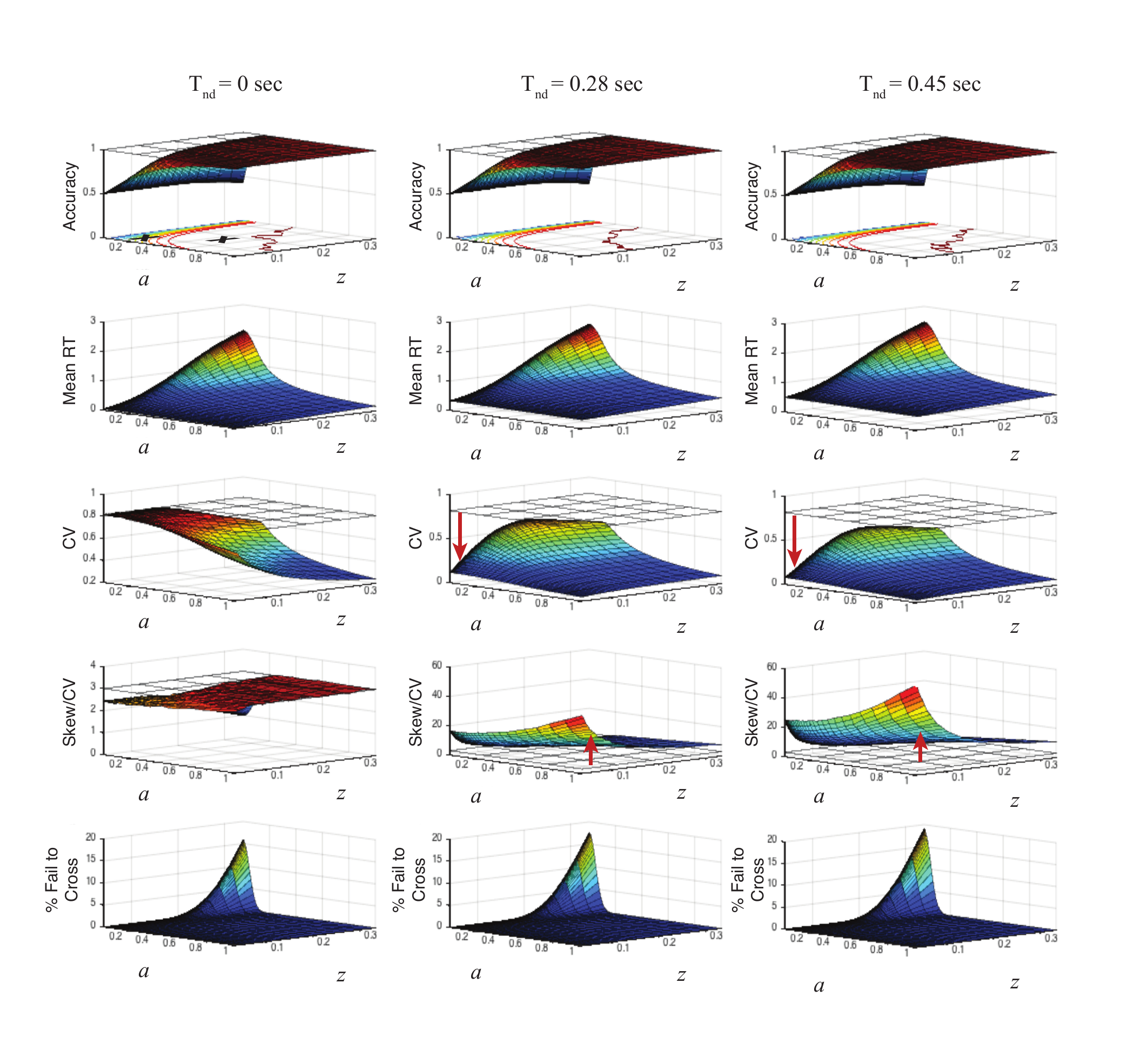}
\vspace{-1.5cm}
\caption{Effect of  deterministic non-decision time $\subscr{\rm T}{nd}$ in
the extended DDM. Left column: $\subscr{\rm T}{nd} =0$ sec. Middle column:
$\subscr{\rm T}{nd}=0.28$ sec. Right column: $\subscr{\rm T}{nd}=0.45$ sec. Curves plotted on
the drift-threshold  plane in top row denote equally spaced contours
of the accuracy surface; red arrows show the effect of increasing
$\subscr{\rm T}{nd}$ on CV and SCV. 
\label{fig:pat-sims-Ter}}
\end{figure}

Seeking to provide a more complete picture, we conducted
simulations of the extended and pure DD models. 
To obtain the following simulation results we used the RTdist package
for graphical processing unit (GPU) based simulation of the DDM
\cite{StijnMeersTuerl2015} to simulate a large subset of the parameter
space spanning the range of plausible parameter values. We simulated
$1,518,750$ parameter combinations in about $5.5$ hours on a Tesla NVIDIA
GPU, with $1$ msec timesteps up to 5 secs maximum RT, with
$10^5$ trials simulated per parameter combination.   In
Fig.~\ref{fig:pat-sims-Ter},  the noise level
was fixed at $\sigma = 0.1$ and we varied mean drift $a$ and threshold
$z$ over the ranges $[0.1, 1.0]$ and $[0.05, 0.3]$ respectively. Fig.~\ref{fig:pat-sims-Ter} shows accuracy, mean RT, CV, skewness to CV ratio (SCV) and the
percentage of trials that failed to cross threshold within  5 secs. (The latter quantity is  small
except for low drift and high threshold, where it rises to $15-20\%$.)
Note that the left hand column of
Fig.~\ref{fig:pat-sims-Ter} show results for the
pure DDM with $\subscr{\rm T}{nd} = 0$, and thus provide standards for
comparison with other cases. See Appendix~\ref{app:figures} for additional simulation results.

The most profound effect on higher moments of the RT distributions is
due to changes in non-decision latency, $\subscr{\rm T}{nd}$, as shown
in Fig.~\ref{fig:pat-sims-Ter}. Specifically, note the dramatic drop
in the CV of RTs as $\subscr{\rm T}{nd}$ increases from 0 to $0.28$ sec,
and the corresponding increase of skewness to CV ratio (red arrows, row 3).

\begin{figure}[ht!]
\centering
\vspace{-0.75cm}
\includegraphics{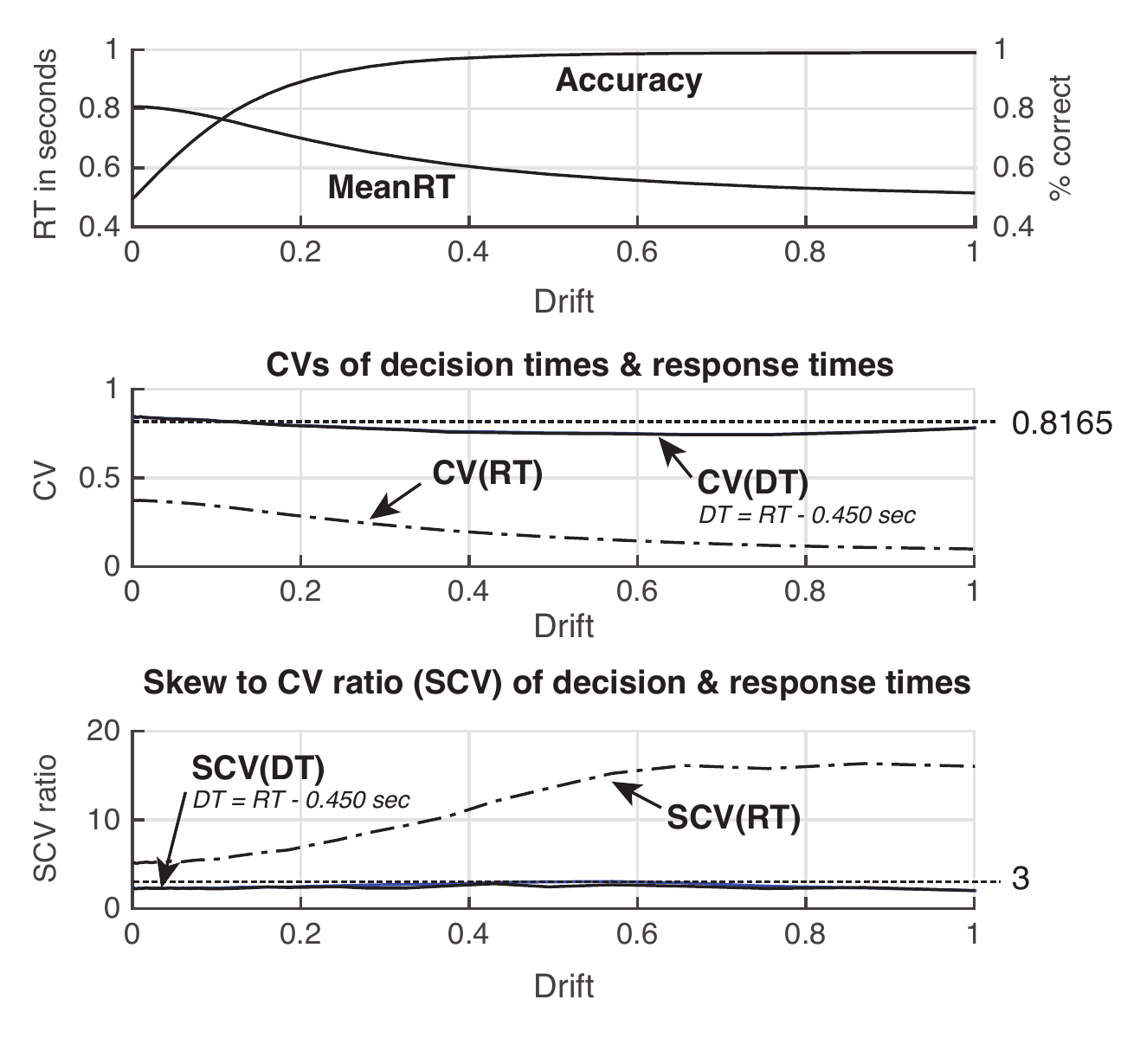}
\vspace{-0.5cm}
\caption{Comparison of CVs and SCVs, as a function of drift, computed
from raw RTs (dot-dashed) and from DTs (solid). DTs have the true,
average non-decision latency of $0.45$ sec subtracted. Representative
levels of extended DDM parameter values were used ($z = 0.06$,
$x_0=0$, $\delta = 0.5 \cdot z$ , $\sigma_a=0.25 \cdot a$, $\expt[\subscr{\rm T}{nd}]=0.45$ sec, and $s_t =0.112$ sec.) 
}
\label{fig:pat-sims-Ter2D}
\end{figure}

Fig.~\ref{fig:pat-sims-Ter2D} shows this phenomenon most clearly,
using  behaviorally plausible values for the extended DDM. When the
correct  expected non-decision latency of $0.45$ sec is subtracted from the RTs,
the CV (middle plot) approaches $\sqrt{2/3} \approx 0.8165$ as drift
approaches 0. Thus researchers may be able to estimate $\subscr{\rm
T}{nd}$ at low accuracy levels when behavior is unbiased toward either
alternative by progressively subtracting from the RT until the CV
approaches $\sqrt{2/3}$  from below  (cf. Proposition~\ref{CVprop} and
Fig.~\ref{fig:cvs}). In contrast, the SCV ratio grows
substantially as drift, and hence accuracy, increase
(Fig.~\ref{fig:pat-sims-Ter}, red arrows, row 4). Researchers may
therefore be able to estimate $\subscr{\rm T}{nd}$ at high drift
levels by subtracting postulated non-decision time from the RT until
the SCV ratio declines to 3. These two heuristics for estimating
$\subscr{\rm T}{nd}$ independently at both low and high levels of
drift may provide robust and easily-computable sanity checks for
constraining the values of $\subscr{\rm T}{nd}$ when using fitting
algorithms.



\section{Conclusion}
\label{s.concl}

We analyzed in detail the first three moments of decision times of the pure and extended DDMs. We derived explicit expressions for unconditional and conditional moments and used these expressions to thoroughly investigate the behavior of the CV and skewness of decision times in terms of two useful parameters: the non-dimensional threshold and non-dimensional initial condition ($k_z$ and $k_x$, Eqn.~\eqref{e.ks}).  These expressions are summarized in Table~\ref{tab:summary}.
The MatLab and R code for these expressions is available at: \verb|https://github.com/PrincetonUniversity/higher_moments_ddm|.

In particular, we computed several limits of interest for the pure DDM.
We established that, for an unbiased starting point ($x_0 = 0$), the CV of decision times is a monotonically decreasing function of $k_z$ and that it approaches $\sqrt{2/3}$ as $k_z \rightarrow 0$ (Proposition~\ref{CVprop} and Fig.~\ref{fig:cvs}).  In the limits of small drift rate and unbiased starting point, we showed that the ratio of skewness to CV approaches $12/5$. Furthermore, for non-zero drift rates and in the limit of large thresholds (high accuracy), we showed that skewness to CV ratio approaches $3$.
We showed that both CV and skewness of decision times diverge  as the starting point approaches either threshold; however, the ratio of skewness to CV is a bounded function of non-dimensional threshold. We also showed that in the limit of large thresholds, these moments match those of first passage times for single-threshold drift-diffusion processes, and we established similar results for conditional CV and skewness of decision times. 
We established that the decision time distribution for the double-threshold DDM converges to he decision time distribution of the single-threshold DDM for large thresholds (Appendix~\ref{app-cumgenfctn}).

\begin{table}[ht!]
\centering
\renewcommand{\arraystretch}{1.4}
\begin{tabular}{|c|c|c|}
\hline
& {\bf 1-threshold} & \bf 2-threshold \\
\hline
{\bf Error rate} & & \\
\hline
 ER &  NA & $ \frac{e^{-2k_x} - e^{-2k_z}}{e^{2k_z} - e^{-2k_z}} $ \\
\hline
{\bf Mean} & & \\
\hline
$\rm \expt[DT]$ & $\frac{\sigma^2}{a^2}(k_z -k_x)$  & $\frac{\sigma^2}{a^2} \left[ k_z \coth(2k_z)
 - k_z e^{-2k_x} \csch(2k_z) - k_x \right]$\\
\hline
$\rm \expt[DT]_+$ & NA   & $\frac{\sigma^2}{a^2} \left( 2k_z \coth(2 k_z) - (k_x + k_z) \coth(k_x+k_z)\right)$\\
\hline

{\bf Variance} & & \\
\hline
$\rm Var$ & $\frac{\sigma^4}{a^4} (k_z -k_x)$ & see equation~\eqref{e.var}\\
\hline
$\rm Var_+$ & NA   & see equation~\eqref{eq:evarc1}\\
\hline
{\bf Coefficient of Variation} & & \\
\hline
$\rm CV$ & $\frac{1}{\sqrt{k_z -k_x}}$ & see equation~\eqref{e.cv}\\
\hline
$\rm CV_+$ & NA   & see equation~\eqref{eq:cvc}\\

\hline
{\bf Skewness} & & \\
\hline
$\rm Skew \times Var^{3/2}$ & $ \frac{3 \sigma^6}{a^6} (k_z -k_x)$ & see equation~\eqref{e.num.skew}\\
\hline
$\rm Skew_+ \times Var_+^{3/2}$  & NA   & see equation~\eqref{eq:skwc1}\\
\hline
\end{tabular}
\caption{Summary of expressions of error rate and  moments of decision time.  \label{tab:summary}}
\end{table}

We then derived analytic and semi-analytic expressions for the moments of decision times of the extended DDM, and numerically investigated the effects of  trial-to-trial variability in starting points and drift rates on the DDM's performance. We observed that variability in drift rate appears to dominate these effects, compared to starting point variability. 

Finally, we investigated the effect of non-decision times (sensory-motor latencies, $\subscr{\rm T}{nd}$) on decision-making performance. 
We observed that CVs of reaction times (${\rm DT} + \subscr{\rm T}{nd}$) decrease and their skewness-to-CV ratios increase as mean $\subscr{\rm T}{nd}$'s increase (Fig.~\ref{fig:pat-sims-Ter}). We propose that the decrease in CVs and increase in skewness-to-CV ratios could be used to estimate non-decision times in low and high accuracy regimes respectively (see Fig.~\ref{fig:pat-sims-Ter2D}). The development of rigorous methods using these metrics to estimate non-decision time is a potential avenue for future research.

It should be noted that difficulties in estimating higher moments of empirical RT data have been highlighted in the literature~\cite{Luce-bk86,RatcliffMoments1993}. However, at least in the context of interval-timing tasks, predictions regarding CV and skewness have proved to be useful in discriminating between different models \cite{Simenetal-JNSci11,SimenVP}. Furthermore, it is possible that future two-alternative perceptual decision task designs could be found that would yield  data amenable to estimation of higher moments, in which case, the expressions we derive here may prove helpful.

More generally, the explicit expressions derived in this paper can be used to quickly identify ranges of parameters that are relevant to fitting specific behavioral data sets, thereby reducing the volumes of multi-dimensional space in which parameter fits need to be run. In principle, the cumulant generating function method outlined in Appendix~A can be used to produce formulae for fourth and higher moments, and although the results will be complex, they and their limiting behaviors may also provide guidance for parameter fitting.

\section*{Acknowledgements}

This work was jointly supported by NIH Brain Initiative grant 
1-U01-NS090514-01 (PH), NSF-CRCNS grant DMS-1430077 and the Insley-Blair Pyne Fund (VS), and  an OKUM Fellowship (PS). The authors thank
Jonathan Cohen and Michael Shvartsman for helpful suggestions.

\bibliographystyle{plain}

\bibliography{ddmbib.bib} 


\section*{Appendices}

\appendix

\section{Error rate and unconditional variance of decision time}
\label{app-mean}

In this section we show that error rate \eqref{e.er} and expected decision time \eqref{e.mdt} are equivalent to the expressions given in the subsection ``The Drift Diffusion Model'' of~\cite[Appendix, Eqns.~(A27-31)]{boga03}. In our notation, the quantities $\tilde z$ and $\tilde a$ defined in~\cite{boga03} are 
\[
\tilde z = \frac{z}{a}, \quad \text{and} \quad \tilde a = \frac{a^2}{\sigma^2}. 
\]
Define $\tilde x_0 = x_0/a$. Note that $k_z = \tilde z \tilde a$ and $k_x = \tilde a \tilde x_0$.  Also note that $\tilde x_0$ and $x_0$ are referred to as $x_0$ and $y_0$, respectively in~\cite{boga03}.

The expression \eqref{e.er} for error rate may be rewritten as follows
\begin{align}
\mathrm{ER} =\frac{e^{-2k_x} - e^{-2k_z}}{e^{2k_z} - e^{-2k_z}} =  \frac{1 - e^{-2k_z}}{e^{2k_z} - e^{-2k_z}} 
- \frac{1- e^{-2k_x} }{e^{2k_z} - e^{-2k_z}}
&= \frac{e^{2k_z} - 1}{e^{4k_z} - 1} - \frac{1- e^{-2k_x} }{e^{2k_z} - e^{-2k_z}} 
\nonumber \\
= \frac{e^{2k_z} - 1}{(e^{2k_z} + 1)(e^{2k_z} - 1)} - \frac{1- e^{-2k_x} }{e^{2k_z} - e^{-2k_z}}
&=  \frac{1}{1 + e^{2k_z}} - \frac{1- e^{-2k_x} }{e^{2k_z} - e^{-2k_z}} \nonumber\\
& =  \frac{1}{1 + e^{2 \tilde{z} \tilde{a}} } - \left[ \frac{1- e^{-2 \tilde{x}_0 \tilde{a} }}{e^{2 \tilde{z} \tilde{a}} - e^{-2 \tilde{z} \tilde{a}} } \right],
\label{e.D2ER}
\end{align} 
which is identical to the ER expression in~\cite{boga03}. Similarly, 
\begin{align}
\mathbb{E}[{\rm DT}] &= \frac{\sigma^2}{a^2} \left[ k_z \coth(2k_z) - k_z e^{-2k_x} \csch(2k_z) - k_x \right] \nonumber \\
&=\frac{\sigma^2}{a^2} k_z \left[ \coth(2k_z) -  \csch(2k_z) + (1- e^{-2k_x}) \csch(2k_z) - \frac{k_x}{k_z} \right] \nonumber \\
&=\frac{z}{a} \left[ \frac{e^{2k_z} + e^{-2k_z} -2}{e^{2k_z} -e^{-2k_z}}  + (1- e^{-2k_x}) \csch(2k_z) - \frac{x_0}{z} \right] \nonumber \\
& = \frac{z}{a} \tanh(k_z) + \frac{2z}{a} \frac{(1- e^{-2k_x})}{e^{2k_z} -e^{-2 k_z}} - \frac{x_0}{a} \nonumber \\
& = \tilde z \tanh(\tilde z \tilde a) +  \frac{2 \tilde z(1- e^{-2 \tilde a \tilde x_0})}{e^{2 \tilde a \tilde z } -e^{-2 \tilde z \tilde a}} - \tilde x_0,
\label{e.D2DT}
\end{align}
which is identical to  the expected decision time expression in~\cite{boga03}.

\section{Unconditional variance of decision time}
\label{app-variance}

The second moment of the decision time $T_2$ is the solution of 
the following linear ODE: 
\begin{equation}
\label{eq:second-moment-ode}
a \frac{d T_2}{dx_0} + \frac{\sigma^2}{2} \frac{d^2 T_2}{d x^2_0} =
 -2 \mathbb{E}[{\rm DT}],
\end{equation}
with boundary conditions $T_2(\pm z) = 0$ (e.g.~\cite[\S5.5.1;
see Eqn.~(5.5.19) for the general $n$'th moment ODE]{gard09}).
To solve Eqn.~\eqref{eq:second-moment-ode} we first rewrite
$\mathbb{E}[{\rm DT}]$ to make dependence on the starting point
$x_0$ explicit:
\[
\mathbb{E}[{\rm DT}] = \alpha_1 - \alpha_2 e^{-2 kx_0} -\frac{x_0}{a} .
\]
Here $\alpha_1 = \frac{z}{a} \coth(2 k_z)$, $\alpha_2= \frac{z}{a}
\csch(2k_z)$ and unlike $k_z, k_x$ defined above, $k =
\frac{a}{\sigma^2}$ is independent of $z$ and $x_0$. A particular
solution to~\eqref{eq:second-moment-ode} is
\[
T_2^p = \frac{x_0^2}{a^2} -\alpha_3 x_0 - \frac{2 \alpha_2}{a} x_0 e^{-2 k x_0},
\]
where $\alpha_3 = \frac{2}{a}(\alpha_1 + \frac{1}{2 k a})$, and the
general solution takes the form
\[
T_2(x_0) = c_1 + c_2 e^{-2 k x_0} + \frac{x_0^2}{a^2} -\alpha_3 x_0 -
 \frac{2 \alpha_2}{a} x_0 e^{-2 k x_0}.
\]
Substituting the boundary conditions $T_2(\pm z) =0$, and solving
for $c_1$ and $c_2$, we obtain
\begin{align*}
c_1 & = \frac{2 z^2}{a^2}  \coth^2(2k_z) + \frac{z}{ka^2} \coth(2 k_z)
 -\frac{z^2}{a^2} + \frac{2z^2}{a^2}  \csch^2(2k_z)\\
&=\frac{z^2}{a^2} + \frac{4 z^2}{a^2} \csch^2(2k_z)+ \frac{z}{ka^2}
 \coth(2k_z), \quad \text{and }  \\
c_2 &= -\frac{4z^2}{a^2 } \csch(2k_z) \coth(2k_z) - \frac{z}{ka^2 }
 \csch(2k_z) ,
\end{align*}
and we therefore find that
\begin{align}
T_2 = \frac{z^2}{a^2} + \frac{4 z^2}{a^2} \csch^2(2k_z) &+
 \frac{\sigma^2 z}{a^3} \coth(2k_z) 
- \frac{4z^2 e^{-2kx_0}}{a^2} \csch(2k_z) \coth(2k_z)
 - \frac{\sigma^2 z e^{-2kx_0}}{a^3} \csch(2k_z)  \nonumber \\
&+ \frac{x_0^2}{a^2} -  \frac{2 z  x_0}{a^2}  \coth(2k_z)
 - \frac{\sigma^2 x_0}{a^3}  -  \frac{2 z x_0 e^{-2 k x_0}}{a^2} \csch(2k_z). 
\label{eq:T2a}
\end{align}
We can now obtain the expression for the variance of decision time:
\begin{align}
\mathrm{Var}&= 
T_2 - \mathbb{E}[{\rm DT}]^2  \nonumber \\
&= \frac{3 z^2}{a^2} \csch^2(2k_z) + \frac{\sigma^2 z}{a^3} \coth(2k_z)
 -\frac{\sigma^2 x_0}{ a^3} -\frac{2z^2 e^{-2 kx_0}}{a^2} \csch(2k_z)
 \coth(2k_z) \nonumber \\
&- \frac{\sigma^2 z e^{- 2kx_0}}{a^3}  \csch(2k_z)
 -\frac{4z x_0 e^{-2kx_0}}{a^2 } \csch(2k_z) -\frac{z^2 e^{-4 k x_0}}{a^2 }
 \csch^2(2k_z).
\label{eq:vara1}
\end{align}
Equivalently, we may write
\begin{align}
\mathrm{Var} &= \frac{\sigma^4}{a^4} \left[3 k_z^2 \csch^2(2k_z)
 - 2 k_z^2 e^{-2 k_x} \csch(2k_z) \coth(2k_z) - 4 k_z k_x e^{-2k_x}
 \csch(2k_z) \right. \nonumber \\ 
 & \left. - k_z^2 e^{-4 k_x} \csch^2(2k_z) + k_z \coth(2k_z) 
 - k_z e^{-2k_x} \csch(2k_z) - k_x \right] .
\label{e.var-repeat}
\end{align}

\section{Method for computation of conditional moments}
\label{app-cumgenfctn}

The \emph{moment generating function} $\map{M_X}{\mc H}{\real_{>0}}$
of a random variable $X$ is defined by
\[
M_{X}(\alpha) : = \expt[e^{\alpha X}] ,
\]
provided the expectation exists for each $\alpha$ in some neighborhood
of zero, i.e.,  for each $\alpha \in \mc H$, where $\mc H \subset
\real$ is some interval containing zero. The moment generating
function is a special case of the \emph{characteristic function}
defined on the complex plane (see~\cite[\S5.7, Theorem~12]{GG-DR:01}),
 and from it the \emph{cumulant generating function}
$\map{K_X}{\mc H}{\real}$ of $X$ can be obtained by taking the natural
logarithm:
\begin{equation}
\label{eq:mgf0}
K_X(\alpha) = \log M_X(\alpha).
\end{equation}
The $n$-th cumulant $\kappa_n$ of $X$ is defined as $\kappa_n =
\frac{d^n K_X(\alpha)}{d \alpha^n} \big|_{\alpha=0}$, or equivalently
$K_X(\alpha) = \sum_{n=1}^\infty \frac{\kappa_n \alpha^n}{n !} $.
It can then be shown that 
\[
\kappa_1 = \mu_1, \quad \kappa_2 = \supscr{\mu}{cen}_2, 
\quad \kappa_3 =\supscr{\mu}{cen}_3, \quad \text{and } 
\kappa_4 = \supscr{\mu}{cen}_4 - 3 \kappa_2^2 ,
\]
where $\mu_n=\expt[X^n]$ and $\supscr{\mu}{cen}_n = \expt[(X -
\expt[X])^n]$ denote the $n$th non-central and central moments. Thus,
successive moments of the distribution from which $X$ is drawn can be
generated from $M_{X}(\alpha)$. For further details and derivations of
moment generating functions, see  \cite[Chap 4, \S6]{Gut:07} and
\cite[\S2.6]{gard09}.

We now derive the moment generating function for DTs of the 
DDM~\eqref{e.dd1}. We define 
$\map{\subscr{M}{DT}}{\mc A}{\real_{>0}}$, $\map{M_+}
{\mc A}{\real_{>0}}$, and $\map{M_-}{\mc A}{\real_{>0}}$
by
\begin{equation}
\label{eq:mgfs1}
\subscr{M}{DT}(\alpha) = \expt[e^{\alpha \tau}], \quad
M_+(\alpha) = \expt[e^{\alpha \tau} | x(\tau) =z],
 \quad \text{and } M_-(\alpha) = \expt[e^{\alpha \tau}
 |x(\tau) =- z],
\end{equation}
where $\mc A \subset \real$ is some interval containing zero in which
the above expectations exist. $\subscr{M}{DT}(\alpha)$, $M_+(\alpha)$
and $M_-(\alpha)$ are, respectively, the moment generating functions
for unconditional decision times (for all responses) and for decision times conditioned on
correct responses and on errors. Expressions for these functions are
well known in the literature~(e.g. \cite{ANB-PS:02}). Here, for
completeness, we derive them from first principles. 

We begin by deriving an expression for $M_+(\alpha)$.  We note that
for a given set of parameters $a, \sigma, z$, and $\alpha$,
$M_+(\alpha)$ depends only on $x_0$. Let $\tau(x_0)$ denote the decision time (DT)
starting from initial condition $x_0$.  Define
$\map{g}{\real}{\real_{>0}}$ as the map from initial condition $x_0$
to  $M_+(\alpha) \prob(x(\tau)=z)$, i.e.,
\begin{equation}
\label{eq:gx0}
g(x_0) = \expt[e^{\alpha \tau(x_0)} \bs 1(x(\tau(x_0))=z)] , 
\end{equation}
where $\bs 1 (\cdot)$ is the indicator function.

Consider the evolution of the DDM~\eqref{e.dd1} starting from $x_0$
at $t=0$ for an infinitesimal duration $h \in \real_{>0}$. Let $X_h :=
x(h) = x_0 + a h +\sigma  W(h)$. It follows that
\begin{align*}
g(x_0) & = \expt_{X_h} \expt_{\tau(X_h)} [e^{\alpha (h + \tau(X_h))}] \\
& = e^{\alpha h} \expt_{X_h} [g(X_h)] \\
& = e^{\alpha h} \Big( g(x_0) + \frac{d g}{d x_0} a h + \frac{1}{2}
 \frac{d^2 g}{ d x_0^2} \sigma^2 h \Big) + O(h^2),
\end{align*}
where $O(h^2)$ represents terms of order $h^2$ and higher.
Rearranging terms and setting $h \to 0^+$, we obtain the following
ODE for $g$
\begin{equation}
\label{eq:mgf-ode}
\frac{\sigma^2 }{2} \frac{d^2 g}{ d x_0^2}  + a \frac{d g}{d x_0} + \alpha g=0,
\end{equation}
with boundary conditions $g(z) = 1$ and $g(-z)=0$. The solution
to~\eqref{eq:mgf-ode} is of the form $g(x_0) = \zeta_1 e^{\lambda_1 x_0}
 + \zeta_2 e^{\lambda_2 x_0}$, where $\lambda_1$ and $\lambda_2$ are
roots of the equation $\sigma^2 \lambda^2/2 + a \lambda + \alpha =0$,
 i.e., 
\[
\lambda_1 = \frac{-a -\sqrt{a^2 -2 \alpha \sigma^2}} {\sigma^2},
\quad \text{and} \quad 
\lambda_2 = \frac{-a +\sqrt{a^2 -2 \alpha \sigma^2}} {\sigma^2}.
\]
Substituting the boundary conditions, we get two simultaneous equations
\[
\zeta_1 e^{\lambda_1 z} + \zeta_2 e^{\lambda_2 z} =1, \quad \text{and }
 \quad \zeta_1 e^{-\lambda_1 z} + \zeta_2 e^{-\lambda_2 z} =0, 
\]
the solution to which is
\[
\zeta_1 = \frac{e^{\lambda_1 z}}{e^{2\lambda_1 z} - e^{2 \lambda_2 z}},
 \quad \text{and} \quad 
\zeta_2 = - \frac{e^{\lambda_2 z}}{e^{2\lambda_1 z} - e^{2 \lambda_2 z}},
\]
and consequently,
\[
g(x_0) =  \frac{e^{\lambda_1 (z+x_0)}- e^{\lambda_2 (z+x_0)}}{e^{2\lambda_1 z} - e^{2 \lambda_2 z}} = \frac{ e^{-a(z+x_0)/\sigma^2}}{e^{-2 a z/\sigma^2}} 
\frac{\sinh(\frac{(z+x_0)\sqrt{a^2 -2 \alpha \sigma^2}}{\sigma^2}}{\sinh(\frac{2z\sqrt{a^2 -2 \alpha \sigma^2}}{\sigma^2})} =
 e^{\frac{a(z-x_0)}{\sigma^2}}
\frac{\sinh(\frac{(z+x_0)\sqrt{a^2 -2 \alpha \sigma^2}}{\sigma^2})}{\sinh(\frac{2z\sqrt{a^2 -2 \alpha \sigma^2}}{\sigma^2})} .
\]

Thus, recalling the definition \eqref{eq:gx0} of $g(x_0)$, the 
moment-generating function conditioned on correct decisions is 
\begin{equation}
\label{eq:mgfc}
M_+ (\alpha)=\expt[e^{\alpha \tau} | x(\tau) =z] = \frac{e^{\frac{a(z-x_0)}{\sigma^2}}}{\prob(x(\tau)=z)} 
\frac{\sinh(\frac{(z+x_0)\sqrt{a^2 -2 \alpha \sigma^2}}{\sigma^2})}{\sinh(\frac{2z\sqrt{a^2 -2 \alpha \sigma^2}}{\sigma^2})},
\end{equation}
and substituting this in the definition \eqref{eq:mgf0} yields the
cumulant generating function \eqref{eq:cgfc} used in \S\ref{s.cond}.

Similarly, we may obtain analogous expressions for incorrect decisions
\begin{equation}
\label{eq:mgfe}
M_- (\alpha)=\expt[e^{\alpha \tau} | x(\tau) =-z] = \frac{e^{\frac{-a(z+x_0)}{\sigma^2}}}{\prob(x(\tau)=-z)} 
\frac{\sinh(\frac{(z-x_0)\sqrt{a^2 -2 \alpha \sigma^2}}{\sigma^2})}{\sinh(\frac{2z\sqrt{a^2 -2 \alpha \sigma^2}}{\sigma^2})},
\end{equation}
and for all decisions, correct and incorrect:
\begin{equation}
\label{eq:mgfa}
\subscr{M}{DT} (\alpha)=\expt[e^{\alpha \tau}] = e^{\frac{-a(z+x_0)}{\sigma^2}}
\frac{\sinh(\frac{(z-x_0)\sqrt{a^2 -2 \alpha \sigma^2}}{\sigma^2})}{\sinh(\frac{2z\sqrt{a^2 -2 \alpha \sigma^2}}{\sigma^2})} +e^{\frac{a(z-x_0)}{\sigma^2}}
\frac{\sinh(\frac{(z+x_0)\sqrt{a^2 -2 \alpha \sigma^2}}{\sigma^2})}{\sinh(\frac{2z\sqrt{a^2 -2 \alpha \sigma^2}}{\sigma^2})}. 
\end{equation}
It should be noted that in the limit $ z \to \infty$
\[
\subscr{M}{DT} (\alpha) = \exp \left(\frac{az}{\sigma^2} \left( 1 - \sqrt{1 -\frac{2 \alpha \sigma^2}{a^2}}\right) \right),
\]
which is the moment generating function of the Wald distribution~\cite[Eq. 2.0.1]{ANB-PS:02}, i.e., the decision time distribution of the single-threshold DDM. Consequently, the decision time distribution of the double-threshold DDM converges to the 
the decision time distribution of the single-threshold DDM as $z \to \infty$.

\section{Proof of Proposition~\ref{CVprop}} 
\label{app-proof-CVprop}

We first show that the CV for the single-threshold DDM provides an
upper bound for the double threshold case. Canceling the $\sqrt{1/k_z}$
terms in the inequality \eqref{e.cv00}, squaring, rearranging and
dividing by $2 e^{-2 k_z}$ shows that this is equivalent to
\beq
(1-e^{-2 k_z})^2 > (1 - e^{-4 k_z} - 4 k_z e^{-2 k_z}) \Leftrightarrow
 e^{-2 k_z} > 1 - 2 k_z ,
\label{e.bd}
\eeq
which clearly holds for all $k_z \neq 0$.

We next evaluate the limit of $F(k_z)$ as $k_z \to 0$ by
expanding the numerator of Eqn.~\eqref{e.cv00} in Taylor series:
\beqn
\sqrt{\frac{1}{k_z} \left[1 - \left(1 - 4 k_z + \frac{16 k_z^2}{2}
 - \frac{64 k_z^3}{3!} \right)  - \left( 4 k_z (1 - 2 k_z + \frac{ 4 k_z^2}{2}) \right)
 + \mathcal{O}(k_z^4) \right]} = \sqrt{\frac{8}{3}k_z^2 + \mathcal{O}(k_z^3)} .
\eeqn
Expanding the denominator likewise, we have
\beq 
F(k_z) = \frac{ \sqrt{ \frac{8}{3} k_z^2 + \mathcal{O}(k_z^3) } }
{ [ 1 - (1 - 2 k_z + \mathcal{O}(k_z^2))] } \to 
\sqrt{\frac{2}{3}} \mbox{  as  } k_z \to 0 .
\label{e.lm}
\eeq

The exponentials in the numerator and denominator of $F(k_z)$ decay
rapidly, so that it differs from $\sqrt{1/k_z}$ by less than $0.24\%$ for 
$k_z \ge 4$, implying that the slow monotonic decay $\sim k_z^{\frac{1}{2}}$
dominates for large $k_z$; see Fig.~\ref{fig:cvs}. However, the
behavior for smaller $k_z$ is more subtle and requires computation of
all terms in the Taylor series.

To prove monotonic decay throughout we use the fact that $F(k_z) > 0$
and show that the derivative of 
\beq
F^2(k_z) = \frac{ \left( \frac{1-e^{-4 k_z}}{2 k_z} - 2e^{-2 k_z}
 \right) }{(1-e^{-2 k_z})^2} 
\label{md1}
\eeq
is strictly negative for all $k_z >0$. Henceforth, for convenience, we
set $y = 2 k_z$ and compute
\beqr
\frac{d}{dy} [ F^2(y) ]
& = & \frac{ (1-e^{-y}) \left( - \frac{1}{y^2} + \frac{e^{-2y}}{y^2}
 + \frac{2e^{-2y}}{y} + 2e^{-y} \right) - 2e^{-y} \left(
 \frac{1-e^{-2y}}{y} - 2e^{-y} \right) }{(1-e^{-y})^3} \nonumber \\
& = & \frac{ 
\frac{-(1-e^{-y})(1-e^{-2y})}{y^2} - \frac{2 e^{-y}(1-e^{-y})}{y}
 + 2e^{-y}(1+e^{-y})}{(1-e^{-y})^3} .
\label{md2}
\eeqr
Since $(1-e^{-y})^3 > 0$ it suffices to show that the numerator
of Eqn~\eqref{md2} is negative, or, multiplying by $y^2 e^{3y}$ and
rearranging, that
\beq
1+e^{3y}+2ye^{2y} \eqdef L > e^{y} + e^{2y} + 2ye^{y} + 2y^2e^{y}
 + 2y^2e^{2y} \eqdef R .
\label{md3}
\eeq

We expand both $L$ and $R$ in Taylor series, obtaining
\beqr
L & = & 1 + \left(1 + 3y + \frac{(3y)^2}{2!} + \frac{(3y)^3}{3!}
 + \ldots + \frac{(3y)^j}{j!}  + \ldots \right) + 2y \left( 1+2y
 + \frac{(2y)^2}{2!} + \ldots + \frac{(2y)^{j-1}}{(j-1)!}
 + \ldots \right)  \nonumber \\
& = & 2 + 3y + \frac{9y^2}{2} + \frac{27y^3}{6} + \ldots + 2y
 + (2y)^2 + \frac{(2y)^3}{2!} + \ldots + \frac{(3y)^j}{j!}
 + \frac{(2y)^j}{(j-1)!} + \ldots  \nonumber \\
& = & 2 + 5y + \frac{17}{2} y^2 + \frac{17}{2} y^3
 + \frac{145}{24} y^4 + \frac{403}{120} y^5 + \ldots
 + \left( \frac{3^j + j2^j}{j!} \right) y^j + \ldots ;
\label{md4} \ \ \mbox{and} \\
R & = & 1 + y + \frac{y^2}{2!} + \frac{y^3}{3!} + \ldots
 + \frac{y^j}{j!} + \ldots + 1 + 2y + \frac{(2y)^2}{2!}
 + \frac{(2y)^3}{3!} + \ldots + \frac{2^j y^j}{j!} + \ldots
  \nonumber \\
& + & 2y + 2y^2 + \frac{2 y^3}{2!} + \frac{2 y^4}{3!} + \ldots
 + \frac{2 y^j}{(j-1)!} + \ldots + 2y^2 + 2y^3 + \frac{2y^4}{2!}
 + \ldots  + \frac{2 y^j}{(j-2)!} + \ldots  \nonumber \\
& + & 2 y^2 + 2^2 y^3 + \frac{2^3 y^4}{2!} + \ldots
 + \frac{2^{j-1}y^j}{(j-2)!} + \ldots \nonumber \\
& = & 2 + 5y + \frac{17}{2} y^2 + \frac{17}{2} y^3 + \frac{145}{24} y^4
 + \frac{403}{120} y^5 + \ldots + \frac{1 + 2^j + 2j^2 + 2^{j-1}
 j(j-1)}{j!} + \ldots \, .
\label{md5}
\eeqr
Note that the first 6 terms of $L$ and $R$, up to $\mathcal{O}(y^5)$,
are identical, and the 4 succeeding coefficients of $L - R$ up to
$\mathcal{O}(y^9)$ are strictly positive (specifically, $1/45, \,
1/30, \, 11/420$ and $1/70$). To show that all succeeding
coefficients are likewise positive, we make pairwise comparisons of
the six terms in the numerator of the general coefficient of $L - R$:
\beq
3^j + j2^j - [1 + 2^j + 2j^2 + 2^{j-1} j(j-1)] = [j2^j - 2j^2]
+ [j2^{j-1} - (1 + 2^j)] + [3^j - j^2 2^{j-1}] .
\label{md6}
\eeq
It can be checked that
\beqr
j2^j > 2j^2 & \Leftrightarrow & 2^j > 2j \ \ \mbox{for} \ \
 j \ge 3 ,  \label{md7a} \\
j2^{j-1} > 1 + 2^j & \Leftrightarrow & j > 2 + \frac{1}{2^{j-1}}
 \ \ \mbox{for} \ \ j \ge 3 , \label{md7b} \\
3^j > j^2 2^{j-1} & \Leftrightarrow & \left( \frac{3}{2} \right)^j
> \frac{j^2}{2} \ \ \mbox{for} \ \ j \ge 10 ;
\label{md7c}
\eeqr
thus, all coefficients of terms greater than $\mathcal{O}(y^5)$
are strictly positive, completing the proof. \hspace{0.5cm} $\Box$

\section{Additional Figures}\label{app:figures}

\begin{figure}[ht!]
\vspace{-0.75cm}
\includegraphics[width=\textwidth]{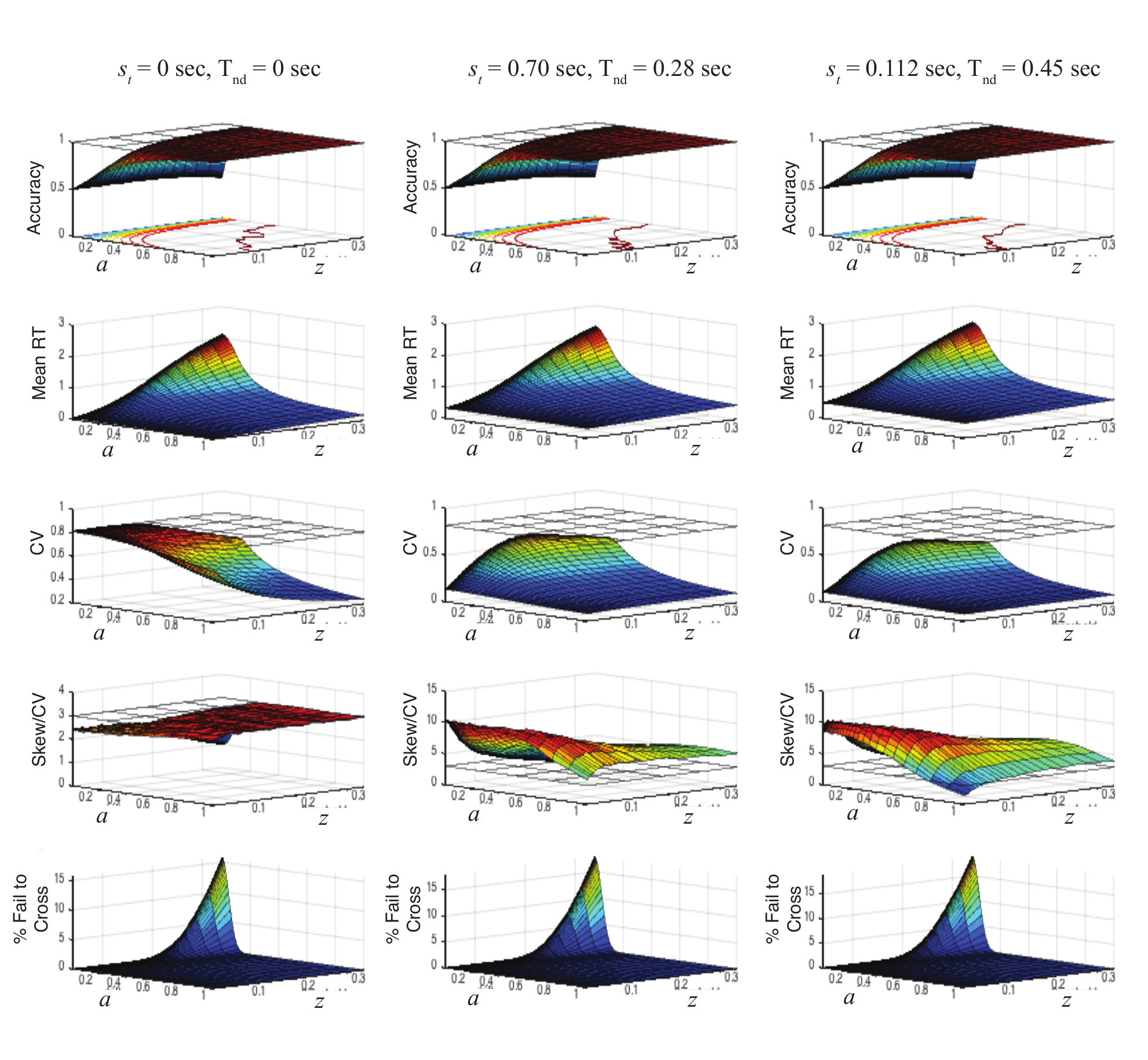}
\vspace{-1.25cm}
\caption{Effects of range $s_t$ of uniformly distributed $\subscr{\rm T}{nd}$. High values of $s_t$ (25\% of $\subscr{\rm T}{nd}$)
produce little effect compared with that of increasing $\subscr{\rm
T}{nd}$.
}
\label{fig:pat-sims-sTer}
\end{figure}

\begin{figure}[ht!]
\vspace{-0.75cm}
\includegraphics[width=\textwidth]{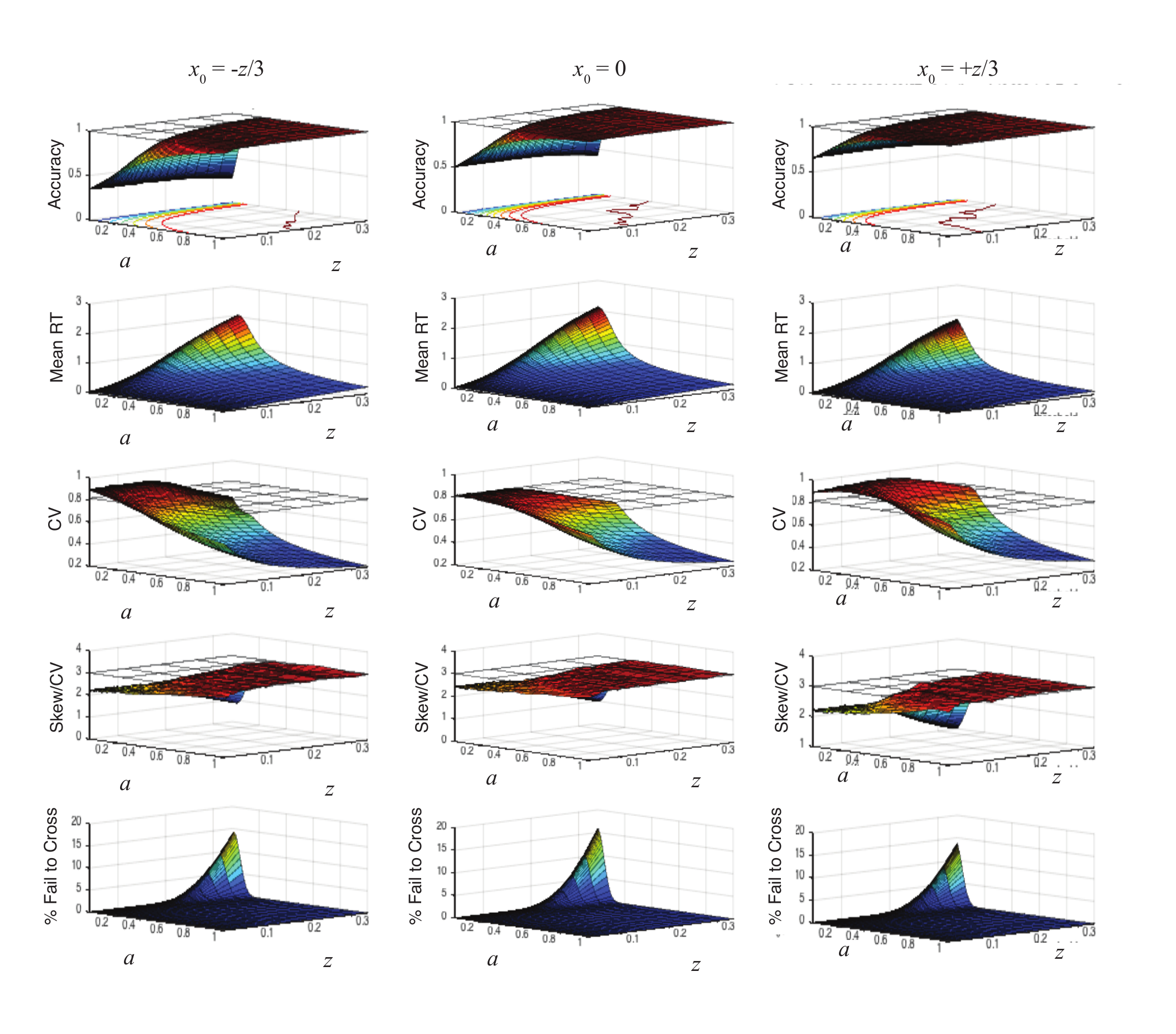}
\vspace{-1.5cm}
\caption{Effects of mean starting point $x_0$. Positive and negative biases from $0$ produce increases in the CV of decision times,
raising it above $\sqrt{2/3}$, but with very little change in SCV.
Note that $\subscr{\rm T}{nd} = 0$ in all plots. 
}
\label{fig:pat-sims-StartingPoint}
\end{figure}

\begin{figure}[ht!]
\vspace{-0.75cm}
\includegraphics[width=\textwidth]{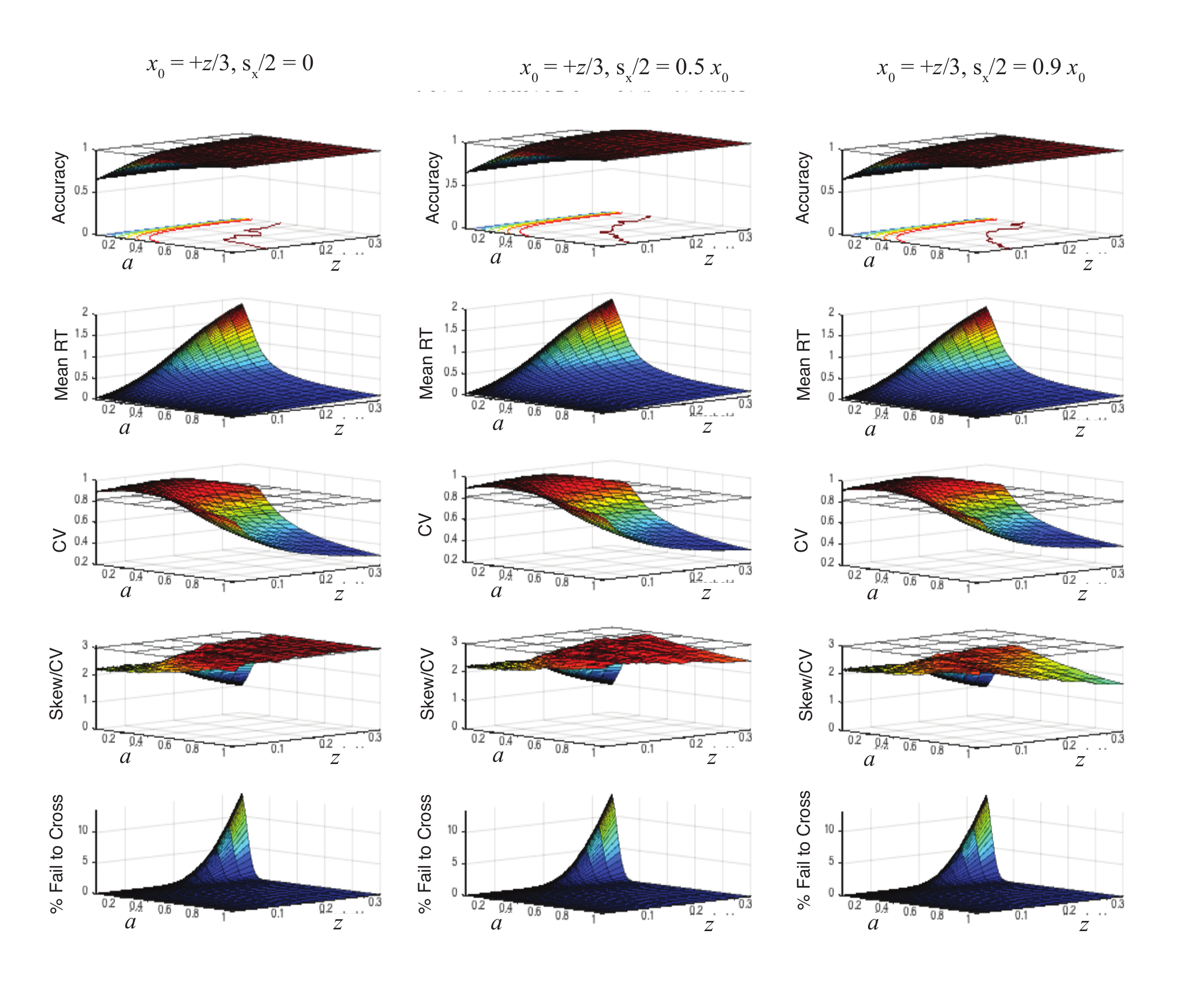}
\vspace{-1.25cm}
\caption{Effects of range of starting point distribution
$s_x$. Increasing $s_x$ has little
effect, except at unrealistic parameter combinations of high threshold
and high drift, where a decrease in SCV is evident. 
Mean starting
point in all plots is $+z/3$, 
2/3 of distance from lower to upper
threshold biased toward the correct response.  
Indeed, $s_x$ is used to explain differences in conditional decision times, and any apparent lack of effect in Fig.~\ref{fig:pat-sims-StartPtWidth} is due to the fact that only unconditional quantities are plotted. 
}
\label{fig:pat-sims-StartPtWidth}
\end{figure}

\begin{figure}[ht!]
\vspace{-0.75cm}
\includegraphics[width=\textwidth]{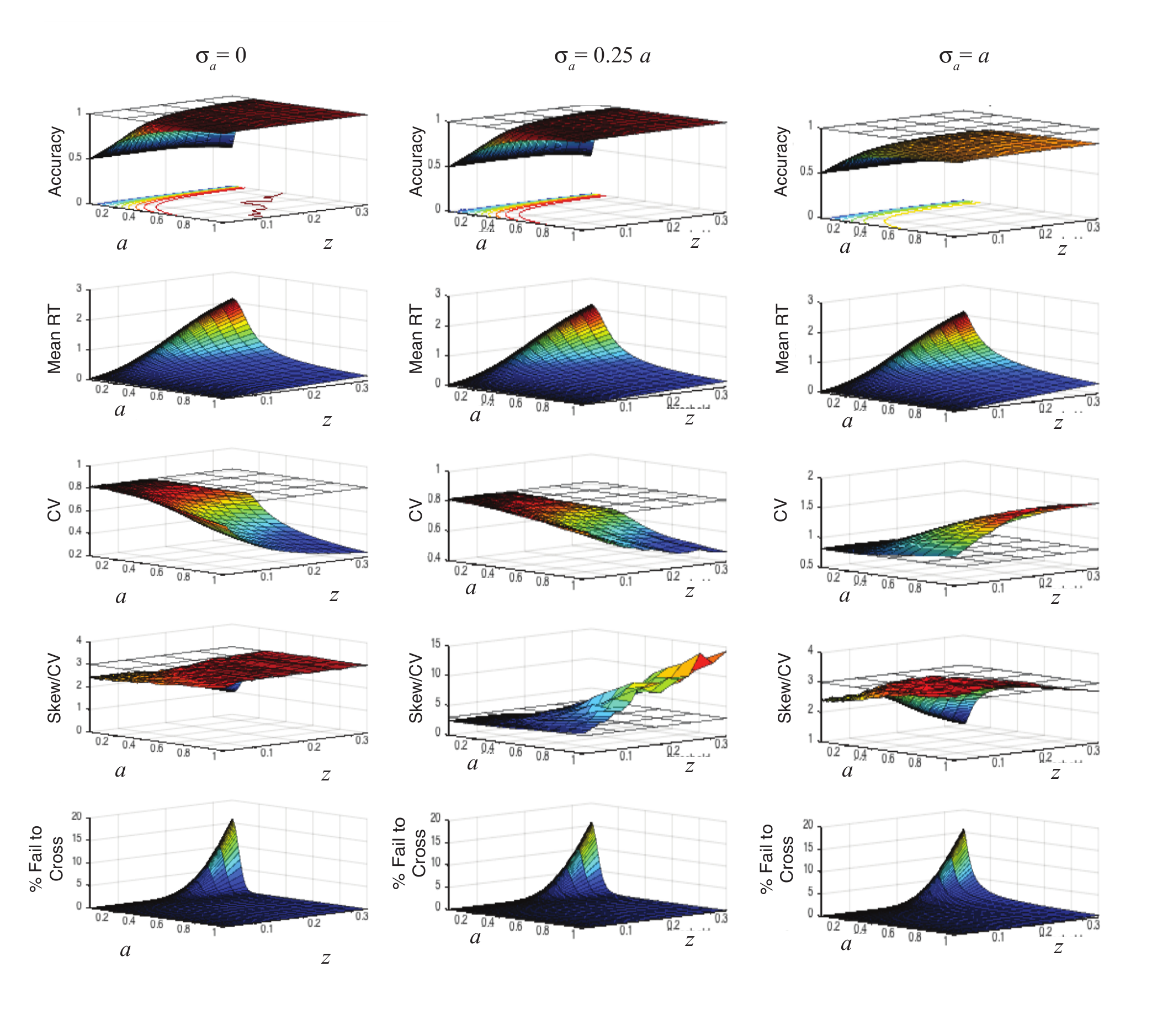}
\vspace{-1.5cm}
\caption{Increases in drift variability $\sigma_a$ across
trials reduce accuracy, increase CVs, and initially raise and then
lower SCV ratios; overall mean RTs are little changed. 
}
\label{fig:pat-sims-Eta}
\end{figure}

In this section we present some additional simulations for the extended and pure DD models. 
\begin{table}[htb!]
\begin{center} \begin{tabular}{|c|c|c|c|c|c|}
	\hline
   & $x_0$ & $s_x$ & $\sigma_a$ & $\expt[\subscr{\rm T}{nd}]$ (sec) & $s_t$ (sec) \\
	\hline\hline
Fig.~\ref{fig:pat-sims-Ter} & 0 & 0 & 0 & $0, \, 0.28, \, 0.45$ & 0 \\
	\hline
Fig.~\ref{fig:pat-sims-sTer} & 0 & 0 & 0 & $0, \, 0.28, \, 0.45$ & $0.25 \ \expt[\subscr{\rm T}{nd}]$ \\
	\hline
Fig.~\ref{fig:pat-sims-StartingPoint} & $-z/3, \, 0, \, +z/3$ & 0 & 0 & 0 & 0 \\
	\hline
Fig.~\ref{fig:pat-sims-StartPtWidth} & $+z/3$ & $0, \, z/3, \, 0.6 z$ & 0 & 0 & 0 \\
	\hline
Fig.~\ref{fig:pat-sims-Eta} & 0 & 0 & $0, \, 0.5 a, \, a$ & 0 & 0 \\
	\hline
\end{tabular}
\end{center}
\caption{Parameter values for extended DDM simulations.  Here $s_x$ is the range of the starting point distribution, $\sigma_a$ the standard deviation of drift rate, and $s_t$ the range of the non-decision time distribution. These parameters are given as fractions of threshold, mean drift and mean non-decision time respectively. }
\label{tab:extparams}
\end{table}
Simulations were performed using the RTdist package for graphical processing unit (GPU) with the same details as outlined in \S\ref{s.extddm}. 
In Figs.~\ref{fig:pat-sims-sTer}-\ref{fig:pat-sims-Eta},  the noise level
was fixed at $\sigma = 0.1$ and we varied mean drift $a$ and threshold
$z$ over the ranges $[0.1, 1.0]$ and $[0.05, 0.3]$ respectively. Each
figure shows accuracy, mean RT, CV, skewness to CV ratio (SCV) and the
percentage of trials that failed to cross threshold within  5 secs. (The latter quantity is similar in all cases: it remains small
except for low drift and high threshold, where it rises to $15-20\%$.)
Other parameters chosen for these figures are listed in
Table~\ref{tab:extparams}. Note that the center
column of Fig.~\ref{fig:pat-sims-StartingPoint} and  the left hand columns of Fig.~\ref{fig:pat-sims-Eta} show results for the pure DDM with $\subscr{\rm T}{nd} = 0$, and thus provide standards for comparison with other cases. 

Figs.~\ref{fig:pat-sims-StartingPoint} and
\ref{fig:pat-sims-StartPtWidth} show that deviations in mean starting
point in either direction lead to increases in CV, but with little
effect on SCV ratios. Introducing trial-to-trial variability raises
CVs for $x_0 = 0$, and yields lower SCV ratios for high thresholds
and drift rates.
 Fig.~\ref{fig:pat-sims-Eta} shows that trial-to-trial
variability in drift rates reduces accuracy, that CVs increase
substantially for high variability, and that SCV ratios initially
increase and then decrease with variability.

The remaining figures show the effects of variability in $\subscr{\rm
T}{nd}$, of starting point and its variability, and of variability in
drift rates. In  Fig.~\ref{fig:pat-sims-sTer} we keep the ratio
$s_t/\subscr{\rm T}{nd}$ constant at $0.25$ and use the same values of
$\subscr{\rm T}{nd}$ as in Fig.~\ref{fig:pat-sims-Ter}, revealing
similar effects to those of Fig.~\ref{fig:pat-sims-Ter}, except for
SCV, which increases as $\subscr{\rm T}{nd}$ increases.

\end{document}